**Enterprise Architecture as a Dynamic Capability for Scalable and Sustainable Generative AI Adoption: Bridging Innovation and Governance in Large Organisations**

by

Alexander Ettinger

IB9NLD

A dissertation submitted in partial fulfilment
of the requirements for the degree of
Masters of Business Administration
(Global Online MBA)
at the Warwick Business School
2025

Academic Supervisor: Mo Moeini (mo.moeini@wbs.ac.uk)

Alexander Ettinger

alexander.ettinger@warwick.ac.uk

Word Count: 14800 words[1]

---

[1] Excluding captions, references and footnotes

## Acknowledgements

First, I want to sincerely thank my academic supervisor, Mo Moeini, for his guidance and support throughout this journey. His advice and encouragement made a big difference and helped me complete this research on time.

I also want to thank my colleagues at SAP LeanIX. A special thanks to Nandor Orvos and Julian van Megen for trusting me with this MBA journey, and to Dominik Rose, Wim Frankena, and Andre Christ for helping shape the objectives and problem statements for this dissertation. I'm also grateful to my other colleagues who took the time to help refine the interview guide, it really meant a lot.

Finally, I want to thank my wife, Alix, and my young son, Timothée, for being so patient and understanding, especially during the more stressful parts of the MBA. Your love and support kept me going.

**Table of Contents**

# List of Tables

# List of Figures

## List of Appendices

## Usage of (Generative)AI for this Dissertation

- Microsoft Stream utilises AI-driven speech-to-text technology to generate transcripts from recorded Microsoft Teams meetings. OpenAI GPT4o (OpenAI, 2025) was then used to translate one of the interviews & key quotes from French to English for example.
- OpenAI GPT4o (OpenAI, 2025) was used to post-process the interview transcripts. For example, to remove all shapes generated by Microsoft Stream in the word documents.
- OpenAI GPT4o (OpenAI, 2025) was used in the literature review to help expand search terms and improve search quality by considering additional/similar keywords leading to better search result in Google Scholar, Science Direct etc. Refer to Table 2Table 3.
- OpenAI GPT4o (OpenAI, 2025) was used to generate charts & trends for Figure 2.
- OpenAI GPT4o (OpenAI, 2025) was used for the (later rejected) sentiment analysis of the interview transcripts (refer to the Data Analysis section and Appendix C. GPT4o Sentiment Analysis Method).
- OpenAI GPT4o (OpenAI, 2025) was also used to synthesize and create summaries of text. For example, for Table 8. Summary of Literature Review Topics in Appendix D. Summary of Literature Review Topics and RQ Validity.
- OpenAI GPT4o (OpenAI, 2025) and specific AI research assistants including Consensus and ScholarGPT, were also used to find pertinent forward & backward citations, generate Harvard style citations and provide quick access to scholarly resources.

# Abstract

Generative Artificial Intelligence (GenAI) is a powerful new technology with the potential to boost innovation and reshape governance in many industries. Nevertheless, organisations face major challenges in scaling GenAI, including technology complexity, governance gaps and resource misalignments. This dissertation explores how Enterprise Architecture Management (EAM) can meet the complex requirements of GenAI adoption within large enterprises.

This research, following a Systematic Literature Review (SLR) and the qualitative analysis of 16 semi-structured interviews with experts, examines the relationships between EAM, Dynamic Capabilities and GenAI technology adoption. The SLR identified key limitations within existing EA frameworks, particularly their inability to fully address the unique requirements presented by GenAI. The interviews were analysed using the Gioia methodology, and discovered critical enablers and barriers to organisational adoption of GenAI across industries.

The findings indicate that EAM, when theorised as sensing, seizing, and transforming Dynamic Capabilities, can enhance the adoption of GenAI by improving strategic alignment, governance frameworks, and by promoting organisational agility. However, the study also underscores the necessity of tailoring EA frameworks to address GenAI specific challenges, including low data governance maturity, and finding the right balance between innovation and compliance.

The findings led to several conceptual frameworks that have been developed to provide practicable guidance to EA leaders for aligning GenAI maturity with organisational readiness.

This dissertation contributes to both academic literature and industry practices by developing the understanding of EA's role in bridging innovation and governance in disruptive technology environments.

# Chapter 1 – Introduction

This chapter provides an overview of the thesis structure, starting with the motivation and problem statements that frame the research. It then outlines the objectives and research questions, followed by a summary of the methodology. Finally, the chapter concludes with an outline of the remaining chapters.

## 1.1 Motivation

Generative Artificial Intelligence or GenAI, epitomised by advancements such as ChatGPT, is revolutionizing economic and technological paradigms while fundamentally altering perceptions of creativity and human-machine collaboration. Early economic projections underscore its substantial potential, with McKinsey (2023) projecting that GenAI's contribution to the global economy could exceed $4 trillion, reflecting its transformative potential. In response, tech giants have significantly increased their investments, viewing GenAI as a strategic "big bet" (FT, 2024a). However, these firms also acknowledge the duality of GenAI, recognizing both its opportunities and inherent risks (FT, 2024b). Despite the enthusiasm, substantial barriers persist, including inconsistent adoption rates and limited return on investment (ROI), which are exacerbating frustrations (Economist, 2024). These challenges, coupled with high Total Cost of Ownership (TCO), have elevated discussions on proving GenAI's value to the boardroom level.

A notable disparity exists between the rapid personal adoption of GenAI, exemplified by ChatGPT's user growth, and the slower, more fragmented pace of organisational adoption. A study by SAP LeanIX (2024) reveals that while 80% of companies utilise GenAI, only 20% have implemented a cohesive governance framework. This suggests that many GenAI initiatives remain isolated pilot projects, lacking integration with broader enterprise strategies and governance practices (Rittelmeyer et al., 2021). Organisations are navigating the delicate balance between the transformative potential of GenAI and the regulatory and ethical challenges it presents, such as those addressed by the EU AI Act. The enforcement of these regulations in 2026, regardless of an organisation's risk level or type, represents a critical development.

This landscape underscores the urgent need for scalable, compliant, and value-driven frameworks. Although Enterprise Architecture (EA) is integral to aligning technology with strategy, its role as a Dynamic Capability in addressing GenAI's complexities remains underexplored. This gap presents a compelling opportunity to examine how EA can facilitate GenAI's integration, governance, and scaling.

1.2 Problem Statement and Research Question

Enterprise Architecture is pivotal in aligning business and IT, leveraging dynamic capabilities to enable technological and organisational transformation (van de Wetering, 2022). By enabling firms to reconfigure assets and sustain competitive advantage (Grant, 1991), EA serves as a critical resource. However, traditional EA frameworks struggle to address GenAI's unique challenges, including its rapid innovation cycles and broad applicability (Rittelmeyer et al., 2021). These limitations demand new approaches to scalability, governance, and value realization.

Shanks et al. (2018) integrate dynamic capabilities into EA frameworks, focusing on IT and business-driven changes. However, their work does not extend to disruptive technologies like

GenAI, leaving a gap in understanding how EA can sense, seize, and transform opportunities in this context (Rittelmeyer et al., 2021; SAP LeanIX, 2024). This research explores how EA can evolve to meet these demands, positioning it as a Dynamic Capability that supports the scalable and sustainable adoption of GenAI.

**RQ:** *How can Enterprise Architecture Management, as a Dynamic Capability, facilitate the scalable and sustainable adoption of Generative AI within large organisations?*

Dynamic Capability (DC) theory provides a useful framework for understanding the role of Enterprise Architecture in the adoption of GenAI. This dissertation applies the core dimensions of DC theory, sensing, seizing, and transforming, to explore how EA can actively identify new opportunities, evaluate their relevance to the organisation, and adapt resources and frameworks strategically. This approach aims to facilitate the scalable and sustainable adoption of GenAI.

By integrating these theoretical constructs with empirical data, the study develops a deeper understanding of EA's potential to bridge the gap between governance and innovation. This is particularly important given the unique challenges presented by GenAI.

1.3 Overview of Research Methodology

In late 2024, 16 semi-structured interviews were conducted with experts in Enterprise Architecture and GenAI to investigate how EA frameworks and governance models can enable scalable and sustainable GenAI adoption. Participants were selected based on diverse industry representation and their depth of technical knowledge, whether gained through roles in EA teams, managing GenAI governance, or directly implementing GenAI solutions.

A Systematic Literature Review (SLR), completed earlier that year, informed the research design and interview framework, ensuring alignment with current academic and practitioner insights. The qualitative analysis followed the Gioia methodology, facilitating the structured identification of first-order concepts, second-order themes, and aggregate dimensions. This systematic approach ensured the study captured the nuances of GenAI adoption while addressing the research questions with methodological rigor.

1.4 Summary of contributions

A notable research implication is the demonstration of how EA's sensing, seizing, and transforming capabilities can tackle issues like fragmented governance and scaling innovation. This contributes to the theoretical understanding of EA in disruptive contexts. From a practical standpoint, the study presents conceptual frameworks that help organisations align EA maturity with readiness. These frameworks offer actionable insights to assess current states, prioritise investments, and strategically scale GenAI adoption.

1.5 Dissertation Structure

This thesis is structured into seven chapters. Chapter 2 introduces the intersection of Generative AI (GenAI) adoption and Enterprise Architecture (EA), offering a conceptual perspective and providing foundational context on the factors influencing GenAI adoption. Chapter 3 presents a systematic literature review, exploring the diverse implications of GenAI adoption and the critical role of EA in supporting its integration. Chapter 4 outlines the research methodology, while Chapter 5 presents the findings from the semi-structured interviews, organised using a defined conceptual framework. Chapter 6 discusses the research questions in depth, analysing the implications of the findings, addressing study limitations, and identifying avenues for future research. Finally, Chapter 7 concludes the study with a synthesis of its contributions.

# Chapter 2 – Background

This chapter begins by defining Generative AI, Governance Frameworks and Enterprise Architecture and concludes by presenting the theoretical foundations underpinning this research.

## 2.1 Defining Generative Artificial Intelligence

Generative Artificial Intelligence (GenAI) is a subset of Artificial Intelligence (AI) that leverages advanced deep learning to create novel and practical outputs, unlike traditional AI's rule-based automation (Bankins et al., 2024). GenAI builds on machine learning, representing one of its most dynamic applications. García-Peñalvo (2023) maps the GenAI landscape, categorising generated content, tasks, techniques, and application domains (Figure 1).

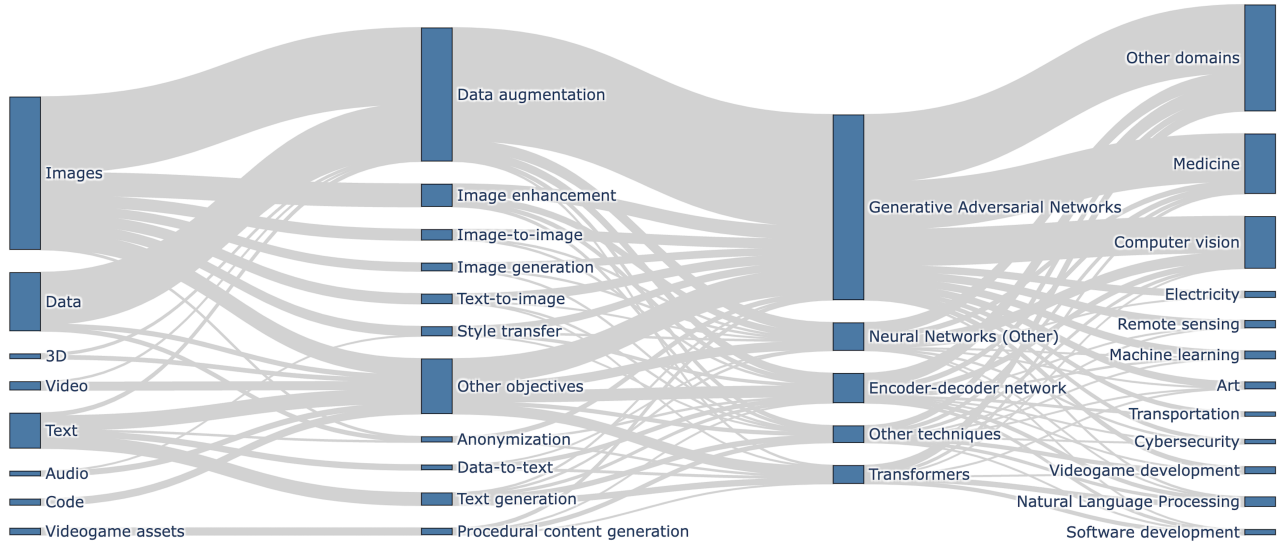

Figure 1. The connections between the type of generated content, the task, the AI technique employed, and the application domain (Source: García-Peñalvo et al., 2023)

The rising prominence of models like Generative Adversarial Networks (GANs) has expanded GenAI's reach beyond AI specialists, including EA. As shown in Figure 2, GenAI gained widespread attention by November 2022, driven by the rapid growth of Large Language Models (LLMs) since 2019. For EA, understanding these trends is crucial to aligning strategies with technological shifts, leveraging emerging capabilities, and positioning organisations to harness GenAI's transformative potential effectively.

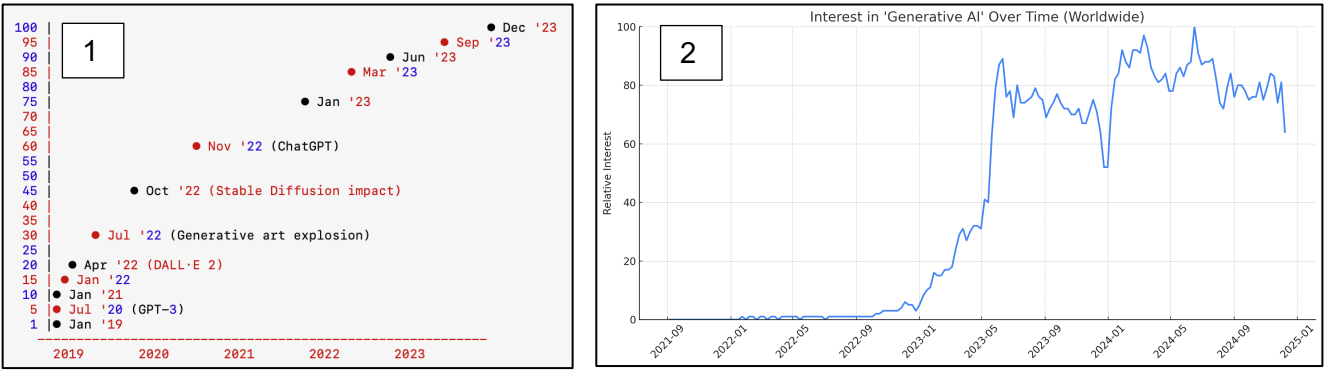

Figure 2."Generative AI" trends: 1. Conceptual trend line approximation based on OpenAI LLM data (Open AI, 2025) 2. Based on actual usage data sourced from Google Trends. Y axis is relative interest.

2.2 **Governance frameworks**

Governance frameworks in large organisations are structured systems that align strategic goals, operational processes, and compliance with both internal and external regulations. They incorporate established standards like COBIT and ISO 17799, which aid in risk management and standardise procedures. This fosters transparency and adaptability in decision-making (Robles et al., 2023).

GenAI governance frameworks ensure the ethical, transparent, and responsible use of GenAI technologies. They address unique challenges such as algorithmic accountability and trust (Schneider et al., 2024b). These frameworks are particularly relevant under regulations like the EU AI Act, which sets a global standard for AI governance (Gstrein et al., 2024).

2.3 **Defining Enterprise Architecture Management and EA Frameworks**

Enterprise Architecture (EA) serves as a strategic framework for aligning an organisation's business strategy, processes, information and IT infrastructure to achieve its objectives. By offering a holistic integrated view across the organisation, EA facilitates informed decision-making, optimised resource allocation and effective adaptation to evolving business needs. (Kotusev et al., 2021). Enterprise Architecture Management (EAM) takes this a step further by incorporating governance, strategic planning, development, communication and controlling (Weiss et al., 2012) to guide the continuous evolution of EA, enabling a more efficient integration of emerging technologies like Cloud and AI into organisational structures (Coltman et al., 2015). To summarise, EA provides the foundational framework for aligning GenAI with organisational goals, while EAM ensures its sustainable integration through governance and continuous adaptation.

In the context of GenAI, two key dimensions emerge: EA for (Gen)AI, which applies EA frameworks to integrate GenAI solutions into organisations, addressing challenges like scalability and governance (Stecher et al., 2020); and (Gen)AI in EA, where GenAI transforms EA strategies with enhanced modelling, workflows and technology optimisations (Rittelmeyer et al., 2021).

## 2.4 Defining Dynamic Capability theory

This research is grounded in two key theoretical underpinnings. The first draws from organisational theories that emphasise the importance of innovation and adoption, such as Rogers' (1995) Diffusion of Innovation (DoI) theory, which positions EA as a critical facilitator of technology adoption and organisational success in a competitive environment (Weiss et al., 2012). The second, Dynamic Capabilities (DC), is drawn from the RBV theory (Penrose, 1959), which views EA capabilities as intangible, valuable, rare, inimitable and firm embedded organisational resources leading to sustained competitive advantage. Dynamic Capabilities theory has been defined as "appropriately adapting, integrating, and reconfiguring internal and external organisational skills, resources, and functional competences toward changing environment." (Teece et al., 1994, p.2).

DC routines are repeatable organisational processes that help adapt to changing environments (Teece, 2007). They include:

a) Sensing: Identifying opportunities and threats through market analysis, technological scanning, and understanding customer needs.
b) Seizing: Selecting and capturing opportunities by aligning initiatives with organisational strengths and allocating resources effectively.
c) Transforming: Reconfiguring resources and structures to implement innovations, ensuring flexibility and avoiding organisational inertia.

## Chapter 3 – Literature Review

This chapter presents the findings of a Systematic Literature Review conducted in the fall of 2024, outlining the defined concept spaces, domains of interest, methodology, data collection, and finally the analysis of key papers to inform the subsequent research methodology.

3.1 **Review Methodology**

The author has chosen the Systematic Literature Review (SLR) methodology (Tranfield et al., 2003; Denyer et al., 2009), which has advantages compared to traditional unstructured methods, and that align well with the research objectives. This includes the need to capture an up-to-date literature review, with a systematic approach ensuring a review that is both comprehensive and current, whilst minimising biases by choosing to follow a transparent search and selection process. Furthermore, this method is very effective at identifying and categorising key themes, leading to a more structured data collection and thematic analysis. Lastly, in the highly dynamic research environment of GenAI, this systematic approach will help synthesise existing studies and identify under-research areas, allowing for future research directions proposals.

3.2 **Concept Space**

Concept spaces and Venn diagrams are effective tools for visualising intersections and systemising complex relationships (Novak, 2004). The author identified three primary concept spaces: (1) EAM, (2) GenAI, and (3) Technology Adoption, developed during internal workshops with SAP LeanIX's CEO and SVP Product Management. These align with a (1) People/Discipline, (2) Technology, and (3) Process-driven approach.

As shown in Figure 3, two sub-spaces were identified during exploratory analysis: Dynamic Capabilities, intersecting (1) and (3); and the trade-off between Innovation and Governance, linked to Technology Adoption (3) and informed by Innovation Diffusion theory. The broader AI technology space was also considered for its relevance to GenAI. This research focuses on the intersections of (1), (2), and (3), where overlaps are limited but hold potential for transferability, aiming to provide new insights into these domains.

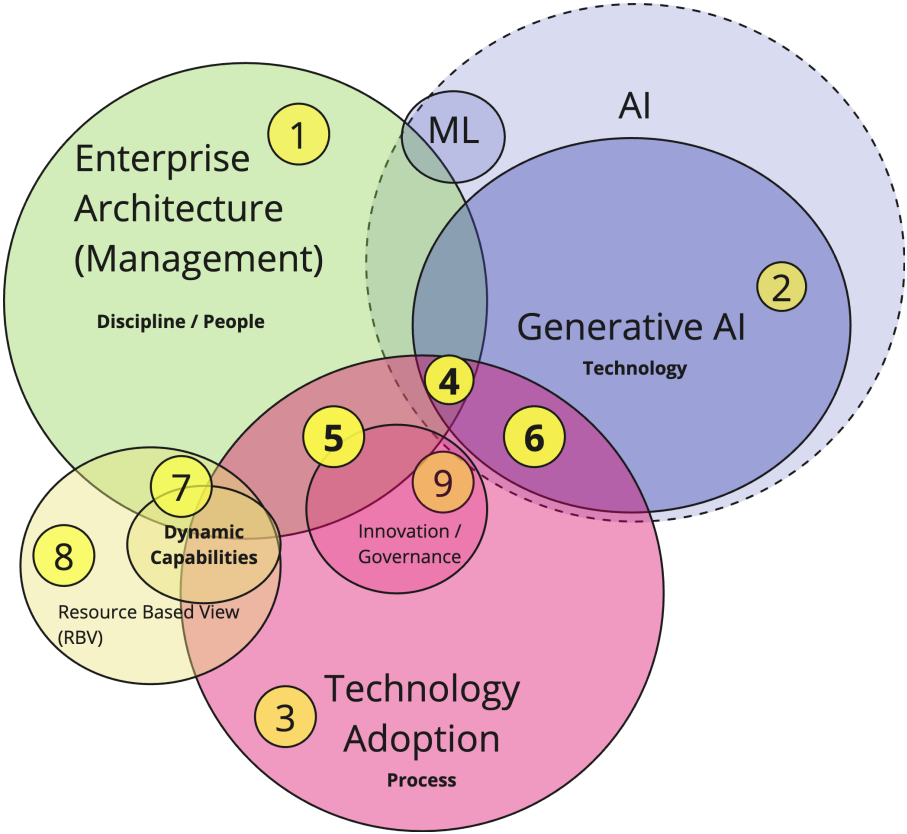

Figure 3. Concept Space Venn Diagram

3.3 **Structured Literature Review**

The data collection for this Systematic Literature Review (SLR) was conducted in the fall of 2024, ensuring the inclusion of the most up-to-date research on GenAI and as a preparation for the research method. This review investigates the diverse implications of GenAI adoption, examining its perceived benefits, associated challenges, and the critical role of EA in supporting its integration. Notably, no prior SLR has specifically addressed this topic, with existing reviews primarily focusing on AI adoption before the emergence of GenAI in 2021 (Stecher et al., 2020; Rittelmeyer et al., 2021).

*3.3.1 Research Objectives*

The research objectives were defined based on concept spaces and analysis outlined in section 3.2, now summarised in Table 1. The primary domains are defined in the Background section of the thesis. These objectives are organised around the intersections illustrated in the Venn diagram. The results characterise the multi-dimensional issues of this research, with each objective linking to both theoretical and practical gaps in research.

Table 1. Research objectives based on concept space analysis (The Ref Nr. Is referring to the defined Venn diagram intersections)

| Ref Nr. | Research objective |
|---|---|
| 4 | Research of the role of EA in large scale GenAI adoption |
| 5 | Research of the role of EA in novel technology adoption |
| 6 | Research on the role of GenAI in driving innovation / governance frameworks |
| 7 | Research on the role of EA to drive organisational wide Dynamic Capabilities to enhance technology adoption |

Figure 4 illustrates conceptually interconnected research domains, with Domain (4) serving as the foundation for integrating the other three aims. This central domain underscores a critical theoretical and practical gap, emphasising the need for a cohesive framework to support large-scale GenAI adoption. Domains (5), (6), and (7) build upon this foundation, interacting to provide a multi-dimensional perspective on how EA facilitates innovation, governance, and organisational agility. Exploring the interplay among these domains offers a comprehensive understanding of EA's foundational role in driving transformative technological adoption and dynamic organisational capabilities.

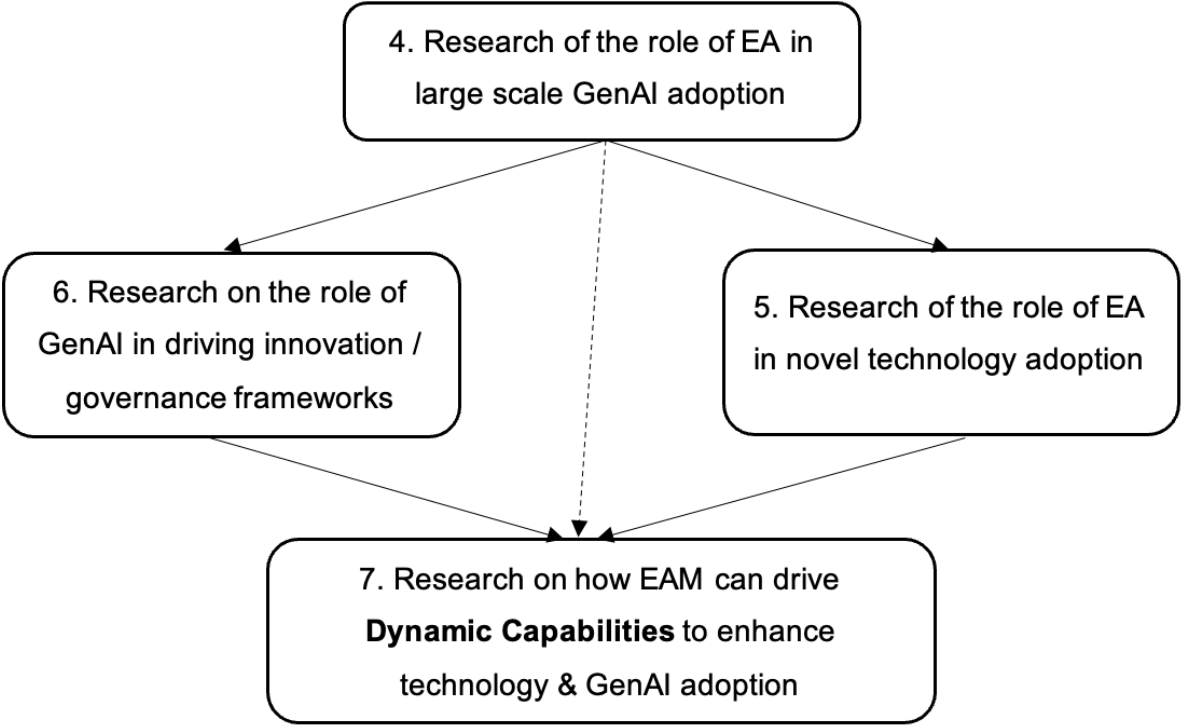

Figure 4. Conceptual Map of Research Objectives

This research, starting from the outlined conceptual interactions, aims to conduct a comprehensive review of GenAI adoption in private organisational settings, with a novel focus on its relationship with EA. A structured approach, addressing both theoretical and practical challenges, will begin by categorising existing research into key themes and identifying areas and gaps where EA has been underexplored. This approach offers an initial conceptual framework to be validated in later qualitative research.

### *3.3.2 Conceptual Boundaries*

The conceptual boundaries, inclusion and exclusion criteria are summarised in Figure 6.

a. GenAI adoption in Private vs Public Organisations in Europe

This research is conceptually limited to exploring GenAI adoption within private organisations, excluding public entities that operate under distinct regulatory, bureaucratic, and policy-driven conditions (Boyne, 2002). Public organisations often require tailored governance models and innovations that lack the urgency, agility, and competitive drivers characteristic of the private sector. In private organisations, EA is leveraged as a unique resource for competitive advantage (Grant, 1991) and as a DC to foster innovation (Teece et al., 1997). In contrast, EA adoption in public organisations faces greater resistance and its benefits are less apparent (Sundberg et al., 2023).

The study also integrates intra- and inter-organisational governance considerations, with inter-organisational governance addressing external stakeholders, including government bodies (Kongsten et al., 2024). The EU AI Act, introduced by the European Commission, directly impacts private organisations, shaping GenAI adoption trajectories (Gstrein et al., 2024). Focusing on private European organisations is thus logical, given their direct exposure to these regulatory frameworks. Although leading AI firms are predominantly US-based, European regulatory differences, especially regarding privacy, create unique challenges for cross-border collaborations and compliance (Cath, 2018). These distinctions underscore the need for region-specific research into GenAI adoption.

b. Transferable AI research to GenAI

Artificial Intelligence Governance (AIG) has emerged as a rapidly expanding research area, as noted by Gstrein (2024). Despite its growth (refer to Figure 5), scholars and regulators struggle to keep pace with the increasing volume of publications, principles, and standards in this dynamic domain (Cath, 2018). Since 2021, GenAI specific research has expanded upon foundational AI topics like data privacy and security, making many earlier insights transferable and relevant for this study. However, GenAI's unique characteristics significantly broaden the scope of AI adoption and governance, further evolving the field and underscoring its importance.

c. Applicability of Dynamic Capability and Innovation Diffusion theories

Building on the selected domains of EA, Innovation, Technology Adoption, and Governance, the RBV (Grant, 1991) and DoI theory (Rogers, 1995) offer valuable theoretical frameworks, additionally capturing the current environmental dynamics with an emphasis on the Dynamic Capabilities of firms (Teece et al. 1997). These perspectives help elucidate the factors driving GenAI adoption and the role of EA as a critical organisational resource in facilitating this adoption.

d. Grey literature

The rapid advancements in GenAI and GenAI literature (refer to Figure 7), underline the fields dynamic and evolving nature. The inclusion of grey literature, especially pre-published and yet to be peer-reviewed articles, helps bridge the gap and capture the most recent development and practical applications. In addition, grey literature also captures larger adoption studies, filling niche gaps, and provides practitioner perspectives that align well with the overall research aims and complementing academic papers.

e. Industries

While studies indicate significant variations in GenAI adoption across industries (The Economist, 2024; Jaffe et al., 2024) its unique characteristics enable use cases that transcend industry boundaries. GenAI's transformative potential, particularly its capacity to disrupt traditional business models, ensures that no industry remains unaffected by its impact (Dencik et al, 2023). Therefore, at these early stages of research, it is crucial to capture the role of EA across industries to also inform future research. Focusing on broader organisational adoption strategies and frameworks rather than individual applications allows for a more generalised understanding applicable across industries.

### *3.3.3 Inclusion Criteria: Cover Period and Search Terms*

The review spans literature published between January 2021 and October 2024, coinciding with the rise of large-scale language models in late 2021 (see Figure 5). This period ensures comprehensive coverage of studies across domains and organisational sizes, providing a holistic view of GenAI adoption. Pre-2021 literature is selectively included for context, such as GDPR's influence on regulatory frameworks shaping GenAI adoption. Dynamic trackers on platforms like Google Scholar captured critical papers published up to October 2024.

Search terms, refined with OpenAI's GPT4o (OpenAI, 2025), aligned with the research objectives and organisational boundaries. Multiple databases were used: Google Scholar for broad sources, including preprints, and Scopus and Web of Science for high-quality, peer-reviewed journals in English. This approach ensured both breadth and depth, reflecting the study's conceptual boundaries.

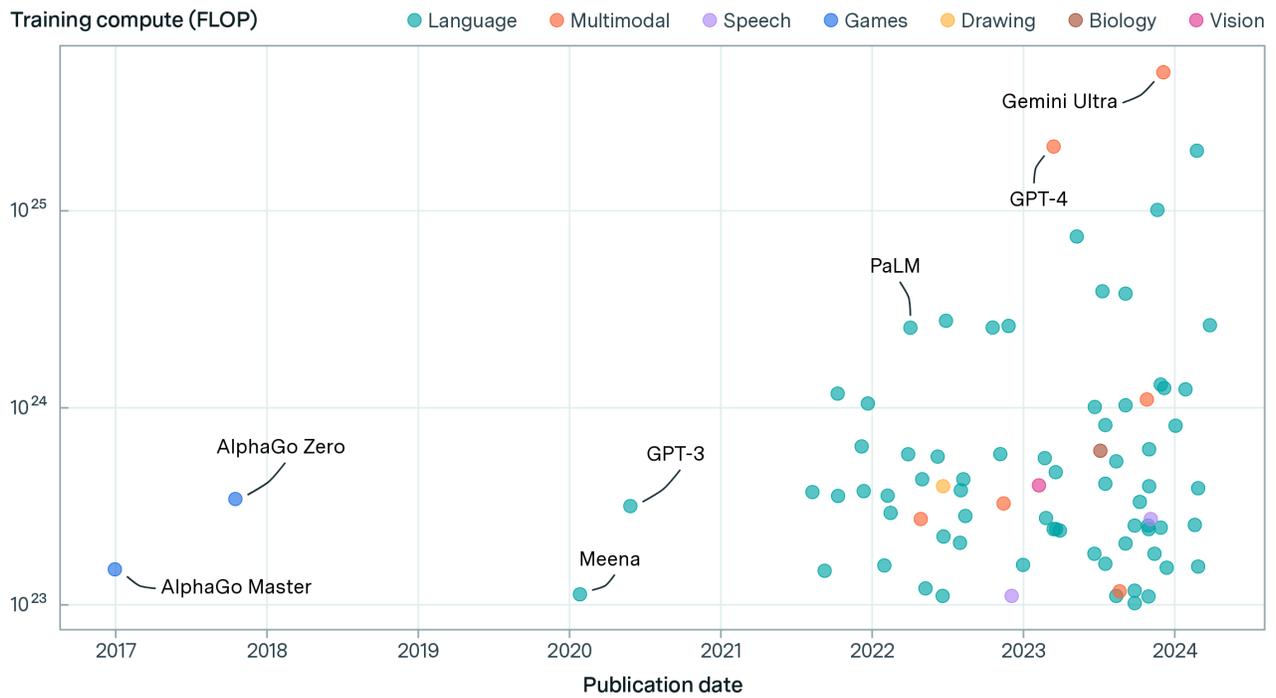

Figure 5. Large-scale models by domain and publication date (Source: Rahman, 2024)

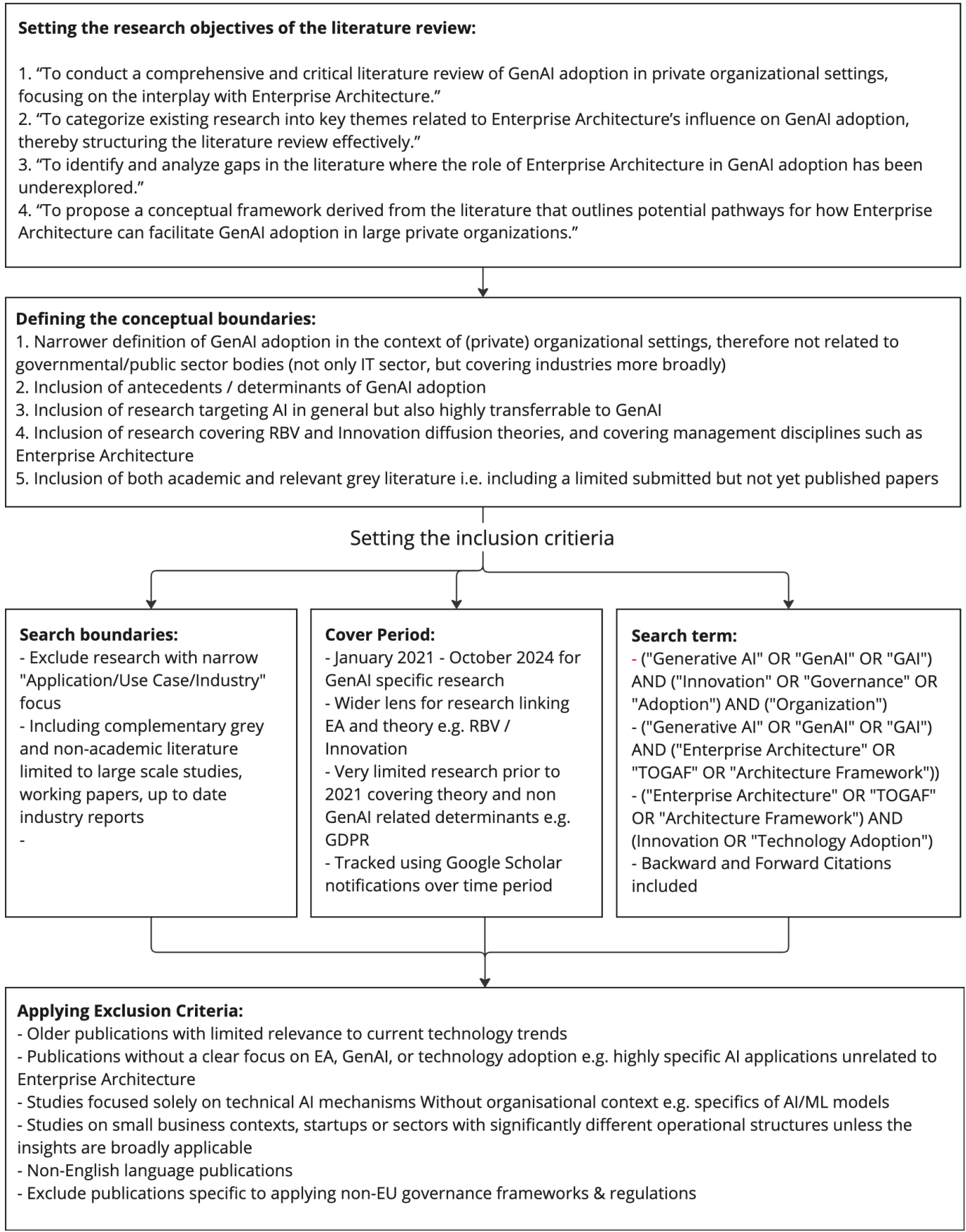

Figure 6. Structured literature review method

## 3.4 Review data collection analysis

### 3.4.1 Literature highlights

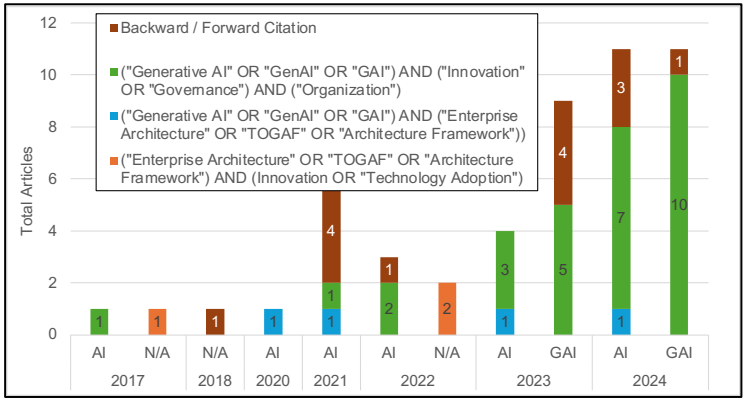

Total Articles by publication year, search phrase and AI/GAI view

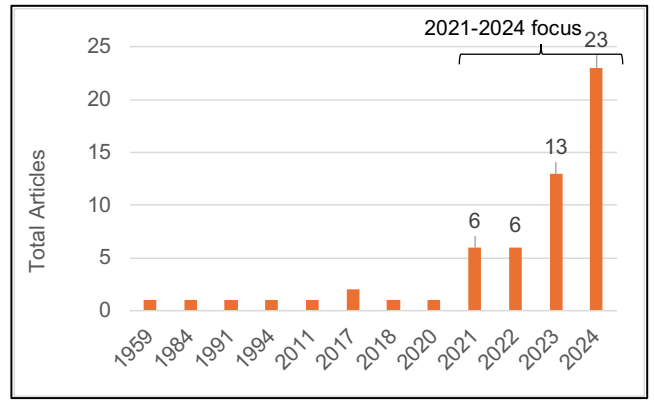

Total number of articles by publication year

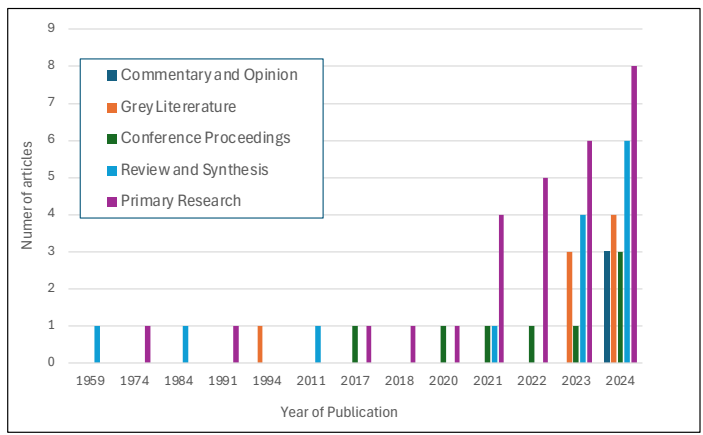

Total Articles by type and year of publication

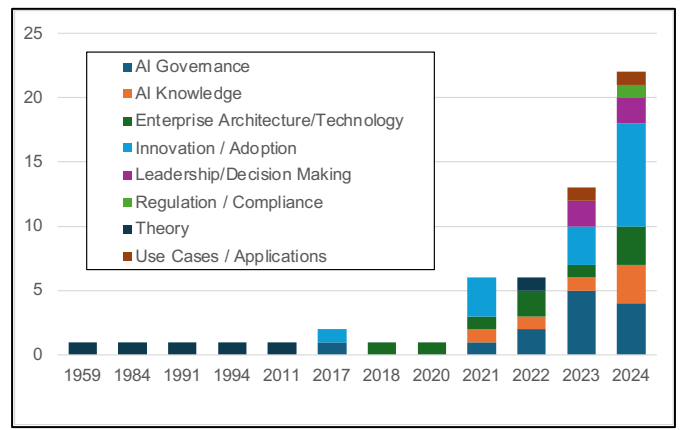

Total Articles by publication year and categories

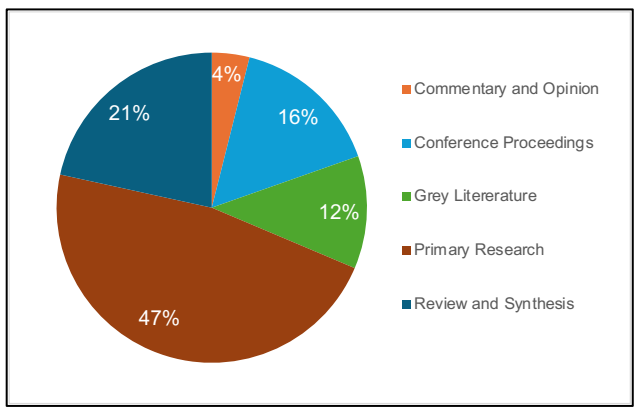

Percentage of articles by primary literature type

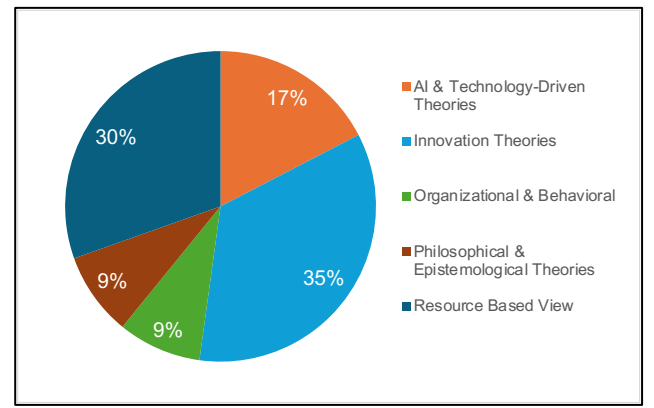

Percentage of articles by Theory (where applicable)

Figure 7. Data Collection Analysis

The data in Figure 7 provides an analysis of the 57 academic articles (66 total, including non-academic sources) reviewed for the systematic literature review (SLR), structured using the method in Section 3.3. Key insights include:

a) Publication Trends: 80% (48) of the articles were published between 2021 and 2024, reflecting the surge in GenAI and AI advancements. Most articles involve primary research, including journal articles, conference papers, and workshop proceedings, with fewer commentary pieces (3) appearing in 2024.

b) Search Terms: The dominant search term, "Generative AI" AND ("Innovation" OR "Governance") AND ("Organisation"), yielded most of the included articles. Backward and forward citation tracking supplemented fewer results from other terms. Articles cantered on GenAI primarily explored adoption barriers, innovation processes, and business model innovation, while EA-related studies showed steady interest since 2018.

c) Focus Areas: Since 2021, articles increasingly examine innovation and adoption processes, paralleling GenAI's growth. GenAI governance studies also show a recent upward trend, while EA-related topics maintain consistent academic engagement.

d) Theoretical Foundations: 35% of articles align with Dynamic Capabilities and RBV, and 30% with Innovation Diffusion theories, underscoring their relevance in understanding GenAI adoption and governance.

Future analysis could delve into methodological trends and their alignment with the dissertation's empirical and theoretical focus.

Table 2. Results of search and final selection of detailed literature review papers

| Detailed Search Terms | Total Search Size | Final Selection |
|---|---|---|
| ("Generative AI" OR "GenAI" OR "GAI") AND ("Innovation" OR "Governance Framework") AND ("Organization" OR "Organisation") AND ("Private Sector") | 280 | 31 |
| ("Generative AI" OR "GenAI" OR "GAI") AND ("Enterprise Architecture" OR "TOGAF" OR "Architecture Framework")) AND (Private Sector) | 190 | 5 |
| ("Enterprise Architecture" OR "TOGAF" OR "Architecture Framework") AND ("Technology Adoption") AND (Organization OR Organisation) AND ("Private Sector") | 278 | 6 |
| Backward / Forward Citation (G-AI) | N/A | 10 |
| Backward / Forward Citation (EA) | N/A | 3 |
| Backward / Forward Citation (Other) | N/A | 1 |
| **Total** | **748** | **56** |

## 3.5 Detailed Review

This section summarises the papers from each identified theme, reflecting a logical progression of understanding of GenAI adoption and its interactions with EA, as shown in Table 3. Following an overview of GenAI, the review is structured into four themes: 1. Scaling and Diffusion of GenAI Innovations followed by 2. Governance and Strategic Alignment for GenAI Integration, then 3. Dynamic Capabilities in EA for GenAI Adoption and finally 4. The indirect influence of GenAI in EA to accelerate GenAI adoption. A summary of each topic and relevancy to the research question is detailed in Appendix D.

Table 3. Structure of detailed Literature Review

| Section | Description |
|---|---|
| Overview of GenAI | Overview of GenAI using Zittrain's generativity definition. |
| Theme 1: Scaling and Diffusion of GenAI Innovations | Realized benefits, value and business model innovations with GenAI. |
| Theme 2: Governance and Strategic Alignment for GenAI Integration | Unique determinants of GenAI adoption, complexity, end-user perceptions and Governance Frameworks |
| Theme 3: GenAI innovation and adoption can be reinforced by **EA** Dynamic Capabilities | Overview of DC theory, GenAI fostering innovation, DC and Enterprise Architecture |
| Theme 4: The indirect influence of GenAI in **EA** to accelerate GenAI adoption | GenAI "in" Enterprise Architecture |

*3.5.1 Overview of GenAI using Zittrain's generativity definition*

Zittrain (1974) defines generativity as a technology's capacity to produce unprompted change driven by diverse, uncoordinated audiences. Four key traits of generative technologies are identified: leverage, adaptability, ease of mastery, and accessibility. These traits are highly applicable to GenAI, underscoring its transformative potential.

**Capacity for Leverage**

GenAI systems, such as large language models (LLMs) based on Transformer architecture, demonstrate versatility by supporting diverse applications. These include natural language processing (NLP), software development, and content creation. GenAI enhances productivity by streamlining complex tasks, much like the PC revolution. For example, in this dissertation, ChatGPT aided in transcript translation, search term refinement, and chart generation, highlighting its ability to improve workflows.

**Adaptability**

GenAI's adaptability is evident in its capacity to be fine-tuned for various industries, including healthcare and finance. Advanced architectures, such as Retrieval-Augmented Generation (RAG), integrate external data for specific use cases, ranging from customer support to medical diagnostics (Microsoft, 2024). Emerging trends, like autonomous agents and seamless integrations, present challenges for Enterprise Architects (EAs) managing adaptive systems within organisational infrastructure (Riemer & Peter, 2024).

**Ease of Mastery**

GenAI's user-friendliness democratises access, allowing even non-experts to deploy and utilise AI. Similar to low-code platforms and early PCs, GenAI supports both novices and experts, fulfilling Zittrain's assertion that generative tools should cater to varied skill levels. The growing accessibility of GenAI aligns with regulatory frameworks like the EU AI Act, drawing strategic attention at the executive level (Gstrein et al., 2024).

**Accessibility**

GenAI leverages cloud-based services and open-source platforms, such as Hugging Face, to lower barriers to entry. While this accessibility expands innovation potential, it also raises challenges, such as balancing internal versus external usage (Retkowsky et al., 2024). These developments reinforce Zittrain's principle that broad availability is crucial for unlocking generativity's full potential.

**Theme 1: Overview of GenAI - Scaling and Diffusion of GenAI Innovations**

*3.5.2 Realised Benefits and Value from GenAI Adoption*

GenAI demonstrates transformative potential in efficiency, productivity, and innovation across industries. Wamba (2024), whilst analysing the adoption matrix of ChatGPT and the perceived benefits for non-adopters and adopters, highlights operational efficiency gains at DHL and Zalando via ChatGPT adoption, while Jaffe et al. (2024) notes enhanced user productivity through Co-Pilot tools in a survey of 60 large firms.

Brynjolfsson et al. (2023) report 34% productivity improvement among low-skilled workers using GenAI agents, with limited impact on experts, showcasing its role in bridging skill gaps. Orchard (2023) illustrates GenAI's capacity to drive premium services and novel revenue streams in finance, supported by Singh et al. (2024), who link GenAI to enhanced exploratory and exploitative innovation. A global IBM survey of 400 executives (Dencik, 2024) underscores quality improvement and capability maturity over cost reduction as critical drivers of adoption (refer to Figure 8).

Lastly, Shanks et al. (2018), using RBV and DC theories, stress EA's role in aligning governance with innovation. These studies underscore the importance of tailoring GenAI adoption strategies to organisational contexts, with factors like skill and task selection offering a more nuanced view of benefit realization (Jaffe et al., 2024; Brynjolfsson et al., 2023).

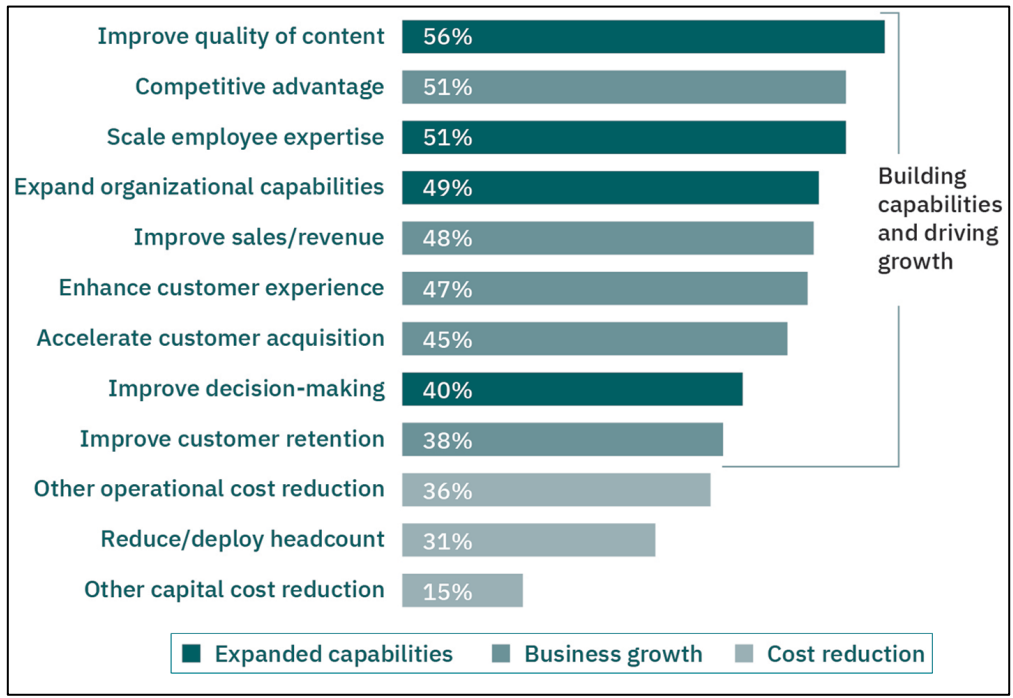

Figure 8. Automating tasks is not the primary focus for CIOs (IBV, 2023)

### 3.5.3 Business Model Innovations with (Gen)AI

AI is transforming business models by enabling new value propositions, scalability, and ecosystem integration through agile customer co-creation and data-driven operations (Sjödin et al., 2021). Addressing challenges in scaling AI projects, Sjödin (2021) developed a co-evolutionary framework based on observations of B2B providers. Early adopters, particularly financial institutions, excel with centralised models, though decentralisation has exposed governance challenges and highlighted adaptability needs (McKinsey, 2024; Müller et al., 2024). Kanbach (2023) explores GenAI's disruption of traditional structures, requiring process, structure, and skillset adaptation. Holmström (2024) stresses that GenAI's success in driving business model innovation depends on strategic alignment, operational feasibility, and robust governance. Linde et al. (2021) underscore the importance of structured governance in GenAI-influenced digital business models, as shown in Swedish manufacturing. These findings highlight the need for adaptive governance frameworks to navigate GenAI's unique opportunities and challenges, explored further in the next theme.

**Theme 2: Governance and Strategic Alignment for GenAI Integration**

*3.5.4 Unique determinants & barriers to deployment of GenAI*

The adoption of GenAI in organisations is influenced by several critical factors.

Firstly, a significant determinant is the technological complexity and compatibility of these systems, with high levels of customisation and associated costs presenting notable challenges (Agrawal, 2023). In industries such as services, cybersecurity and privacy risks add further complexity, requiring a careful balance between leveraging innovation and maintaining security (Gupta & Rathore, 2024). Ethical concerns, encompassing fairness, accountability, and transparency, along with institutional pressures, also play a crucial role in moderating GenAI adoption and its overall impact on organisational performance (Rana et al., 2024).

Secondly, Kongsten et al. (2024) utilise the Technological, Organisational, and Environmental (TOE) framework to examine GenAI adoption dynamics (refer to Figure 9). Technological factors are closely linked to the DoI theory (Rogers, 1995), focusing on innovation complexity. Organisational elements highlight the need for new processes and skill development, while environmental considerations emphasise the influence of legal and regulatory structures.

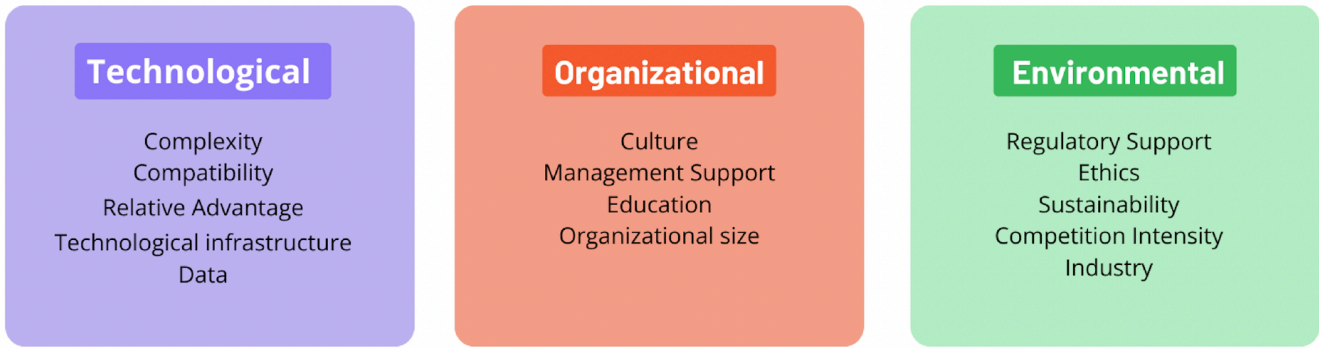

Figure 9. The TOE framework on GenAI adoption (Kongsten et al., 2024)

Furthermore, the adoption of GenAI faces barriers such as the opacity of black-box models, limiting interpretability and trust, which hinders large-scale deployment (Arize, 2021). Explainable AI (XAI) and GenXAI address this by enhancing interpretability, using taxonomies to clarify decision-making processes and mitigate technical complexity (Samek, 2017; Schneider, 2024a).

Operationally, managing LLMs via LLMOps requires specialised infrastructure distinct from traditional MLOps, supporting scalability (Jaffe et al., 2023). Aligning Product IT with EA adds structural challenges, complicating GenAI adoption (Kaidalova et al., 2017).

Finally, training large models like GPT-4 raises sustainability concerns due to significant energy use and costs (Orchard et al., 2023; Barbierato, 2024). As with ethical issues, these challenges are often attributed to LLM providers, shifting responsibility away from adopting organisations.

*3.5.5 Technology Complexity of GenAI & End User Perceptions*

GenAI is a complex, multi-dimensional entity where computational infrastructure, foundational models, and consumer-facing applications are interconnected, requiring robust end-to-end (E2E) observability (Ferrari et al., 2023). This complexity has been amplified by the rapidly expanding AI landscape, making it increasingly challenging to track and adapt to advancements.

The opaque and non-deterministic nature of GenAI introduces additional challenges around transparency and governance. The "black box problem" continues to undermine trust and usability (Ochmann et al., 2021; Castelvecchi, 2016). Gutiérrez (2023), drawing on actor-network theory, highlighted the intricate network of human and non-human actors within enterprises, further complicating governance and oversight.

Moreover, resource misalignment and inadequacies exacerbate these issues. Bughin (2024) analysed the role of Responsible AI adoption (RAI), emphasising governance frameworks as tools to manage the uncertainties associated with (Gen)AI. However, end-users often perceive such frameworks as coercive or overly costly, presenting yet another hurdle to effective adoption.

*3.5.6 (Gen)AI Governance Frameworks*

Effective governance of GenAI necessitates explainability, both as a regulatory requirement and as a tangible operational asset (Ferrari et al., 2023). The EU AI Act represents a significant milestone in European AI governance, marking a transition from reactive to proactive regulatory approaches. Gstrein's (2024) interdisciplinary analysis highlights the Act's potential as a global standard for ensuring product safety and fundamental rights. However, challenges remain in aligning governance with the rapid pace of technological evolution and ensuring effective enforcement, which relies heavily on EU wide cooperation. Advanced governance models, such as polycentric and layered approaches, have been proposed to address these complexities. The layered model emphasises technical foundations, including data governance and algorithm accountability, within broader societal contexts (Gasser & Almeida, 2017). Adaptive frameworks are essential to enhance trust and accessibility while mitigating GenAI's opacity (Reuel & Undheim, 2024).

In organisational contexts, governance frameworks must accommodate the complexities of GenAI, integrating human-AI collaboration, multi-agent systems, and ethical compliance (Schneider et al., 2024b; Gahnberg, 2021). Eglash et al. (2024) propose fractal scaling principles to balance global standards with local empowerment, fostering adaptability and decentralised governance. These principles challenge bottom-up governance models, including established IT frameworks such as COBIT and ISO 17799, which support risk management and standardisation but require adaptation for AI contexts (Robles et al., 2023; Pittaras & McGregor, 2023).

**Theme 3: GenAI innovation and adoption can be reinforced by EA Dynamic Capabilities**

The literature is summarised in Table 4.

### *3.5.7 Overview of Dynamic Capabilities theory*

Dynamic Capabilities (DC) theory, based on the Resource-Based View (RBV) of the firm, offers a comprehensive framework for understanding how firms achieve and sustain competitive advantage in dynamic environments. Edith Penrose's foundational work on the Theory of the Firm established the RBV, viewing firms as bundles of resources whose growth and success depend on managerial capabilities and innovation.

Building on this, Teece and Pisano (1994) introduced DC theory, emphasising the ability of firms to reconfigure resources in response to changing environments. Teece et al. (1997) formalised this concept, linking DCs to competitive advantage and highlighting their role in addressing market volatility. Teece (2007) further expanded this by identifying the micro-foundations of DCs, sensing opportunities, seizing them, and transforming resources, necessary for sustainable performance.

Pavlou (2011) bridged theory and practice by examining the mechanisms of DCs, providing empirical evidence for their effectiveness. Together, these contributions position DC theory as an evolution of RBV, emphasising the dynamic, innovation-driven processes that enable firms to adapt and thrive in uncertain and competitive markets.

Weiss et al. (2012) highlighted the critical need for theoretical foundations to better integrate DC concepts into EA practices, positioning DCs as a bridge between resource configuration and organisational agility.

### *3.5.8 Enterprise Architecture and Dynamic Capabilities*

Drawing from DC theory (Teece & Pisano, 1994), the literature emphasises the role of EA in fostering Dynamic Capabilities that enhance organisational agility and operational ambidexterity. DC's driven by EA enable firms to adapt to rapidly changing digital technologies (coined operational "digital ambidexterity"), significantly impacting business value and resilience during disruptive periods like the COVID-19 pandemic (van de Wetering, 2022).

Weiss et al. (2012) argue that foundational theories are essential for effective EAM, particularly in navigating the interplay between artefacts and decision-making. They underscore the importance of using cognitive fit and boundary object theories to address gaps in communication and alignment across stakeholders. This perspective supports the shift towards more dynamic EA practices that integrate actor-stakeholder relationships, artefact usability, and adaptive capabilities (Kotusev & Kurnia, 2020).

Therefore, EAM is conceptualised as a dynamic managerial capability that requires continuous resource renewal and adaptation to support iterative technologies such as GenAI (Pattij et al., 2022). The findings suggest that traditional EA frameworks, such as TOGAF [2] and Zachman, may be inadequate for GenAI, underscoring the need for more adaptive and experimental approaches.

Moreover, research shows that EAM (as a DC) plays a pivotal role in enabling organisations to adopt GenAI effectively. EA must evolve from artifact-centric approaches to dynamic, capability-driven practices that enhance business-IT alignment and strategic flexibility (Shanks et al., 2018; Stecher et al., 2020). To support GenAI adoption, EA models need to capture organisational readiness and align with business objectives, as traditional frameworks often lack the nuance required for complex AI technologies (Rittelmeyer et al., 2022). New EA metamodels are proposed to enhance transparency and support holistic business-IT alignment, which is critical for managing ML and GenAI projects (Takeuchi et al., 2024). Sustainability and trusted use of AI technologies are also highlighted as key imperatives for future EA frameworks (Kamalabai et al., 2024).

*3.5.9 GenAI fostering innovation, new capabilities and the role of EA*

---

[2] TOGAF (The Open Group Architecture Framework) is a globally recognised framework for developing, managing, and aligning enterprise architecture with organisational objectives, ensuring effective integration of business and IT.

Sedkaoui and Benaichouba (2024) highlight GenAI's transformative potential for innovation but note its limited application-focused research, stressing the urgency of robust frameworks and "preventative ethics frameworks and guardrails…to ensure responsible development" (p.17). This aligns with the need for responsible innovation and EA (Anderson, 2016).

While GenAI supports creativity and exploratory advancements, environmental factors significantly moderate its adoption and outcomes (Singh et al., 2024). Despite investments in process improvements and revenue growth (McKinsey, 2023), most initiatives remain experimental with limited returns (Holmström et al., 2024). Ramaul et al. (2024) argue that capturing value from GenAI innovations is more challenging than value creation, underscoring adoption complexities.

An empirical study of 6,000 micro-businesses found early GenAI adopters gained agility and innovation (Gupta et al., 2024). However, governance gaps and ethical challenges remain barriers, with organisational innovativeness key to successful scaling (Rana et al., 2024).

Empirical research on the role of EA in organisational innovation is limited. However, Anderson (2016) leveraged the TOGAF framework to define a model aimed at enhancing firms' innovation capabilities, aligning architecture principles with strategic innovation goals. Similarly, Missah (2015) explored the integration of EA into business processes, emphasising how EA facilitates innovation by optimising the alignment between business architecture and technical architecture. Missah highlights that EA not only streamlines internal processes but also provides a structural foundation for innovation, enabling organisations to formalise their strategic objectives and enhance flexibility in adapting to emerging technological trends.

Table 4. Summary of EA literature

| Theme | Title of Literaure | Authors | Year |
|---|---|---|---|
| EA & (Gen)AI | Challenges in Integrating Product-IT into Enterprise Architecture – a case study | Kaidalova et al. | 2017 |
| EA & (Gen)AI | Enterprise Architecture's effects on organizations' ability to adopt artificial intelligence - a resource based perspective | Stecher | 2020 |
| EA & (Gen)AI | Effects of Artificial Intelligence on Enterprise Architectures - A Structured Literature Review | Rittelmeyer & Sandkuhl | 2021 |
| EA & (Gen)AI | TOGAF-based Enterprise Architecture Framework for Utilizing Artificial Intelligence | Fitriani et al. | 2023 |
| EA & (Gen)AI | Enterprise architecture-based metamodel for machine learning projects and its management | Takeuchi et al. | 2024 |
| EA & (Gen)AI | Sustainable Enterprise Architecture: A Critical Imperative for Substantiating Artificial Intelligence | Kamalabai et al. | 2024 |
| EA & (Gen)AI | Navigating the Generative AI-Enabled Enterprise Architecture Landscape: Critical Success Factors for AI Adoption and Strategic Integration | Denni-Fiberesima | 2024 |
| EA & DC's | The role of enterprise architecture-driven dynamic capabilities and operational digital ambidexterity in driving business value under the COVID-19 shock | van de Wetering | 2022 |
| EA & DC's | Enhanced digital transformation supporting capabilities through enterprise architecture management: A fsQCA perspective | Pattij et al. | 2022 |
| EA Theory | Towards a Reconstruction of Theoretical Foundations of Enterprise Architecture Management | Weiss et al. | 2012 |
| EA Theory | The theoretical basis of enterprise architecture: A critical review and taxonomy of relevant theories | Kotusev | 2020 |
| EA Value & Innovation | Business Innovation with Enterprise Architecture | Missah | 2015 |
| EA Value & Innovation | Enterprise Architecture for Innovation Realization and Sustainability | Anderson | 2017 |
| EA Value & Innovation | Enterprise architecture: A competence-based approach to achieving agility and firm performance | Hazen et al. | 2017 |
| EA Value & Innovation | Achieving benefits with enterprise architecture | Shanks et al. | 2018 |
| EA Value & Innovation | Enterprise architecture adoption in government: a public value perspective | Sunderb et al. | 2023 |

*3.5.10 Philosophical implications for management disciplines*

Lastly, challenges in GenAI adoption are rooted in philosophical issues, offering a novel perspective (Velthoven & Marcus, 2024). Misunderstandings of how GenAI operates lead to flawed policies, impacting its adoption in management disciplines. Philosophical implications revolve around aligning its complexity with human understanding, governance, and organisational processes. Misaligned frameworks (Velthoven & Marcus, 2024; Riemer et al., 2024) and lack of interpretability (Graziani et al., 2022) hinder trust and adoption. Riemer et al. (2024) suggest conceptualising GenAI as "style engines" to improve understanding and governance. Graziani et al. (2022) emphasise the need for clear taxonomies of interpretable AI in multi-disciplinary settings.

Institutional pressures necessitate robust governance (Rana et al., 2024), while AI literacy is critical for navigating GenAI's complexities (Annapureddy et al., 2024). Dual narratives, technosolutionist and generative, illustrate its dynamic influence on policy (Agbon, 2024).

**Theme 4: The indirect influence of GenAI in EA to accelerate GenAI adoption**

*3.5.11 GenAI "in" Enterprise Architecture*

GenAI "in" EA reflects the transformative impact GenAI can have on the EA discipline itself. While practitioner research in this area remains limited, several theoretical studies highlight its potential. For instance, GenAI is reshaping EA practices by necessitating advanced competencies such as digital literacy, human-AI collaboration, and the strategic integration of AI technologies (Denni-Fiberesima, 2018). Tailored frameworks, including those influenced by TOGAF, aim to offer structured guidance for incorporating AI into enterprise strategies, emphasising digital transformation and operational efficiency (Fitriani et al., 2023). However, a significant research gap persists in understanding the specific EA layers most affected by AI, signalling the need for further studies to map GenAI's comprehensive impact on EA structures and processes (Rittelmeyer et al., 2021). These insights underscore EA's dual role in adapting to AI-driven transformations and providing structured methodologies for effective AI integration.

3.6 **Conclusion and Research Gaps**

*3.6.1 Key Findings in Literature*

The literature highlights GenAI as a dual force, driving innovation and operational transformation while posing significant challenges due to its complexity, institutional pressures, and philosophical implications. While early adopters can gain a competitive edge, this requires carefully balancing innovation with compliance and adapting organisational operating models. Regulatory frameworks are evolving, but gaps remain in embedding explainable (Gen)AI and adaptive governance within organisations. EA, increasingly focused on business and innovation, shows potential as a DC to address GenAI's experimental and rapidly changing nature. However, empirical evidence on EA's adequacy for navigating GenAI and enabling tailored, scalable governance frameworks remains sparse, particularly regarding how GenAI influences EA practices and future ways of working.

*3.6.2 Research Gaps*

The literature highlights several critical gaps in understanding the broader adoption of GenAI and the role of EA in facilitating this process. Firstly, there is insufficient empirical research examining large-scale GenAI adoption strategies across industries, particularly their impact on operating models and organisational adaptation. Additionally, the evolution of EA frameworks and AI governance models to support strategic alignment and enable scalable GenAI adoption remains underexplored. There is also limited focus on how EA, as a DC, can enhance GenAI adoption and drive organisational performance. Furthermore, the challenges specific to EA's role in supporting and influencing GenAI initiatives, beyond standard barriers to technology deployment and diffusion, are yet to be fully studied. Finally, research on the role of GenAI within EA, including its influence on broader organisational adoption and its implications for EAM is scarce.

3.7 **Conceptual Research Model**

Building on the findings from the SLR and the identified research gaps, the conceptual research model outlined in Figure 10 serves as the foundation for guiding the research methodology and data analysis. Rooted in Dynamic Capabilities, the model explores EA's dynamic capabilities in relation to scaling, diffusion, governance, and strategic alignment, and their impact on improving (Gen)AI adoption. The antecedents of (Gen)AI adoption are framed within the Diffusion of Innovation (DoI) theory, while the indirect influence of GenAI within EA on broader (Gen)AI adoption completes the model's structure.

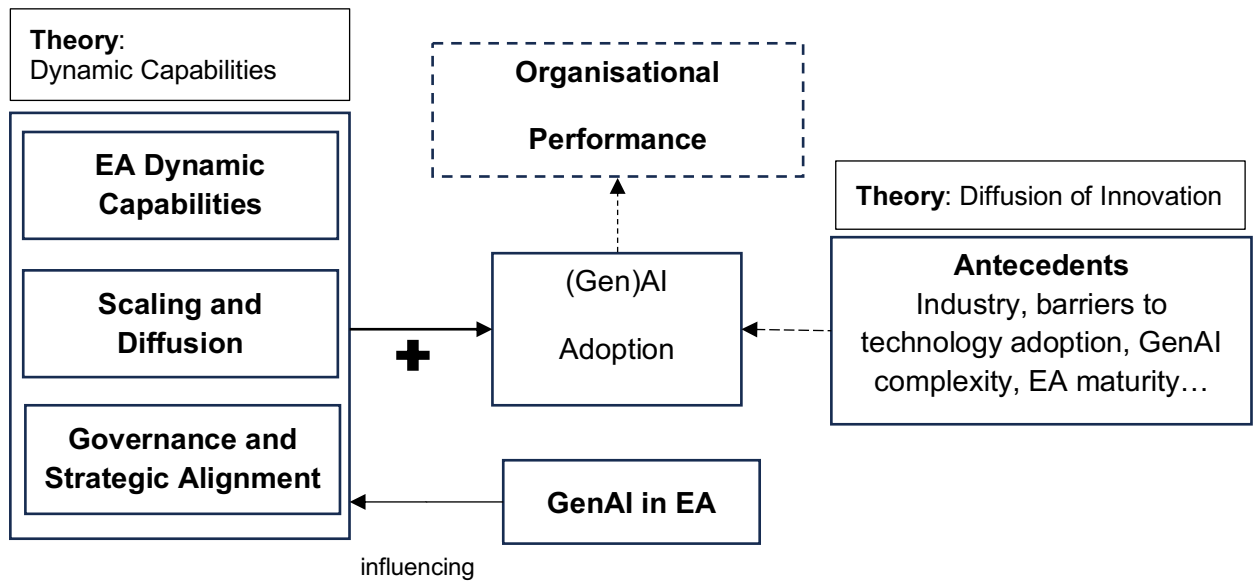

Figure 10. Conceptual Research Model

# Chapter 4 – Method

This chapter outlines the qualitative method utilise d to explore the role of EA in the adoption of GenAI. A total of 16 semi-structured interviews were conducted with professionals holding expertise in EA and GenAI, employing a systematic analytical approach grounded in the Gioia methodology. The chapter presents the qualitative method, research model and setting, and finally the data collection approach.

## 4.1 Preparation

The SLR provided a comprehensive overview of the current state of GenAI adoption in organisations, grounded in Dynamic Capability and Innovation Diffusion theories. The review identified critical gaps in understanding the broader role of Enterprise Architecture frameworks and operating models in scaling GenAI adoption. Consequently, the research focuses on how EA can facilitate strategic, scalable GenAI adoption by developing a framework to guide EA teams and organisational leaders. Given the need to explore complex organisational processes, perceptions, and experiences, qualitative methods were selected as the most suitable approach for this investigation.

## 4.2 Qualitative Method

The qualitative method seeks to explore and understand concepts or phenomena by describing and interpreting their meaning, rather than quantifying their frequency or occurrence (Patton, 2015). This approach is particularly suited to novel research contexts where primary data is scarce and conceptual frameworks remain uncertain. In this study, qualitative methods were chosen due to the complexity of the subject matter and the need to gain nuanced insights into how organisations perceive and adapt frameworks like EA to accommodate dynamic technologies such as GenAI. Additionally, these methods are well-suited for addressing contextual variables, such as governance and organisational culture, which play a critical role in shaping adoption strategies. Qualitative research also provides the flexibility and depth required to explore the intricate interplay of factors within specific organisational and industry settings.

Various qualitative methods were considered, including expert interviews, focus groups, case studies, and observations. Expert interviews were selected as the most appropriate approach due to their ability to provide direct access to individuals with firsthand knowledge and experience. This method enables focused exploration of EA and GenAI, facilitates the emergence of key themes, and allows for the efficient collection of relevant data compared to more time-intensive approaches such as prolonged observations. Preliminary discussions with experts within the author's organisation revealed that organisational maturity in GenAI adoption remains limited, making case studies unsuitable at this stage. Expert interviews, therefore, represent the most practical method to capture insights aligned with the research questions and the current exploratory phase of GenAI technology (Bogner et al. 2019). While triangulation of findings is inherently challenging with this approach, the interview design included a specific focus on concrete GenAI use cases to enhance the quality and reliability of the collected data.

Enterprise Architects and VP/Heads of EA are particularly well-suited candidates for expert interviews, given their combined technical expertise and comprehensive organisational perspective on GenAI adoption. These individuals are uniquely positioned to offer honest assessments of their organisation's current adoption status and insights into potential future implementation roadmaps. Engaging with such experts in sufficiently large organisations ensures a thorough and nuanced understanding of GenAI adoption dynamics and the critical role of EA in scaling these efforts effectively.

Figure 11 illustrates the developed research model, which builds upon the insights gained from the Systematic Literature Review. This model refines the Research Questions and integrates the subsequent stages of expert interviews, results analysis, and a discussion of the findings. These discussions align the interview outcomes with the insights from the literature review to explore their broader implications.

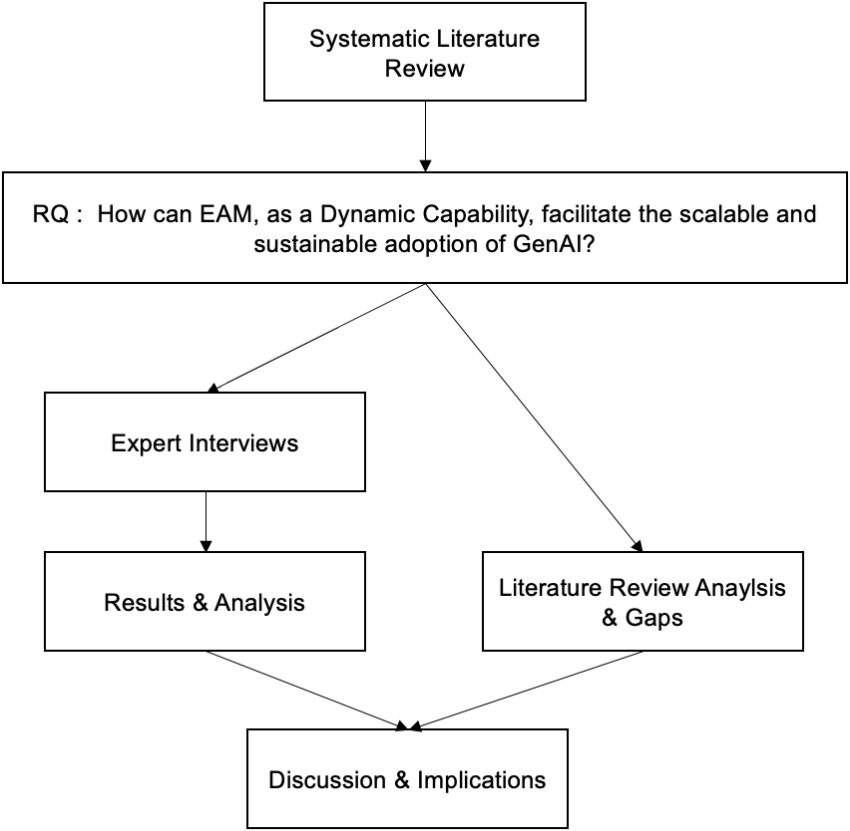

Figure 11. Research Model

### 4.3 Research Setting

Holmström and Carroll (2024) emphasise the significance of industry context in understanding GenAI adoption. To address this, the study engaged with organisations across various industries to provide a holistic perspective on GenAI adoption, considering both GenAI maturity and EA maturity. The focus was on private sector companies located in Europe, aligning with the regulatory environment shaped by the EU AI Act, which influences GenAI adoption differently compared to the U.S., where varying regulatory frameworks and GenAI maturity levels prevail (Gstrein et al., 2024). To ensure a robust foundation for analysis, the author, in agreement with the academic supervisor, committed to conducting a minimum of 10 interviews, with the aim of achieving theoretical saturation. The criteria for expert selection are detailed in Table 5.

Table 5 Selection Criteria

| Selection Criteria | Description |
| --- | --- |
| Interviews | At least 10. |
| Industry | Variety of industries where GenAI is both relevant & in production |
| Company Size | Large Enterprises (100m+ EUR Annual Revenue) |
| Geographical Location | Companies with business or customers in Europe |
| Expert Definition | Demonstrated depth of technical knowledge in Generative AI, either through their role within the Enterprise Architecture team, their involvement in managing GenAI governance, or their direct experience in developing and implementing GenAI solutions. |
| Sector | Private Sector Only |

## 4.4 Data Collection

Prior to data collection, a research ethics application was submitted to Warwick Business School. To ensure interviewees were well-prepared, a concise interview guide was shared alongside the interview invitations (Appendix A). The guides were translated into the interviewees' primary language using GPT-4 (Open AI, 2025), facilitating interviews conducted in both English and French. Before the interviews, participants completed a brief asynchronous online questionnaire using Microsoft Forms (Appendix B), designed to capture their names, roles, experience, and knowledge of GenAI. These close-ended questions were complemented by open-ended discussions during the interviews.

The interview guide underwent several revisions based on feedback from internal test candidates, including the SVP of Product Management and other product managers familiar with the research scope. This iterative process led to simplifying certain questions and clustering them by first- and second-order themes for better clarity and flow.

Participants were primarily selected from the author's firm's customer base, focusing on companies with a demonstrated interest in GenAI governance as identified through the Product Management feedback tool. Invitations to participate were coordinated by the Customer Success Management team, who facilitated contact with prime candidates. Despite a 50% decline rate, largely attributed to end-of-year time constraints, 16 interviews were completed, most lasting an hour or more.

Interviews were conducted remotely via Microsoft Teams, recorded, and transcribed. In some instances, two participants from the same company, typically holding different roles, joined a single session due to time constraints. Table 6 provides an overview of participant details, including industry, participant IDs, locations, years of experience, company size, and interview length.

The study encountered considerable enthusiasm from interviewees regarding both the topic and the anticipated outcomes, reflecting the relevance and timeliness of the research.

Table 6. List of company industries, roles, locations, years of experience, company size and length of each interview

| Industry | Participant ID | Participant(s) Role | Company Location | Years of Experience | Company Size | Interview Length |
|---|---|---|---|---|---|---|
| Consumer Goods | P5 | Domain Architect / Global Director Digital Ecosystem | Germany | 10 | €1000m+ | 57 minutes |
| | P8 | Chief/Group Enterprise Architect | France | 15 | €1000m+ | 53 minutes |
| | P11 | Senior Enterprise Architect | Germany | 5 | €1000m+ | 1hr 17 minutes |
| | P12 | Head of Enterprise Architecture | Germany | 18 | €1000m+ | 59 minutes |
| | P17 / P18 | Co-CIO & Enterprise Architect | Germany | 3 | €500m - €1000m | 1hr 9 minutes |
| | P20 | Director of Technology Transformation | Switzerland | 13 | €1000m+ | 29 minutes |
| Financial Services | P1 | Principle Enterprise Architect / VP IT Strategy and Architecture | US | 30 | €1000m+ | 1hr 29 minutes |
| | P4 | Head of IT Enterprise Architecture | UK | 3 | €100m - €500m | 1hr 21 minutes |
| | P15 / P16 | Enterprise Architect / Solution Enterprise Architect | UK | 6 | €1000m+ | 1hr 22 minutes |
| Technology | P2 | Principle Enterprise Architect | Germany | 15 | €1000m+ | 51 minutes |
| | P10 | Senior Enterprise Architect | US | 10 | €1000m+ | 26 minutes |
| | P19 | Global tech lead AI & Generative AI | Sweden | 6 | €1000m+ | 57 minutes |
| Healthcare | P9 | Lead Enterprise Architect | Germany | 7 | €100m - €500m | 1hr 23 minutes |
| Logistics | P13 / P14 | Group Enterprise Architect / Data Scientist | Germany | 15 | €1000m+ | 58 minutes |
| Real Estate | P3 | Director of Enterprise Architecture | UK | 5 | €100m - €500m | 1hr 9 minutes |
| Security | P6 / P7 | Group Lead EA / Chief EA | Sweden | 15 | €1000m+ | 56 minutes |

4.5 **Data Analysis**

The data analysis was conducted using the Gioia methodology (Corley & Gioia, 2004) and utilised NVivo software to systematically process and organise the interview data. Transcriptions were first created automatically using Microsoft Teams and were subsequently reviewed manually to ensure both accuracy and the anonymisation of participants. Once finalised, these transcripts were imported into NVivo for coding and analysis.

The initial phase involved open coding, during which first-order concepts were identified directly from the participant data. This process emphasised the use of informant-centric terminology, ensuring that the participants' experiences and perspectives were preserved in the analysis. This stage resulted in 395 individual first-order codes, capturing the diversity and granularity of insights from the data.

These first-order codes were then grouped into second-order themes through an iterative refinement process. This step aimed to consolidate related codes and ensure conceptual coherence, ultimately yielding 207 second-order themes. These themes were further synthesised into nine aggregate dimensions, representing higher-level categories such as organisational transformation, governance frameworks, and the integration of GenAI within EA.

The analysis linked these aggregate dimensions to the study's theoretical underpinnings, Dynamic Capabilities, and Diffusion of Innovation theories. The high-level DCs for EA have been developed deductively based on the literature review and then refined based on the data analysis. This final abstraction helped uncover strategic, operational, and governance insights for the adoption of GenAI within organisations.

Figure 12 summarises this analytical process, illustrating the progression from raw transcripts to actionable theoretical dimensions, ensuring a rigorous and systematic approach to understanding the challenges and opportunities of GenAI adoption. Multiple tools were used in this process, including MS Teams, Microsoft Word, Nvivo, Microsoft Excel and finally SAP LeanIX as a way to report on the final analysis.

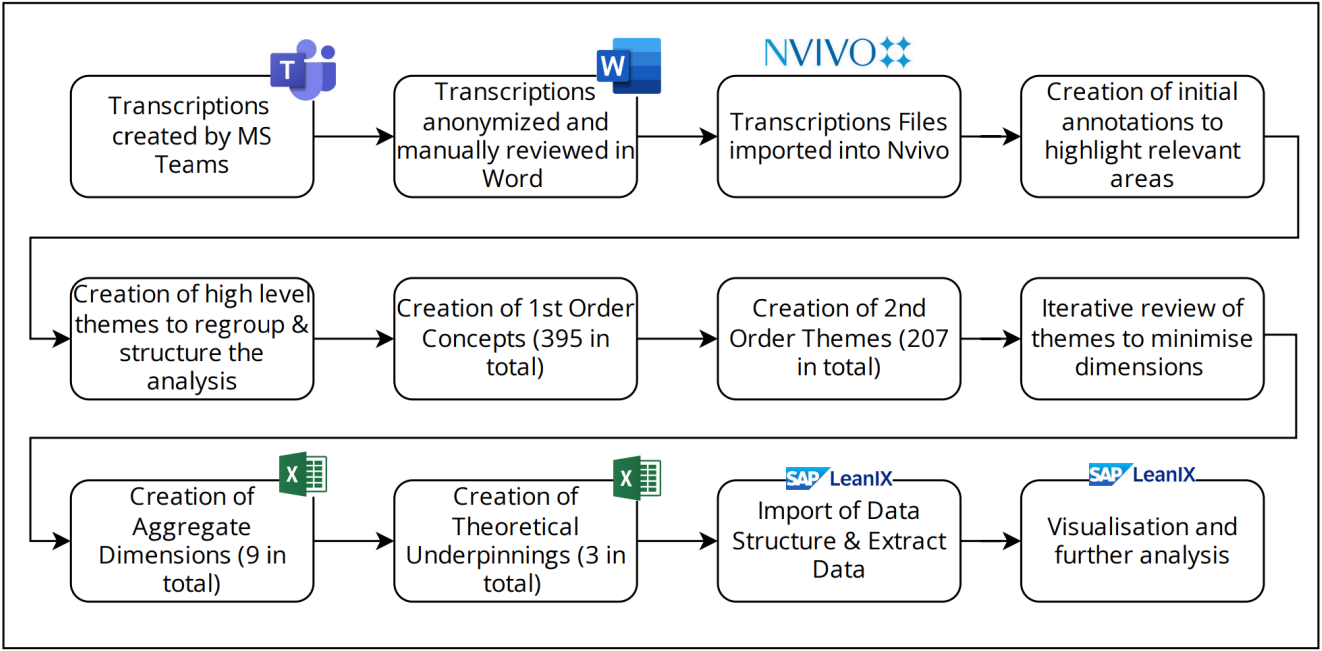

Figure 12. Data Analysis Method

Excerpts of the resulting data structure indicating the full data hierarchy are illustrated in Figure 13, Figure 15 and Figure 15. The import of the data into the SAP LeanIX tool allows for rapid search & reporting capabilities.

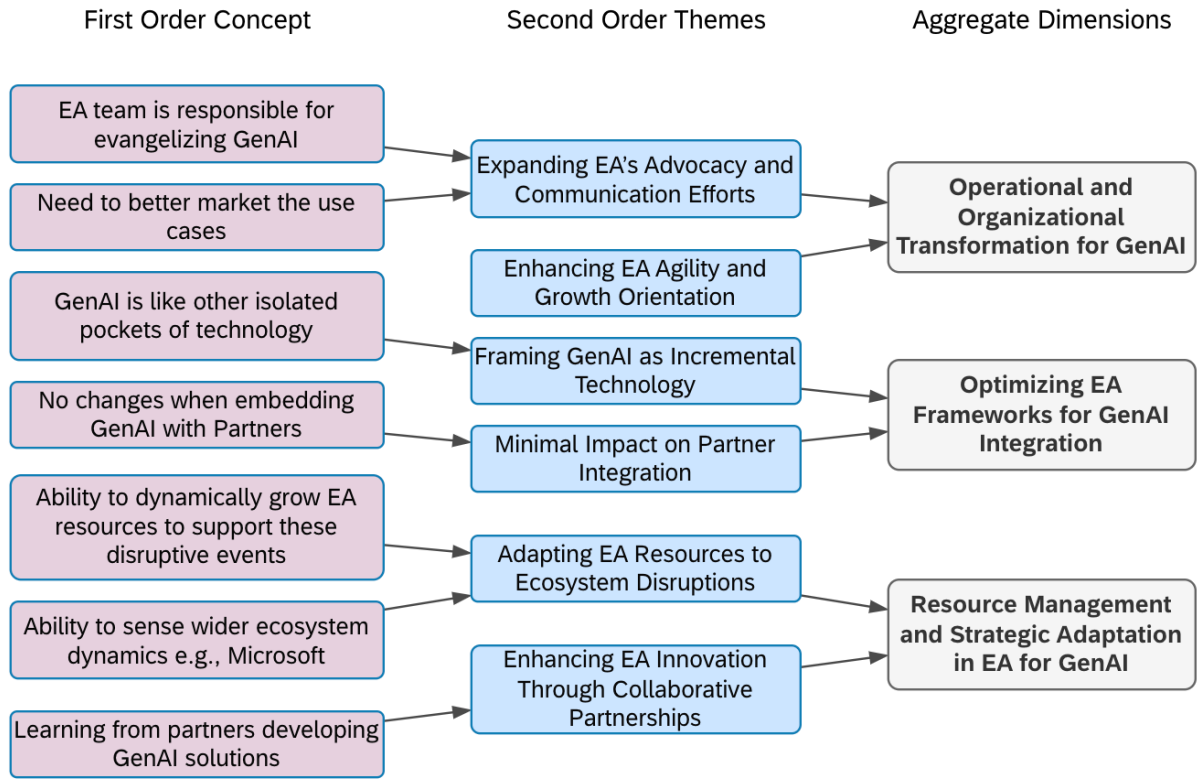

Figure 13. Extract of the Coding Structure

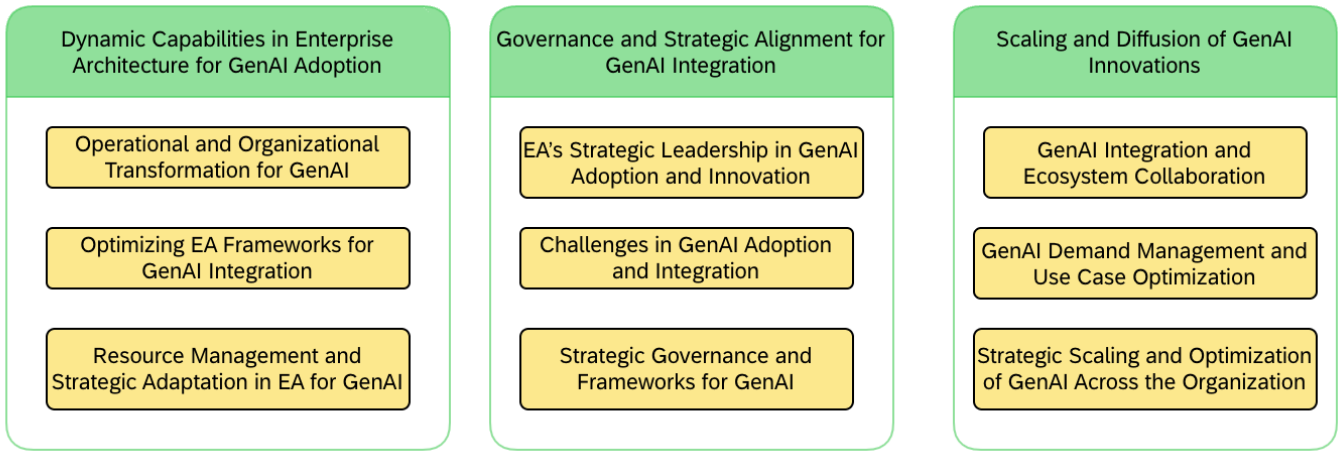

Figure 14. Theoretical underpinning themes and aggregate dimensions

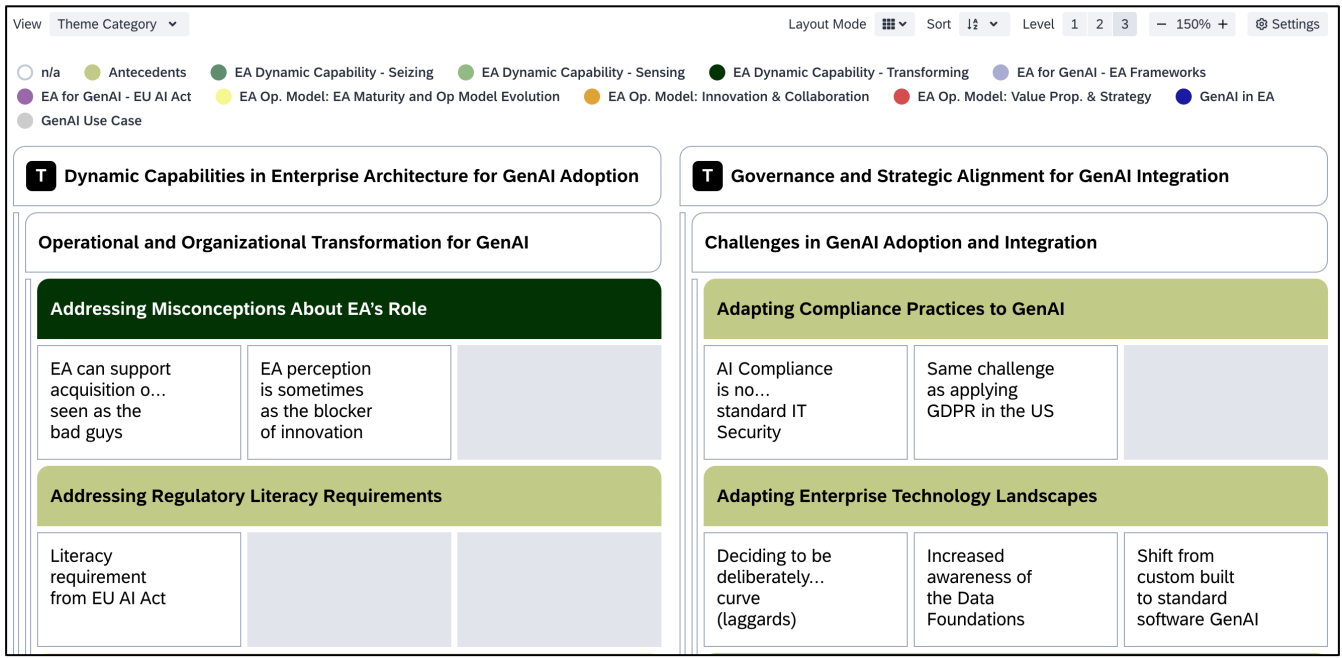

Figure 15. Example report of data extract with theme category heatmapping used in the analysis

Sentiment analysis using Retrieval-Augmented Generation (RAG)-based LLMs has recently proven effective in extracting topics and themes from qualitative data, including semi-structured interviews (Bhaduri et al., 2024).

The author experimented with this approach to analyse sentiment across interview transcripts, focusing on perceptions of GenAI, the maturity of GenAI adoption, and the role of EA in facilitating this adoption. Initial findings from this analysis are presented in Appendix C. While the results provided valuable insights and demonstrated the potential of such techniques for qualitative research, they were ultimately excluded from the final research methodology due to scope limitations.

# Chapter 5 – Results

This chapter presents the findings from 16 semi-structured interviews conducted in November and December 2024. The results are organised into four key factors illustrated in Figure 16: EA for GenAI, GenAI in EA, EA Dynamic Capabilities (DC), and Antecedents. First, the chapter explores how EA, as a DC, is shaped by the influence of GenAI on EA frameworks, operating models, and the implications of the EU AI Act as a regulatory framework. This is discussed alongside critical antecedents such as AI literacy, organisational knowledge, and the distinct and complex properties of GenAI technology. Additionally, it examines how GenAI "in" EA challenges the traditional role of EA. Lastly, the analysis delves into the DC dimensions of EA: sensing, seizing, and transforming, and their impact on GenAI adoption.

The exhaustive list of references for each participant are share in Appendix E, grouped by topic.

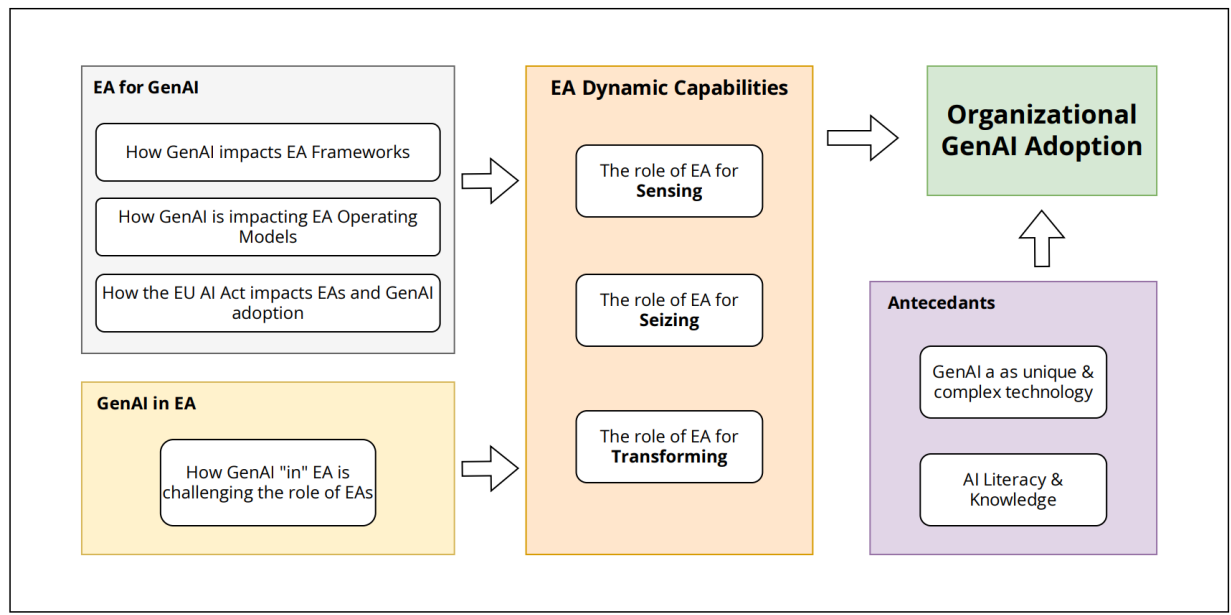

Figure 16. EA factors influencing GenAI adoption

## 5.1 Antecedents for GenAI Adoption influencing EAM

### 5.1.1 GenAI as unique & complex technology

Firstly, the adoption of GenAI continues to face significant hurdles, particularly the absence of large-scale production use cases. Participant 15 noted, *"It's worth saying that none of our (GenAI) solutions in the UKI […] are directly consumer-facing. Most of the use cases we see are internal productivity."* This sentiment was echoed by Participant 20, who remarked, *"We're still in a phase of investigating how we can securely deploy such solutions towards consumers."* Participant 13 added that the slower pace of adoption was tied to development speed: *"I don't think we have it yet. I think it's more or less supportive functionalities […] because we're not that fast."* These insights highlight the emerging phase of GenAI adoption in many organisations, where efforts are still exploratory and disconnected from broader revenue generating strategies. (P1, P13, P15, P20, P3, P4, P8).

Secondly, regulatory constraints and security concerns further complicate GenAI adoption. Participant 4 pointed out the impact of stringent industry-specific regulations, mentioning that *"we can innovate, certainly, but we've got to do that with other constraints and conditions in mind."* This regulatory environment often slows progress and limits experimentation. Participant 17 observed that tools like Microsoft Co-Pilot could serve as a stepping stone to advance AI adoption holistically, addressing dominant *"negative preconceptions"* about GenAI and its integration within organisations. These examples underscore how regulatory and security concerns must be addressed to enable meaningful GenAI adoption. (P4, P17, P19, P5).

Additionally, the extensive applicability of GenAI presents both opportunities and challenges. Participant 1 described GenAI as having a transformative scope compared to earlier AI models: *"The tipping point with GenAI was that now you had something that was much broader in potential scope, because again in these prior models, they're very narrow."* This broad potential, combined with low barriers to entry, has made GenAI accessible to a wide range of organisations. However, this accessibility has also resulted in fragmented use cases and uncoordinated adoption strategies. Participant 13 stressed the importance of flexibility, noting, *"The approach needs to be more flexible and open because we don't know what the status of AI will be in a year from now."* (P1, P10, P11, P13, P19, P20).

Managing the hype and aligning demand with realistic adoption strategies have emerged as critical challenges. Participant 11 remarked, *"[GenAI] will generate a big impact also and it will drive a lot of activities from a business perspective."* Yet, many organisations struggle to find a clear roadmap for integrating GenAI strategically and understanding EA's role in this process (P5, P11, P15, P7, P8, P13).

Furthermore, the foundational challenges of data quality and governance[3] remain essential to GenAI adoption. Participant 3 highlighted the importance of technical readiness, stating, *"You really got to have your data strategy and data maturity right before you shine big torches with (Gen)AI."* Existing data landscapes and low levels of AI maturity exacerbate these challenges, as organisations struggle to align their data infrastructures with GenAI's requirements. Participant 9 further emphasised the limitations imposed by current data foundations, underscoring the need for robust strategies to address these gaps. (P7, P4, P9, P3).

---

[3] Data governance in organisations, especially for GenAI, involves managing data quality, security and lineage to ensure ethical and compliant use throughout the data lifecycle.

GenAI's non-deterministic nature creates ongoing challenges in explainability and transparency, with Participant 19 noting: *"Even (if) you change from GPT 3.5 to 4, generally it's improved […], but does that really improve your use case? You have no clue."*. The lack of transparency in GenAI's advancements makes it harder to assess its value and integrate it into decision-making processes (P1, P3, P4, P15, P20).

Moreover, the rapid pace of technological advancements and the resulting risk of GenAI sprawl further hinders cohesive adoption strategies. Participant 15 noted, *"That's a massive sort of sprawl, with lots of people doing things independently."* Participant 13 also highlighted the importance of agility: *"The approach needs to be more flexible and open because we don't know what the status of AI will be in a year from now."* These observations point to the need for harmonising use cases and balancing innovation with managing technical debt. (P1, P14, P15, P19, P20, P5, P13).

GenAI is frequently perceived as straightforward, however, its implementation involves significant complexity. Participant 20 emphasised the challenges in selecting appropriate technologies and use cases, stating, *"Selecting the right technology, selecting the right use cases is quite challenging."* These insights highlight the critical need for strategic planning and thorough capability assessments to address the complexities of GenAI adoption effectively (P13, P20).

Managing end-user perceptions is another significant aspect of successful GenAI adoption. Participant 20 highlighted the importance of customisation to meet user expectations: *"A chatbot looks different, behaves differently, and maybe has a different tone of voice depending on the context."* However, widespread misconceptions about GenAI's functionality persist. Participant 15 noted that *"end users don't necessarily understand that it's been trained on source material."* This lack of understanding risks creating unrealistic expectations, further complicating adoption efforts. (P20, P8, P15).

Lastly, cultural dynamics also significantly influences adoption outcomes. Participant 20 emphasised the need for tailored change management strategies, stating: *"the fear, uncertainty, and the need for proper change management is much higher than with other things […] also respect the different cultural backgrounds and specifics of different regions."* These words underscore the importance of a nuanced change management approach that accounts for regional and cultural differences to foster trust and ensure successful longer-term adoption (P6, P17, P20).

*5.1.2 AI Literacy & knowledge*

Firstly, the findings indicate that many organisations have not yet prioritised AI literacy as a strategic concern. Participant 4 acknowledged this, stating, "*we haven't really put any thought into [it] at the moment based on where we are*," underscoring the absence of structured initiatives to address literacy needs. Additionally, the EU AI Act's literacy requirements were unfamiliar to a notable proportion of participants, revealing a wider knowledge gap regarding regulatory mandates and their implications.

Secondly, there was considerable debate about the extent to which AI capabilities and knowledge should be internalised. Participant 9 reflected, "*I basically thought for our company, do we need this knowledge internally? Can we monetise it? Can we get an advantage if we have that [knowledge]?"* This highlights the tension between developing in-house expertise and relying on external resources, posing strategic questions about prioritising investments in AI literacy.

Thirdly, the findings reveal differing views on whether EA teams should take responsibility for AI literacy and change management. Some participants argued that these topics lie outside EA's core remit, while others highlighted its potential enabling role. Participant 13 observed, *"there is no AI literacy coming from enterprise architecture […] It is being handled by the central data & analytics team."* Conversely, other participants noted that GenAI expertise and support often sit within Group EA functions, indicating a dual role for EA in facilitating collaboration with other teams (P15, P8).

Participants highlighted the growing need for EA teams to address knowledge gaps, particularly in navigating regulatory changes. Participant 17 remarked, *"EU or the CSRD reporting laws, that's quite a lot we have to do […] and it requires much more knowledge than before for each step we are taking."* This emphasises the increasing complexity of compliance, necessitating enhanced regulatory expertise within EA teams.

The research also reveals notable disparities in AI knowledge within organisations, particularly between IT and business units. Participant 9 highlighted a "*digital divide*" in smaller organisations, where business departments often lack the AI expertise needed to operate independently. In contrast, larger organisations tend to distribute knowledge across multiple teams, which can result in inefficiencies and silos (P17). Participant 4 remarked, *"a lot of people outside of technology will probably be more familiar with or comfortable with using GenAI in certain areas than the organisation is ever going to be,"* illustrating the widening gap between end-user familiarity and organisational preparedness.

Finally, participants underscored the fragmented nature of GenAI knowledge within organisations. Participant 17 noted, "*we have different points where we have our resources and our knowhow about GenAI.*" This fragmentation amplifies difficulties in achieving cross-departmental collaboration and aligning strategies, highlighting the critical need for a cohesive approach to AI literacy and resource management.

## 5.2 How "EA for GenAI" influences EA as a Dynamic Capability

### 5.2.1 How GenAI impacts EA Frameworks

Firstly, the research highlights several limitations in current EA frameworks when addressing the disruptive nature of GenAI. Traditional frameworks like TOGAF face challenges in adapting to GenAI's rapid evolution and unique demands (P12, P13). As Participant 12 noted, "*to understand what your business really wants to achieve […] then define your transitions, stages […] for GenAI. That's an issue I think, because this topic is more in this rapid prototyping kind of stage where […] you need to adopt really fast*." This underscores the inability of TOGAF's structured approach to accommodate the iterative and dynamic requirements of GenAI adoption.

Additionally, the relevance of EA frameworks to GenAI adoption is debated. Participant 13 observed, "*I would not see this actually in the TOGAF of world to be honest. It's more guidance or enterprise architecture management topic.*" Moreover, some organisations bypass formal frameworks entirely, as highlighted by Participant 15: "I *mean, we don't really follow any framework*" and Participant 4: "*We don't formally follow a set and off-the-shelf EA framework like TOGAF.*"

The development of bespoke EA frameworks for GenAI introduces a risk of organisational silos. Participant 14 shared, "*I'm currently making up something myself […] in terms of AI taxonomy or the things that are special about AI.*" While these tailored approaches may foster innovation, they risk creating fragmentation, potentially undermining the cohesive value that EA frameworks are designed to deliver.

A notable gap in existing EA frameworks is the limited emphasis on the data domain. Participant 12 remarked, "*I have not met many enterprise architecture peers […] who have a strong consideration of the data domain in EA.*" This highlights the lack of focus on integrating data management within EA practices, despite the central role of high-quality data in enabling GenAI adoption.

The research indicates a shift in EA priorities from technology-centric models to a stronger focus on business architecture and value articulation (P12, P17). Participant 2 stated, "*it's moving from this technology focus more to the business architecture focus and being able to articulate the value proposition.*" However, this transition demands improved communication and collaboration capabilities within organisations. As Participant 17 emphasised, "*enhance the capability to communicate and collaborate inside the company […] but coming to AI Act and GDPR, it's not possible for a single person to do it.*"

The unique characteristics of GenAI demand adaptations to existing EA frameworks. Participant 10 remarked, "*they're in a different venue that you would see typically across the company for GenAI*" while Participant 11 emphasised, "*it's more how can we fit AI into the frameworks.*" These findings suggest a need for flexible, parallel processes within EA frameworks to address GenAI's distinct requirements (P1, P10, P11, P15).

Adapting EA frameworks also involves revising guidelines and principles to accommodate GenAI (P17, P5). This includes adding new layers to address business, IT infrastructure, and cultural dimensions. As Participant 17 explained, "*you can define the implementation of regulations as part of linking business and IT.*" Additionally, dynamic principles and guidelines are required to ensure agility, as noted by Participant 5: "*We're to put a few new things in place, but that's not so unusual to define new principles and guidelines specifically for these topics*."

Finally, several participants highlighted the importance of continuity and incremental evolution within existing EA frameworks (P10, P11, P5). Participant 10 observed, "*basic concepts of business, architecture and data architecture and infrastructure architecture are all very much there,*" emphasising the foundational stability of current frameworks. However, the more transformative aspects of GenAI adoption were noted to exist outside these established processes.

This suggests that while EA frameworks effectively support operational continuity and incremental improvements, they may lack the agility necessary for addressing the disruptive innovation associated with GenAI. Consequently, organisations may need to establish parallel processes or external mechanisms to meet transformational demands without undermining the stability provided by existing EA structures.

### *5.2.2 How GenAI is impacting EA Operating Models*

**Key Finding:** GenAI adoption exposes both the strengths and limitations of existing EA operating models, emphasising the need for flexibility, cross-functional collaboration, and strategic alignment to balance the demands of innovation, governance, and scalability.

Firstly, heavily decentralised business units create ongoing challenges for GenAI adoption, with fragmented environments impeding scalability and consistency (P6, P19, P10). Participant 6 characterised their organisation as "*extremely federated for most of the data…some data that is becoming more standardised, but the majority of the information we have is like a black box,*" highlighting significant barriers to scaling due to insufficient standardisation. Similarly, Participant 19 emphasised the difficulty of "*adapting across decentralised structures*," further underscoring the complexity of implementing GenAI in such environments.

Secondly, EA's limited strategic influence in decentralised settings further exacerbates challenges in GenAI adoption (P15, P5, P6). Participant 16 commented, "We wouldn't say, we've not got this AI capability, we should probably look at something in this space to try and fill that gap," highlighting a reactive rather than proactive approach to addressing capability gaps. These findings emphasise the need for capability-based planning to position EA as a strategic enabler of GenAI initiatives. Participants also noted a lack of EA involvement early in the GenAI lifecycle (P4), potentially stemming from perceptions that EA processes slow down GenAI implementation (P12).

Interestingly, the method of introducing GenAI into organisations significantly affects EA's role. In many cases, external pressures such as market demands and executive interest have driven adoption (P18, P13, P1, P12, P17). Participant 18 noted, "*Our CEO has a huge interest that we improve in all areas the usage of AI*" yet acknowledged the absence of a coherent model to integrate these advancements. This misalignment often leads to fragmented initiatives, limiting EA's ability to harmonise efforts.

Participant 10 discussed how their organisation established an AI centre of excellence (CoE) to centralise strategies and streamline AI-related functions. However, this approach risks sidelining EA, as centralised efforts may focus narrowly on immediate AI needs rather than broader architectural alignment. As Participant 12 pointed out, EA's role "*depends on how the enterprise architecture practice is positioned in the company.*"

Additionally, EA's role in GenAI initiatives predominantly leans towards governance, with a focus on oversight and guidance rather than strategic leadership. Participants highlighted EA's involvement in approving or providing direction on GenAI requests (P10, P11, P15, P5, P8), often acting as a governance body (P12, P13, P15, P10). Participant 12 explicitly described EA as "*a governance body*" while Participant 10 reinforced this by outlining EA's role in promoting data reuse, a critical component for GenAI adoption. However, this governance-centric approach reflects a broader issue of low EA maturity in many organisations, limiting its strategic influence.

In contrast, some participants shared examples of EA evolving to balance stability with innovation. For instance, organisations are experimenting with start-up-inspired models for GenAI, creating dedicated teams to drive innovation (P11, P12):

P11: *"have a kind of internal startup to lead this because it's a new topic, no one is was able to handle it… we are no more focusing only on the classic enterprise architecture… but also to take a look on disruptive initiatives"*.

Participant 11 noted how GenAI initiatives are reshaping EA teams, resetting expectations, and driving greater investment in EA capabilities (P11, P15). These shifts underscore the potential for EA to transition from a purely governance-oriented function to a more dynamic enabler of innovation, aligning GenAI adoption with broader strategic goals. Participant 19, a GenAI innovation lead, highlighted the potential shift towards a more crowdsourced approach for disruptive technologies, stating, "*The only advantage we have is to say that the only way we can catch up with new technology is by developing everything together.*"

Participant insights highlighted significant differences in how EA teams are structured and their roles in GenAI initiatives. A key distinction is the adoption of federated versus centralised EA operating models (refer to Figure 17, which illustrates these models). In federated models, GenAI knowledge and support are typically housed within Group EA functions but often lack a comprehensive strategy for scaling. Participant 15 described Group EA's emphasis on internal productivity and developing typologies for embedding EA in products, while Participant 8 noted similar activities with no clear direction for broader scaling. Conversely, centralised models tend to prioritise compliance and reporting, as Participant 14 observed, focusing on meeting regulatory obligations rather than fostering innovation.

Additionally, EA teams exhibit diverse strategies for engaging with GenAI. Some deliberately adopt a generalist approach, focusing exclusively on internal GenAI applications while delegating customer-facing products to specialised teams. Participant 10 noted, "*Where we put GenAI and products we sell our customers, that's a whole different team that does that, and I'm not engaged in that piece at all.*" This separation highlights a strategic decision by certain EA teams to prioritise a broad focus rather than evolving into a tactical function dedicated solely to GenAI, reflecting organisational priorities and resource limitations.

Adapting EA operating models for GenAI integration necessitates dynamic governance frameworks and tailored strategies (P1, P11, P13, P6). Participants (P10, P11, P15, P3, P5, P9) highlighted the importance of mechanisms such as AI steering committees to ensure effective oversight. Participant 10 described their organisation's approach: "*They go through an incubation process… and then through an AI board that reviews all of the different scenarios and options.*" The development of GenAI-specific typologies and capabilities also emerged as a priority. Participant 8 emphasised the importance of structured typologies, stating: "*the fact of structuring implementation typologies to be clear about the type of things you are doing… you are ultimately enriching this solution with GenAI capabilities*." Furthermore, Participant 7 identified a dual focus on efficiency and competitive advantage as critical drivers of capability development, underscoring the strategic imperatives behind these initiatives.

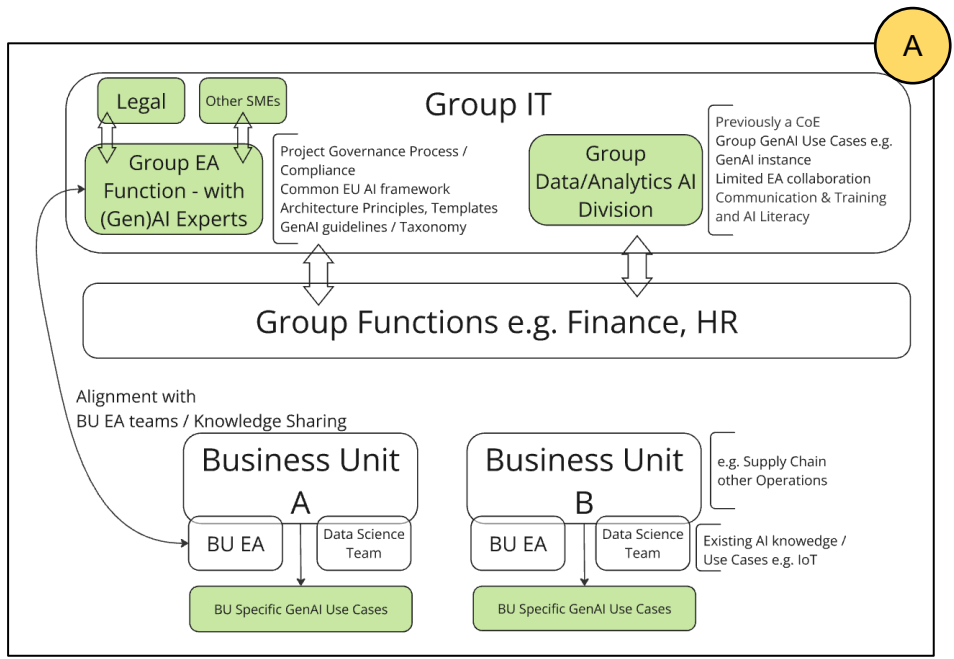
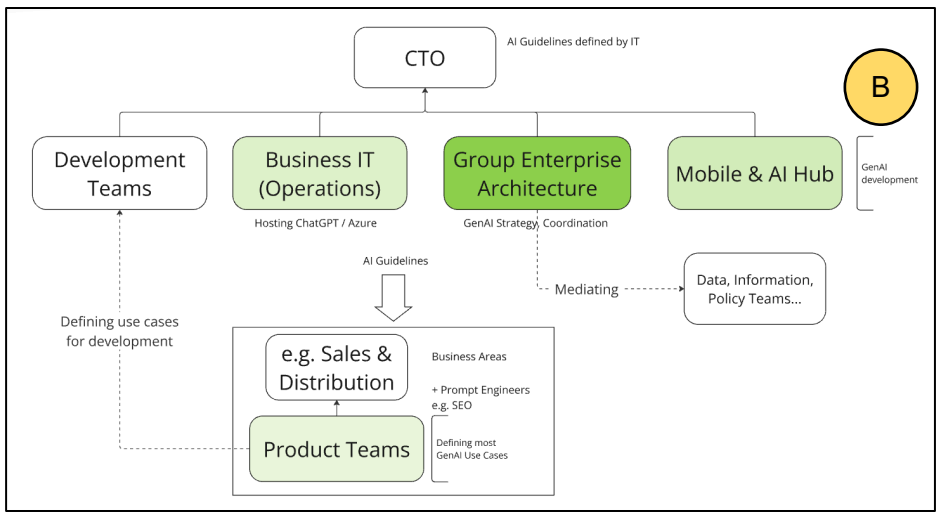
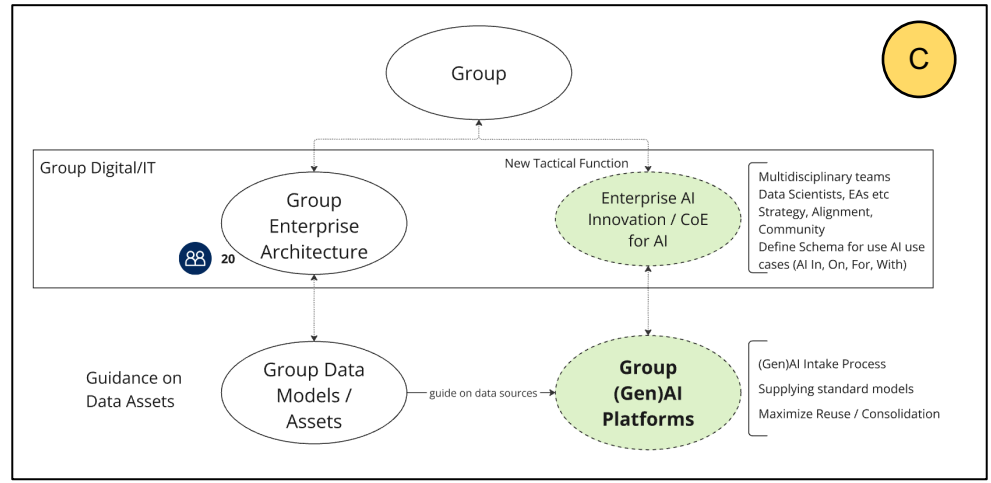
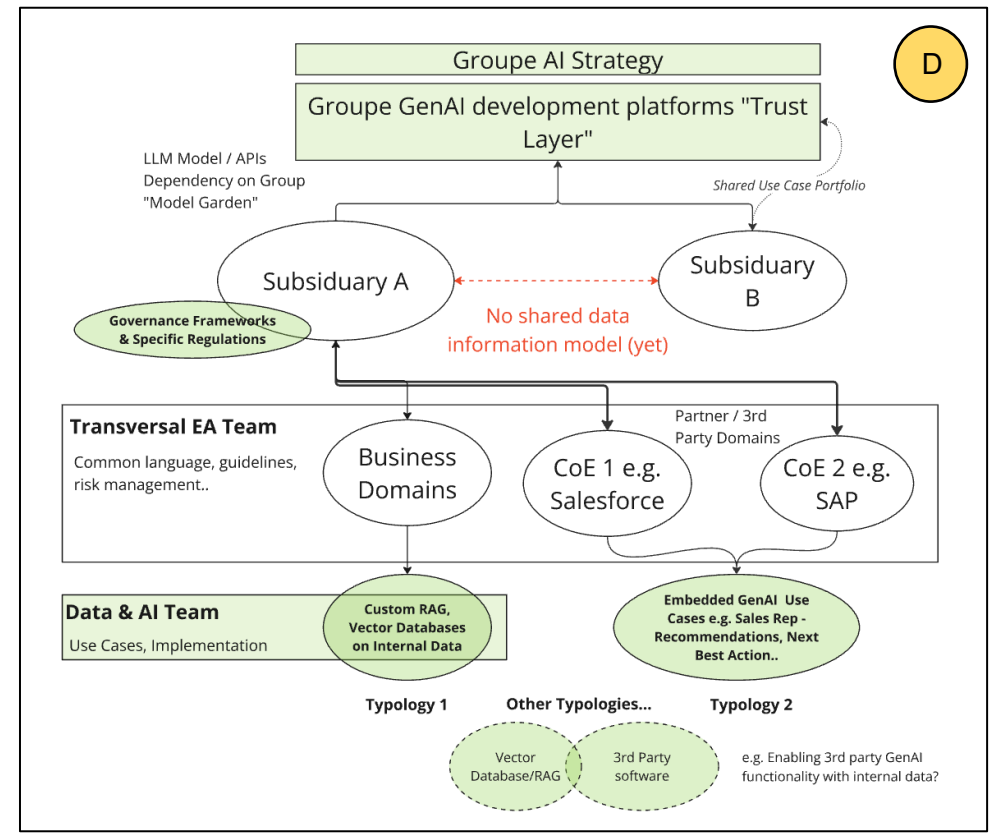

Figure 17. Extract of the qualitative analysis illustrating four (A, B, C, D) different EA operating models & structures for implementing GenAI

### 5.2.3 How the EU AI Act impacts EAs and GenAI adoption

**Key Finding:** EA is in a good position to support regulatory frameworks like the EU AI Act, which is gaining recognition as a global standard for AI governance. However, the requirements of the act demand certain adjustments to operating models, the development of new capabilities, and greater adaptability in compliance strategies. While EA is acknowledged as a key driver of regulatory compliance, there are ongoing debates about the extent to which these efforts should shape EA's strategic priorities and resource allocation.

The EU AI Act is driving significant changes in operating models and governance processes. Participant 13 highlighted its influence, stating, *"From [the EU AI Act] coming…we are changing our operating model. We have changed it (the intake process) …there is a big AI swim lane in it."* The act is increasingly being treated as a global regulatory standard, shaping practices in organisations worldwide (P5, P13).

However, compliance with high-risk use cases under the Act presents substantial challenges. Participant 5 noted, *"[High-risk applications under the EU AI Act] bring a lot of documentation and technical setup … that would require us to do a lot of new work and establish some additional capabilities that we don't have in the organisation today."*

Additionally, the Act has introduced more extensive governance and compliance processes for GenAI use cases, including rigorous assessments and reviews (P5, P7, P9). Participant 5 emphasised the increased focus, stating, *"Governance, risk management, compliance on this topic…in light of the EU AI Act, is very prominent in this topic, much more so than in…other previous topics."*

Adapting to the regulatory demands of the EU AI Act requires organisations to adopt flexible and innovative strategies. Participant 17 highlighted the complexity, stating, "*It's quite hard for [the business] to identify if my tool is using AI according to the EU AI Act, and that's much more complex than GDPR.*" This complexity has driven some organisations to explore modular, platform-based approaches to compliance (P19).

Knowledge gaps within EA teams further complicate adaptation, particularly in smaller or localised teams and where AI risk frameworks remain misaligned with the Act's requirements (P12, P8, P15). Participant 15 observed, "*AI risk frameworks are in place, but they have not yet been adapted to meet the requirements of the EU AI Act.*"

A critical issue is understanding the scope and implications of GenAI use cases. Participant 17 identified a key challenge: "*The major challenge is to define what is AI and what isn't, and to identify if someone…is using GenAI.*" This highlights the importance of identifying and monitoring GenAI use cases early while implementing measures to ensure compliance (P5, P7, P17).

EA plays a pivotal role in ensuring baseline compliance with the EU AI Act by supporting audits and advancing the broader compliance agenda (P11, P13, P17, P7, P9). Participant 9 highlighted the structured processes in place: "*We have a process established that clearly defines which AI use cases are presented to management and which use cases can be decided on project level.*" EA's ability to maintain an end-to-end inventory of GenAI use cases is particularly valuable in promoting transparency and regulatory alignment (P17). Collaboration is a key element of these efforts, with EA teams working closely with legal and enterprise risk departments to conduct comprehensive assessments of AI use cases. As Participant 5 explained, "*We're assessing them also with legal, and basically legal looks at these use cases from an EU AI perspective as well.*"

## 5.3 GenAI In EA

**Key Finding:** EA, as a Dynamic Capability, offers substantial potential for harnessing GenAI through high-value use cases that can optimise and elevate its organisational impact. However, the adoption of GenAI within EA faces similar challenges to broader implementation efforts, with data quality emerging as a critical foundational barrier.

Participants highlighted the transformative potential of GenAI in EA (P17, P2,), particularly for large organisations with high portfolio complexity (P9) and smaller organisations constrained by limited resources (P11). Participant 17 remarked, *"The potential for EA is quite big, I think, because EA is a field that isn't affected that much at the moment by GenAI,"* underscoring untapped opportunities.

Emerging applications include EA chatbots (P15, P17, P4, P5, P9), text-to-diagram functionalities (P11), regulatory compliance assessments (P1, P15), landscape optimisation and rationalisation (P7), and architecture reviews (P1). These tools are seen as pivotal in democratising EA, automating low-value tasks, and freeing resources for strategic decision-making (P1, P11, P17, P4). Participant 11 described the democratising effect: *"A project manager without having a lot of technical skills is able to provide a high-level diagram quickly."*

While current EA tools demonstrate relatively low-value use cases, trends like GraphRAG point to a promising trajectory for future applications (P1).

GenAI has the potential to significantly enhance EA's organisational value by increasing the accessibility of EA tools and reshaping perceptions of its role (P17, P9, P11, P2). Participant 17 highlighted this opportunity, stating, *"That's a point where we can get more importance in the company because we see that we have to put all the information together to get a single source of truth."*

Additionally, GenAI can support decentralised decision-making. Participant 2 observed, "*I cannot sit in each and every decision board, but if it's decentralised and with the help of GenAI, I believe that decentralised decision-making can be supported in the sense of what the global EA team is thinking or what the intentional architecture is about*."

Future EA roles may also evolve, with Enterprise Architects potentially acting as product owners of their AI agents to ensure alignment with organisational goals and values (P2). This shift could further drive adoption of EA tools and practices (P11).

Lastly, as with broader GenAI adoption, data quality is a critical challenge for its application within EA. Participant 15 underscored the reliance on robust data discovery processes. Other participants highlighted how the value of GenAI varies depending on the maturity of an organisation's technology stack. Achieving high-quality data will be pivotal in unlocking the full potential of GenAI within EA.

5.4 **EA Dynamic Capabilities**

The findings are organised according to the three routine dimensions DCs.

*5.4.1 Sensing*

**Key Finding:** EA demonstrates substantial potential to enhance sensing routines as a Dynamic Capability, allowing organisations to proactively address business demands, market trends, and ecosystem dynamics. However, these capabilities are frequently limited by resource constraints, differing levels of EA maturity, and misaligned priorities. EA's effectiveness in sensing is strongly shaped by its operating model, presenting opportunities to move beyond governance functions into forward-looking trend analysis and strategic decision-making. These findings are summarised in Figure 18.

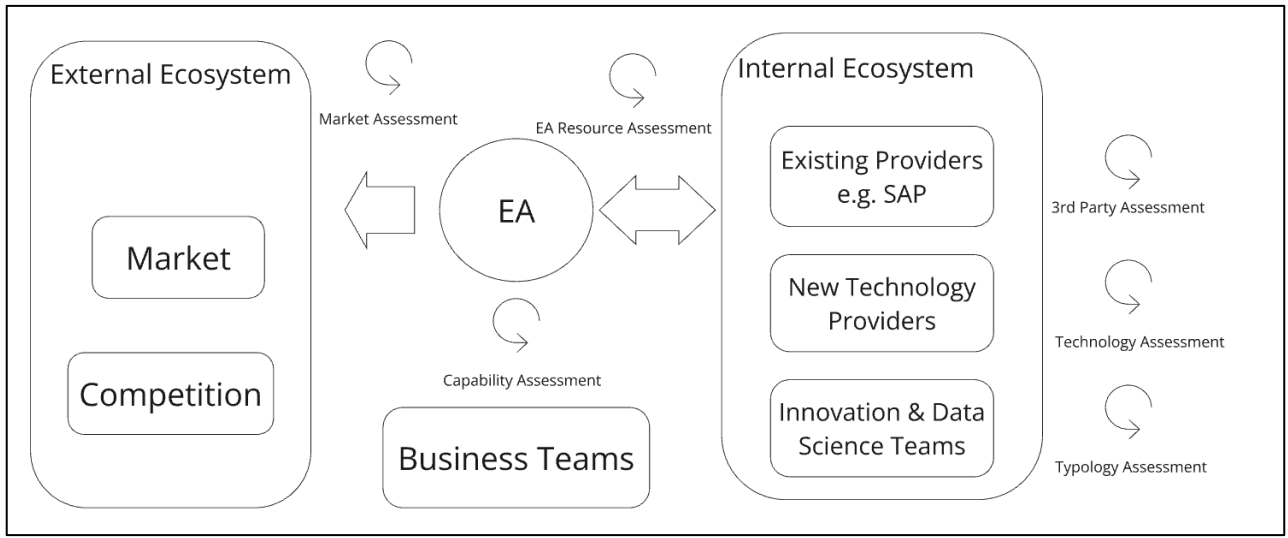

Figure 18. EA sensing conceptual framework

Firstly, EA's role in understanding and responding to business and market dynamics is crucial. Participants highlighted the need for EA to provide a transversal view of the enterprise while engaging with the business and educating stakeholders on GenAI opportunities. For instance, one participant noted, "*It's not strictly a data and AI topic; it's a business topic… architecture takes on meaning by gathering and identifying opportunities within value chains*" (P8). Another stated, "*These things hit the market, and there's lots of uninformed people that want to chase it. That's where EA needs to come in and say, let's actually break this thing down*" (P1).

Additionally, participants emphasised the importance of creating taxonomies for GenAI use cases to assess applicability and identify high-value opportunities. As one participant explained, "*You do a quick scan of your organisation and identify four or five places where this is going to be *beneficial*" (P1). Another added, "*I'm trying to think up dimensions where we capture this so that whatever happens in a few years, we can answer these questions*" (P14).

EA has a clear opportunity to leverage its work around business capabilities[4], aligning GenAI initiatives with strategic priorities. However, challenges include mapping GenAI across all business capabilities and determining its strategic relevance. One participant observed, "*If you have a business capability and want to do GenAI there, but we only have systems of record, is that a smart investment? Is it strategically meaningful?*" (P12).

EA's role extends to harmonising customer experiences and sensing broader ecosystem changes. As one participant put it, "*It's about trying to stay ahead of the curve so we don't just react to the dynamic but instead provide frameworks and guidelines*" (P8). Another emphasised the long-term focus needed: "*How can we try to even form some sense, some standard, where will this take our business in five years?*" (P15).

Participants highlighted the challenge of reallocating EA resources dynamically to meet evolving demands. One participant acknowledged, "*That's something I wish we had already this year*" (P5). This underscores the need for EA teams to remain flexible and proactive in addressing shifting priorities.

EA teams are tasked with maintaining focus on technology roadmaps, though participants noted the potential for increased technical debt if resources are misaligned. While several emphasised the importance of roadmaps, limited input was provided on balancing this with managing technical debt effectively (P15, P20).

---

[4] Business capabilities in the context of EA represent the core functions and competencies that enable an organisation to align strategy with operational execution effectively.

EA can accelerate GenAI adoption by engaging with external ecosystems during early phases, such as filtering priorities through partner ecosystems and leveraging initial proofs of concept. As one participant explained, *"We partner early to filter priorities and accelerate proofs of concept before internal scaling"* (P3).

Participants emphasised the key role of software vendor collaborations, for example, *"Partnering with third-party vendors provides a clearer view of what's required for GenAI adoption"* (P8), and another noted partnership with universities for executive training as an effective strategy (P6).

*5.4.2 Seizing*

**Key Finding:** The structure of GenAI intake processes significantly impacts EA's role in implementation and scaling. Centralised approaches and fragmented models often reduce EA's visibility into the GenAI pipeline, limiting its ability to set guardrails and harmonise use cases. Expanding EA's role in demand management requires addressing gaps in early involvement, resource availability, and knowledge alignment.

Firstly, EA plays a vital role in scaling GenAI by offering structure and oversight for disruptive technologies. Participant 11 emphasised, *"this is where the position of enterprise architecture, in my opinion, is critical, because to avoid the uncontrolled creation of and adoption of GenAI technologies."* In some cases, adopting a start-up model for GenAI within EA has been suggested as a way to accelerate scaling efforts (P20). However, ensuring alignment between EA's governance structures and business priorities remains a challenge (P11, P17, P20).

Secondly, participants highlighted the growing demand for EA involvement in GenAI initiatives, yet there are differing perspectives on the extent to which EA should "shift left" and take on a more proactive role. Participant 16 reflected on this challenge, stating, "*We have a lot of stuff of trying to shift left, and it might be that actually shifting left isn't a great idea… Is it right to keep an AI function separate? Is it better to build an AI CoE and keep your knowledge of that space rather than trying to make everybody experts?*" This reflects the broader debate around balancing EA's involvement in early ideation phases versus focusing on traditional governance responsibilities (P11, P13, P16).

Furthermore, the GenAI demand management process varies widely based on organisational maturity and operating models, as shown in Figure 19. Centralised frameworks offer greater EA visibility and control, while decentralised or experimental models, such as those led by AI Centres of Excellence, often lead to siloed efforts (P10, P13). Participant 11 highlighted the challenges of early-stage initiatives, stating, "*the nature of the initiative is different because, at the very beginning, you don't know what you want in terms of technology*" underscoring the need for adaptable frameworks that address GenAI's unique features, including value-driven triggers and uncertain technological requirements.

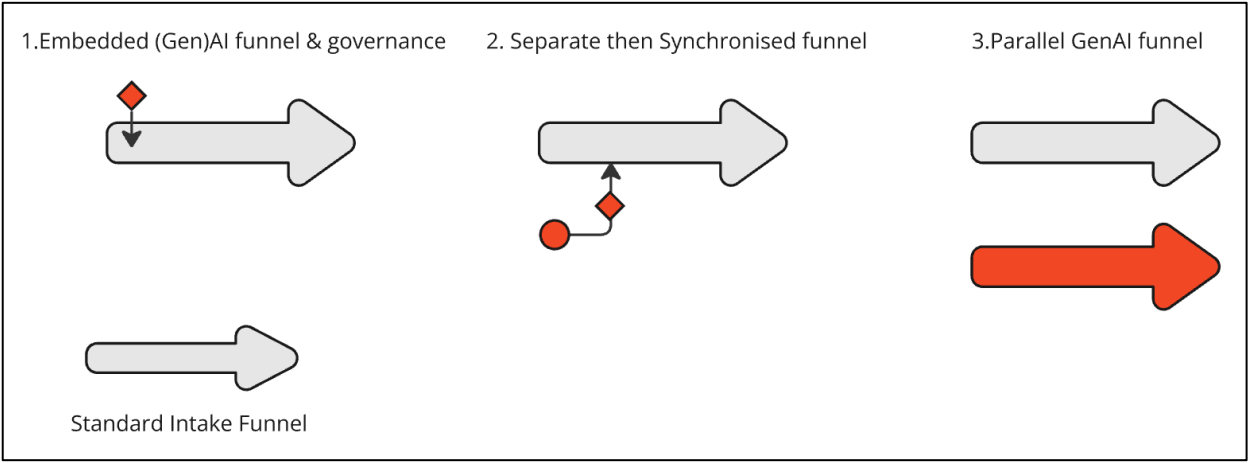

Figure 19. Seizing - Demand Management Frameworks with GenAI

Effectively managing GenAI use cases is essential for scaling adoption and minimising inefficiencies. Participant 8 remarked, "*the challenge of EA is to try to address and approach every [GenAI] use case in the same way*." Adapted EA frameworks have been instrumental in rationalising and structuring use cases, enabling the identification of high-value opportunities aligned with organisational priorities (P8, P12, P15). For instance, capability mapping has allowed EA teams to prioritise initiatives like chatbot optimisation to enhance customer experience (P20). The most predominant scaling frameworks are summarised in Figure 20.

Tools for discovering and controlling AI software usage have further enhanced EA's ability to manage and scale GenAI initiatives. Participant 10 likened these tools to a marketplace, stating, "*think of an Amazon marketplace…we can actually go and buy [GenAI use cases] …and it will configure on your platform for you*." Such innovations accelerate scaling efforts while maintaining centralised oversight.

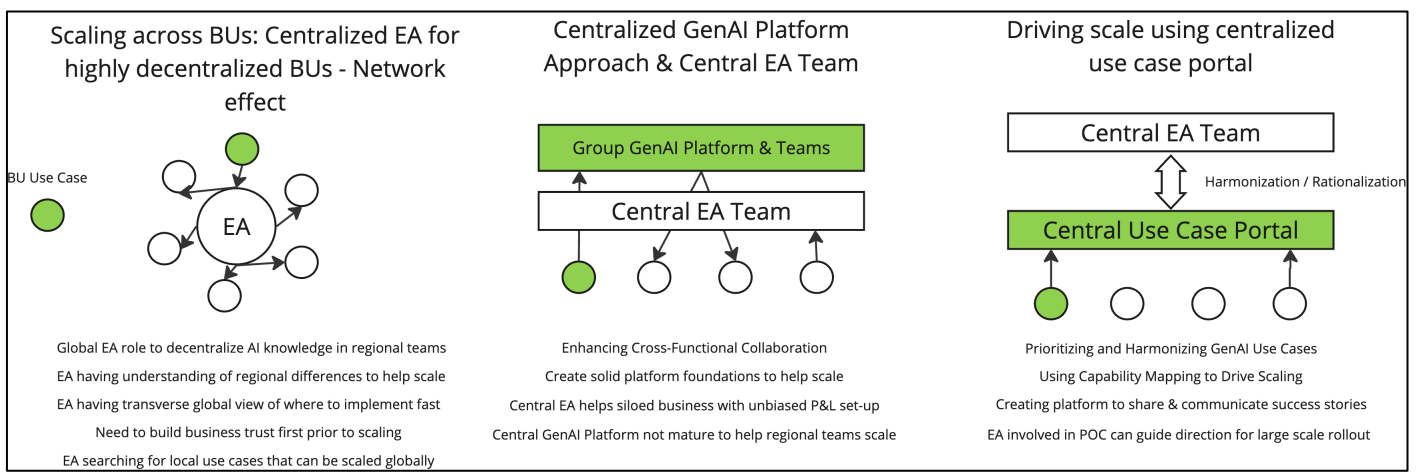

Figure 20. Seizing – Scaling frameworks and related 2nd order themes

The findings reveal that EA's role in scaling GenAI extends beyond governance to include active engagement in proof-of-concept projects and strategic decision-making. Early involvement enables EA to guide transitions to large-scale rollouts while ensuring alignment with organisational goals (P11, P17). Centralised GenAI platforms and frameworks facilitate communication and collaboration, promoting successful practices and consistent application across the organisation (P19).

Scaling GenAI effectively also depends on robust infrastructure. Participant 19 highlighted the importance of preparation, stating, "*In China, we say that if you want to cut the tree, you should sharpen your knife first.*" However, scaling efforts must address EA resource limitations and the need to adapt to regional differences, as observed in initiatives across divisions and countries (P3, P12, P15).

*5.4.3 Transforming*

**Key Finding:** The findings underscore a reconfiguration of the traditional EA role as a vital intangible asset within organisations. GenAI adoption requires enhancing EA's executive advocacy, fostering effective communication, and building stronger execution-focused capabilities. Overcoming misconceptions, reshaping team culture, and dynamically reskilling EA professionals are essential to achieving scalable and strategic GenAI integration.

Firstly, executive advocacy and communication were highlighted as critical for enhancing EA's influence. Participant 4 pointed to the importance of gaining senior leadership's acceptance, stating, "*What I'm currently struggling with is having the acceptance and understanding at more senior levels.*" Participant 11 further emphasised EA's role in evangelising GenAI: "*It's pretty powerful the position today because they are not only managing the incoming demand but also doing great work to go and evangelise a little bit.*" The opportunity to redirect GenAI hype toward actionable initiatives was also noted. Participant 6 suggested EA could guide organisations toward data-centric priorities, while Participant 3 exposed the need for effective communication to market use cases effectively.

Furthermore, misconceptions about EA's role remain a significant barrier, with EA often perceived as an innovation blocker or misunderstood entirely. Participant 15 acknowledged this perception, stating that EA is sometimes seen as "*the bad guys*," despite providing valuable feedback. Participant 19, a GenAI innovation lead, highlighted the lack of awareness, admitting, "*I'm still confused about what enterprise architecture is*," reflecting a broader issue of invisibility outside IT. Similarly, Participant 4 remarked, *"I'd rather we were part of the conversation until we can actually fulfil what we would like to do in that area."*

Cultural challenges within EA teams further complicate their role. Participant 3 observed that some EAs are too reliant on reactive, end-of-initiative firefighting. Reconfiguring team culture is necessary, as highlighted by Participant 12: "*They've changed their hats*," adapting to GenAI demands by assuming new responsibilities, such as platform architecture.

Moreover, workforce transformation is equally important. Participant 20 emphasised the "*need to invest heavily into upskilling, education, and change management*." Participant 6 echoed this, describing efforts to define AI ambassadors within teams to facilitate these transitions.

Finally, leveraging expertise across the organisation emerged as a crucial theme. Participant 19 highlighted the importance of cross-functional knowledge sharing, while Participant 3 emphasised the significant skill development required to maximise GenAI's potential: "*There's definitely a bit that can't be underestimated there in terms of skilling up and being able to really leverage those capabilities.*"

## Chapter 6 – Discussion

This chapter examines the findings in relation to the research question and explores research and practical implications, followed by an outline of study limitations and directions for future research.

6.1 **Research Question**

Firstly, the findings of this study indicate that existing EA frameworks are not fully equipped to handle the complexities posed by GenAI. This aligns with the results from the literature review, advocating adaptive and experimental approaches for disruptive technologies (Shanks et al., 2018; Stecher et al., 2020). GenAI's rapid innovation cycles, broad applicability, and significant scalability require EA frameworks to evolve fundamentally, balancing operational continuity, compliance, and transformative innovation. The findings also present two distinct trends among organisations: early adopters and early majority, each embracing GenAI through differing approaches. In addition, study highlights moderating factors such as organisational size, operating models, and EA maturity as critical in shaping the evolution of these frameworks to address emerging challenges.

Second, the findings highlight a nuanced dichotomy between larger organisations, which often prioritise risk management and regulatory compliance, and smaller organisations and EA teams that tend to adopt more flexible and experimental approaches to GenAI adoption. While smaller organisations are more agile, they frequently lack the resources to fully operationalise GenAI initiatives, resulting in fragmented adoption strategies. Larger firms, on the other hand, face challenges associated with compliance-heavy frameworks, which can stifle innovation, as noted in prior research and surveys (Rittelmeyer et al., 2021; SAP LeanIX, 2024). However, more mature and resourceful organisations are better equipped to adapt their strategies and operating models, for instance, by incubating dedicated teams and resources to support GenAI initiatives. Nevertheless, both organisational models share a common risk of reverting to "business-as-usual" practices, especially when confronted with external regulatory pressures like those from the EU AI Act (Rittelmeyer et al., 2021). This alludes to the importance of evolving governance frameworks that can integrate GenAI strategies without neglecting opportunities for strategic alignment.

Collaboration emerged as a critical yet underutilised factor in advancing EA frameworks. While the literature highlights its importance (Sedkaoui & Benaichouba, 2024), the findings reveal that many organisations restrict collaboration to basic coordination between EA and business units. Genuine collaboration, involving cross-functional synergies and diverse skill sets to address GenAI's disruptive potential, is often lacking. This gap underscores the need for frameworks that go beyond compliance to actively foster innovation. Adopting iterative approaches, such as developing use case backlogs and fostering cross-functional teams, could help organisations overcome inertia and align GenAI adoption with strategic objectives.

The study also emphasises the need for reskilling EA teams and enhancing specific GenAI competencies. Talent redeployment, including drawing expertise from data science or innovation teams, as well as external hires, can bridge knowledge gaps. As van de Wetering (2022) and Pattij et al. (2022) highlight, EA must evolve to address changing business priorities, particularly in high-paced environments.

TOGAF, while foundational, was noted as insufficient for GenAI, echoing Kotusev and Kurnia (2020) and Sandkuhl & Rittelmeyer (2022) reflexions on its inability to support evolving and innovation driven technologies. EA frameworks need to be adaptable, focus on strategic alignment, and include mechanisms to balance compliance with innovation.

Furthermore, the findings show that the EU AI Act serves as a key incentive for firms and EAM to adopt Dynamic Capabilities (DCs) and driving the need for more robust compliance processes. Its global adoption is compelling EA to strengthen collaboration with legal, security, and compliance teams. However, the findings caution that excessive focus on governance risks, sidelining business and strategic architectural priorities. As such, true collaboration must extend beyond coordination, requiring flexibility, diverse skill sets, and closer alignment with evolving business needs (Sedkaoui & Benaichouba, 2024). Striking the right balance remains a significant challenge, as organisations work to reconcile adaptable governance frameworks with the need for agile, innovation-driven strategies.

Secondly, the study highlights the critical role of EA's Dynamic Capabilities, sensing, seizing, and transforming, in driving scalable GenAI adoption. These findings align with Teece's (2007) framework, which positions DC's as essential for reconfiguring organisational assets in response to disruptive innovations. However, this study extends the literature by illustrating how these capabilities are uniquely applied within the context of GenAI.

In the sensing dimension, EA's role in identifying technological trends and assessing organisational readiness emerged as pivotal. However, resource constraints and varying levels of maturity often hinder proactive sensing. As van de Wetering (2022) observes, EA maturity significantly affects its ability to navigate innovation. The study reveals a duality in EA's role: in some cases, EA reacts to overwhelming business demand driven by GenAI's hype; in others, particularly in smaller organisations, EA inadvertently takes the lead due to a lack of AI expertise elsewhere. Enhancing sensing capabilities, with routines such as monitoring market trends, aligning GenAI use cases with business priorities, and creating taxonomies, could position EA as a more strategic enabler in scaling GenAI.

The seizing phase revealed inconsistencies in EA's role within demand management processes, aligning with Shanks et al. (2018), who emphasised the importance of structured routines including demand management for disruptive technologies. The findings highlight the need for EA teams to develop comprehensive demand management capabilities and enhance collaboration with innovation and data science teams. This approach ensures that GenAI initiatives align with organisational objectives and are positioned for sustainable scaling. Tailored frameworks, such as innovation funnels or centralised AI governance processes. echoing Müller's work (2024), were identified as essential for addressing gaps and harmonising GenAI initiatives.

Thirdly, the transforming dimension highlights the need to adjust EA's structures, skills, and culture to align with GenAI's disruptive potential. This aligns with Holmström et al.'s (2024) findings on bridging organisational knowledge gaps to support adoption. The findings reveal that to reposition as a strategic partner, EA must first address misconceptions about its role and improve communication, including its "marketability". This cultural reconfiguration requires substantial investment in upskilling, education, and change management to equip EA professionals with the competencies needed for GenAI integration. Lastly, by leveraging or promoting its central position, EA has the opportunity to champion GenAI as an ambassador for innovation, enabling the development of new business models and value streams while maintaining alignment with longer term organisational goals.

Ultimately, the findings demonstrate that while EA's Dynamic Capabilities are pivotal for GenAI adoption, their effectiveness is moderated by existing organisational structures, EA maturity, and resource availability. By addressing these challenges, EA can transition from a governance heavy function to a strategic enabler of innovation and scalability.

6.2 **Research Implications**

This study makes significant contributions to research at the intersection of EAM and GenAI adoption. It is the first to explore how EA can evolve to enable scalable GenAI adoption in large organisations and how EAM, as a DC, can facilitate sustainable adoption, advancing both theoretical and practical understanding of the topic.

*6.2.1 Adoption of Generative AI in Large Organisations*

This study presents a conceptual framework illustrating critical factors influencing GenAI adoption in large organisations, highlighting their interaction with EA frameworks and organisational dynamics. Unlike earlier studies that broadly explore GenAI adoption (Agrawal, 2023; Gupta et al., 2024) or responsible GenAI governance (Kongsten et al., 2024), this research contextualises these factors within the EA domain, addressing a significant gap in the literature. Furthermore, this study uniquely explores the connection between EA frameworks, operating models, and the core dimensions of DCs in the context of disruptive technologies.

The findings reaffirm the complexity of GenAI adoption, characterised by barriers such as limited large scale use cases, fragmented governance, and inconsistent strategies. These align with the TOE (Technology-Organisation-Environment) framework, widely used to analyse technological adoption determinants (Agrawal, 2023 & Kongsten et al. 2024). Regulatory support and technology maturity, particularly under frameworks like the EU AI Act, emerge as pivotal factors for EA. The need for compliance adds complexity, requiring organisations to translate high level standards into actionable governance mechanisms.

This research uniquely emphasises EA's role in shaping GenAI adoption strategies. While larger organisations often rely on compliance driven, top-down approaches, smaller firms leverage bottom-up strategies, promoting agility and experimentation. The study highlights the need for adaptive frameworks that balance regulatory compliance with opportunities for strategic innovation.

Structured processes to manage GenAI initiatives are essential. EA can create taxonomies for GenAI use cases, assess applicability, identify high-value opportunities, and mitigate risks of sprawl. These approaches foster collaboration across business units and align GenAI initiatives with broader objectives. By tackling these challenges, this research enhances understanding of how large organisations can strategically navigate GenAI adoption through EA, positioning EA at the forefront of scalable GenAI implementation.

### *6.2.2 The Role of Enterprise Architecture in GenAI Adoption*

This study advances understanding of EA as a critical enabler in GenAI adoption, highlighting its potential to function as a DC. Drawing on Teece's (2007) framework, the research demonstrates how EA's sensing, seizing, and transforming capabilities address GenAI's unique demands.

The findings align with Rittelmeyer et al. (2021) observation on the limitations of traditional EA frameworks in high-paced, innovation driven environments. This study expands on their work, showing how EA frameworks can evolve to integrate GenAI specific requirements. Enhanced sensing capabilities, such as leveraging business capability maps, help prioritise investments and mitigate risks of sprawl, aligning with van de Wetering's (2022) emphasis on strategic alignment. Challenges in seizing opportunities, such as siloed demand management processes, limit EA's ability to scale GenAI effectively. The research underscores the importance of integrated approaches to harmonise use cases and optimise scaling strategies, contributing to literature on EA's role in managing disruptive technologies.

Finally, the transformative potential of EA is evident in fostering cross-functional collaboration and addressing organisational inertia. While earlier studies focused on EA's governance role (Shanks et al., 2018), this research uniquely positions EA as a change agent, capable of reshaping team structures, skillsets, and cultural dynamics to support GenAI adoption. By connecting EA's Dynamic Capabilities with practical implications for GenAI integration, this study bridges theoretical gaps in the field.

6.3 **Practical Implications**

This study offers valuable insights for organisations, particularly those operating within Europe and navigating the stringent regulatory requirements of frameworks like the EU AI Act. It provides executive leadership teams with actionable recommendations to align strategic objectives with GenAI adoption, while offering EA leaders guidance on leveraging EA's Dynamic Capabilities: sensing, seizing, and transforming, to support scalable, compliant, and value-driven innovation. Business teams can also benefit from the findings by enhancing collaboration with EA, ensuring alignment across functions, and fostering a shared understanding of GenAI's potential.

The study introduces practical frameworks for EA operating models, demand management, and scaling, offering organisations structured approaches to address the complexities of GenAI adoption. Figure 21, illustrating the "EA & GenAI strategic adoption quadrants," serves as a practical application of the study's findings. By aligning EA's Dynamic Capabilities with organisational readiness, this framework provides a roadmap for organisations to assess their current state, prioritise investments, and strategically advance their GenAI initiatives with EA front and centre. It supports businesses in maturing EA Dynamic Capabilities, improving cross-functional collaboration, and achieving the right balance between regulatory compliance and innovative growth.

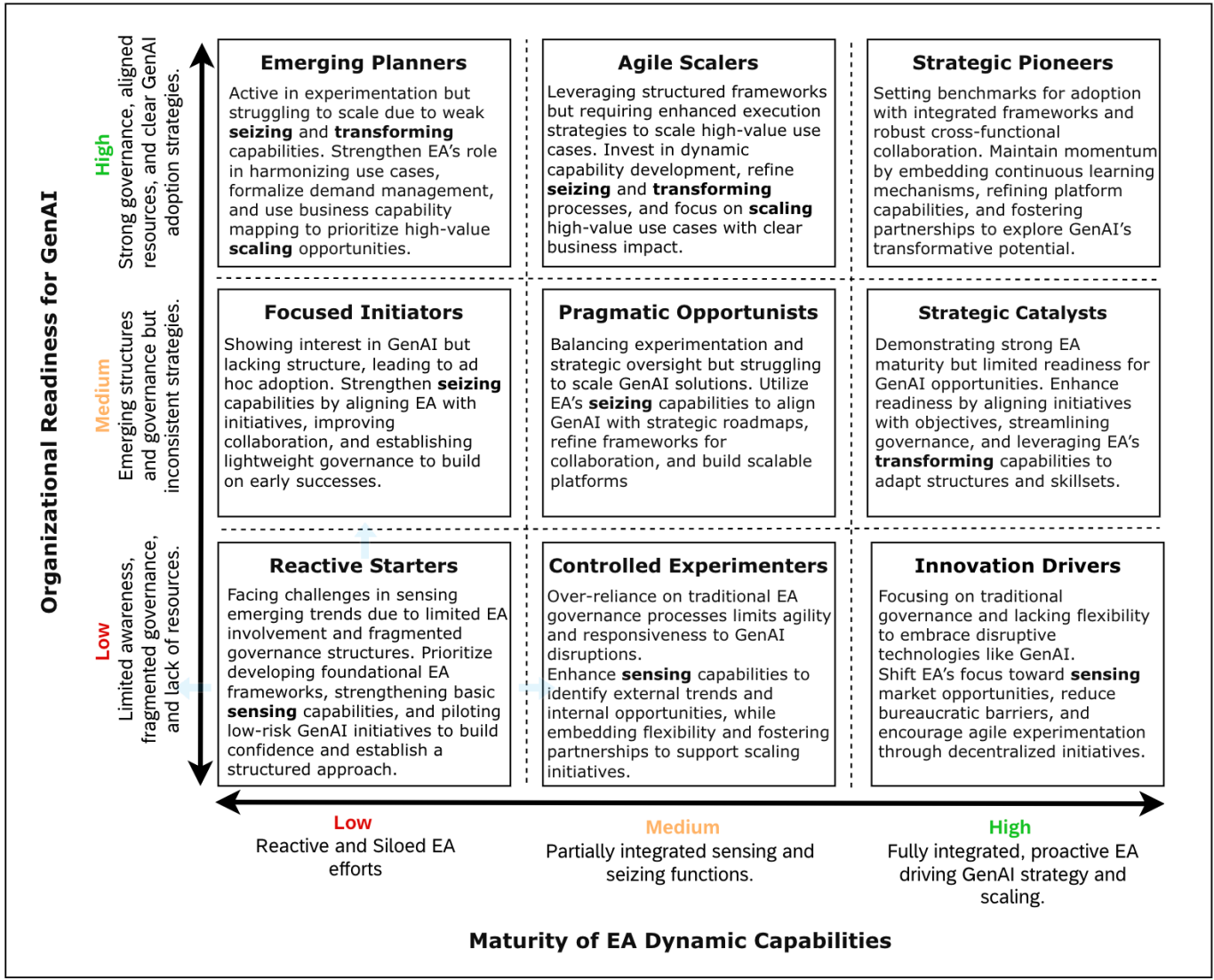

Figure 21. EA & GenAI strategic adoption quadrants: aligning Dynamic Capabilities with readiness for scalable innovation

## 6.4 Limitations and Future Work

This study acknowledges several limitations that may shape the scope and applicability of its findings.

A primary limitation lies in the reliance on a single participant (in most interviews) from each organisation to represent their perspective on GenAI adoption and EA frameworks. While participants were selected based on their expertise and professional roles, their responses may not fully capture the range of perspectives within their organisations. Although patterns of consistency were noted across similar organisational contexts, engaging multiple stakeholders (e.g. with case studies) within the same organisation could provide a more nuanced understanding of EA practices, GenAI strategies and DC maturity.

The study also focuses predominantly on organisations operating within European regulatory contexts, particularly those shaped by the EU AI Act. While this focus provides valuable insights into how regional regulations influence GenAI adoption, the findings may not be directly applicable to regions with different regulatory frameworks or organisational cultures. Expanding the research to explore GenAI adoption and EA practices in other regulatory environments, including industry specific, would help to uncover potential similarities and differences on a global scale.

Another limitation arises from the qualitative approach employed. While qualitative methods allow for in depth exploration of emerging phenomena, they lack the ability to provide the generalizability that quantitative validation offers. This limitation makes it challenging to assess the broader applicability of the proposed conceptual frameworks across varied organisational types and industries. Future research incorporating quantitative methodologies could address this gap, offering more robust insights into the relationships between EA capabilities, GenAI adoption, DC's and organisational outcomes.

Finally, this study captures the early and still evolving landscape of GenAI adoption in organisations. The findings offer a snapshot of initial challenges and opportunities, reflecting the nascent stage of integration efforts. As GenAI technologies advance and embed more deeply into enterprise ecosystems, these insights may need to be revisited and refined to remain aligned with the shifting realities of a rapidly transforming technological environment.

# Chapter 7 – Conclusion

This dissertation examines the intersection of EAM and GenAI adoption, offering insights into how EA can evolve to support scalable, value-driven GenAI integration in large organisations. Drawing on expert interviews and a systematic literature review, the research contributes to both theoretical and practical understanding, emphasising EA's role as a DC: sensing, seizing, and transforming, to address the complexities of disruptive technologies like GenAI.

The study highlights barriers to GenAI adoption, such as fragmented governance, limited large-scale use cases, and regulatory challenges posed by frameworks like the EU AI Act. It underscores the limitations of traditional EA frameworks in high-paced, innovation-driven environments and advocates for adaptive, collaborative approaches to overcome these challenges. The findings reveal EA's critical role in bridging compliance requirements with strategic opportunities, aligning GenAI initiatives with organisational objectives.

Practically, the research introduces frameworks to guide organisations in refining operating models, fostering cross-functional collaboration, and scaling GenAI sustainably. These tools provide actionable insights for improving readiness, enhancing EA practices, and aligning innovation efforts with regulatory and strategic goals.

Ultimately, this study advances the academic discourse on EAM and disruptive technologies while offering practical solutions for organisations navigating GenAI complexities. It lays the foundation for future research, including longitudinal studies to track EA framework evolution and broader analyses comparing adoption strategies across regions and industries.

# Appendices

**Appendix A. Interview Guide**

**Appendix B. Pre-Interview Questionnaire**

**Appendix C. GPT4o Sentiment Analysis Method**

**Appendix D. Summary of Literature Review Topics and RQ Validity**

**Appendix E. Interview Quotes**

# Appendix A Interview Guide

**Introduction**

| 1 | Introduce your current role & experience |
|---|---|
| 2 | Looking back, when and how did the topic of GenAI land in the EA team / the Organisation? |
| 3 | How would you define your current EA operating model? |

**Main Questions** (On average 10 questions)

| Nr. | Theme | Question |
|---|---|---|
| 1 | Operating Models & Innovation | How would you define the most suitable operating model for scaling GenAI within your organisation? |
| 2 | Barriers to adoption | What challenges have you faced in adopting GenAI in your core products (1-2 examples), and how did Enterprise Architecture help address or navigate these challenges? |
| 3 | The role of EA | How do different stakeholders within your organisation perceive and assess the role of Enterprise Architecture in GenAI initiatives? |
| 4 | EA Frameworks Adaption | What changes or adaptations to existing EA frameworks (e.g., TOGAF) have you implemented to address the scalability, transparency, and ethical requirements of GenAI? |
| 5 | Dynamic Capabilities / Sensing | How does EA leverage existing organisational capabilities to enhance GenAI adoption? (e.g., data, infrastructure, human resources). |
| 6 | Dynamic Capabilities / Sensing | How does adopting GenAI compare to other democratized technologies like SaaS, low-code platforms, or open-source tools? |
| 7 | GenAI and EA Strategy | How has the adoption of GenAI influenced the strategic alignment and business objectives of your organisation? And what role has EA played in this Business-AI alignment? |
| 8 | Governance Frameworks | What governance frameworks or best practices have you implemented to support ethical AI, sustainability, data privacy, security etc. within your EA strategy? Are these specifically aligned to the requirements of the EU AI Act? |
| 9 | Dynamic Capabilities / Seizing (Routines) | What structural or procedural changes have been introduced within EA to support the agility and flexibility required for GenAI projects? |

| 10 | Dynamic Capabilities / Seizing (Scaling) | How does EA facilitate the scaling of AI projects beyond pilot phases in terms of governance, resources, and organisational alignment? |
|---|---|---|
| 11 | Standard Taxonomy / Use Cases | Would you say that there's a common language across business units to discuss AI use cases? Has EA had a role in defining a standard naming convention for Gen(AI) use cases? |
| 12 | Dynamic Capabilities / Seizing | What initiatives are you leading to manage the organisational changes required for GenAI adoption, including workforce training, culture shift, and change management strategies? |
| 13 | Dynamic Capabilities / Transforming | More broadly, what role, if any, does Enterprise Architecture play in supporting the development of GenAI-related skills within your organisation? E.g. AI Literacy |
| 14 | GenAI on EA | In what ways, if any, do you see GenAI influencing the practice and application of Enterprise Architecture? How has this impacted the accessibility and usability of EA for individuals with varying levels of expertise? |

**Outro (more open ended):**

| 1 | Looking back, how do you reflect on the way GenAI was introduced / adopted in your organisations? |
|---|---|

## Appendix B. Pre-Interview Questionnaire

The Pre-Interview Questionnaire was shared using MS Forms, and sent to each participant prior to the interview to contextualize the discussion, and especially gain an understanding of the experience, responsibilities and GenAI maturity of the participant and the firm.

**Questions:**

| |
|---|
| How many years of experience do you have in Enterprise Architecture? |
| What is your current role? |
| Please describe the responsibilities of your current role e.g. budget, size of team etc. |
| What is the size of the EA team in your organisation? |
| What is the size of the IT team? |
| How many large scales GenAI Use cases are currently in production? |
| What is the % of business led versus IT led GenAI initiatives? |
| Which processes are driven by your team? |

EA experience on average 11 years, ranging from 3 years to 30 years

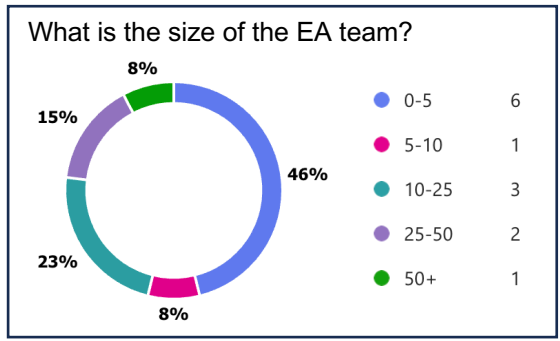
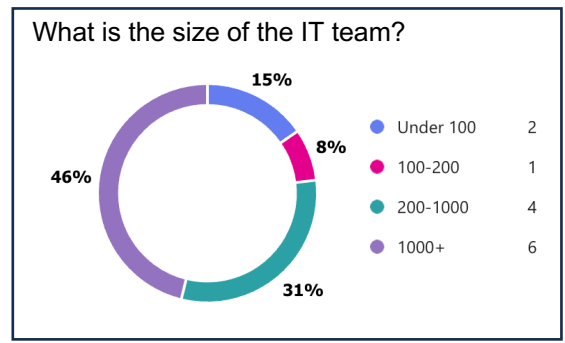
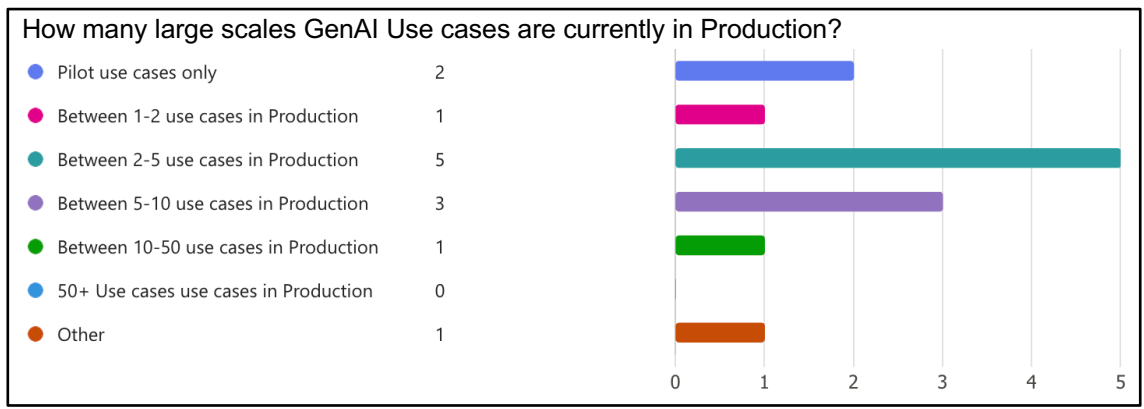

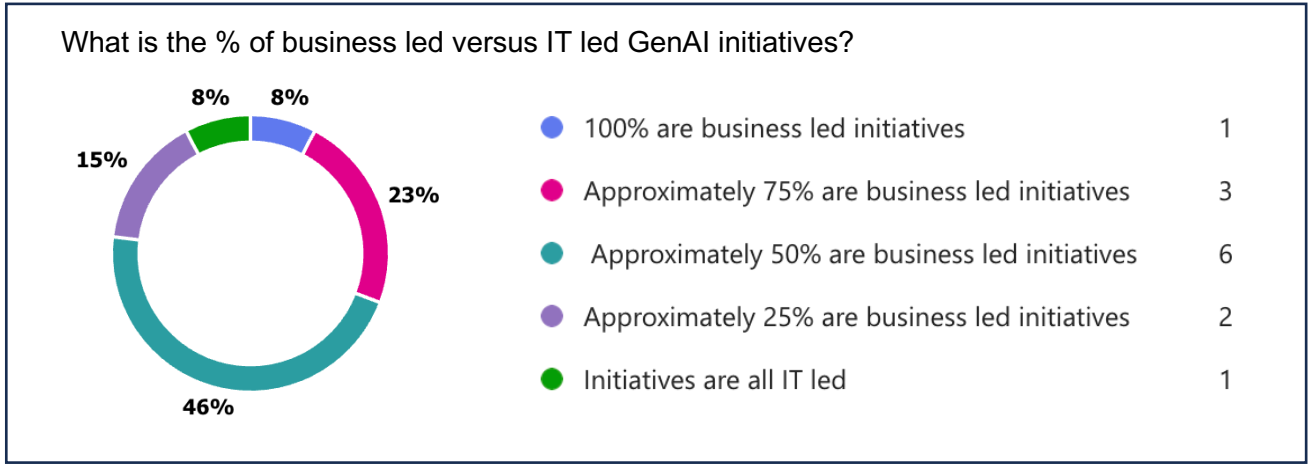

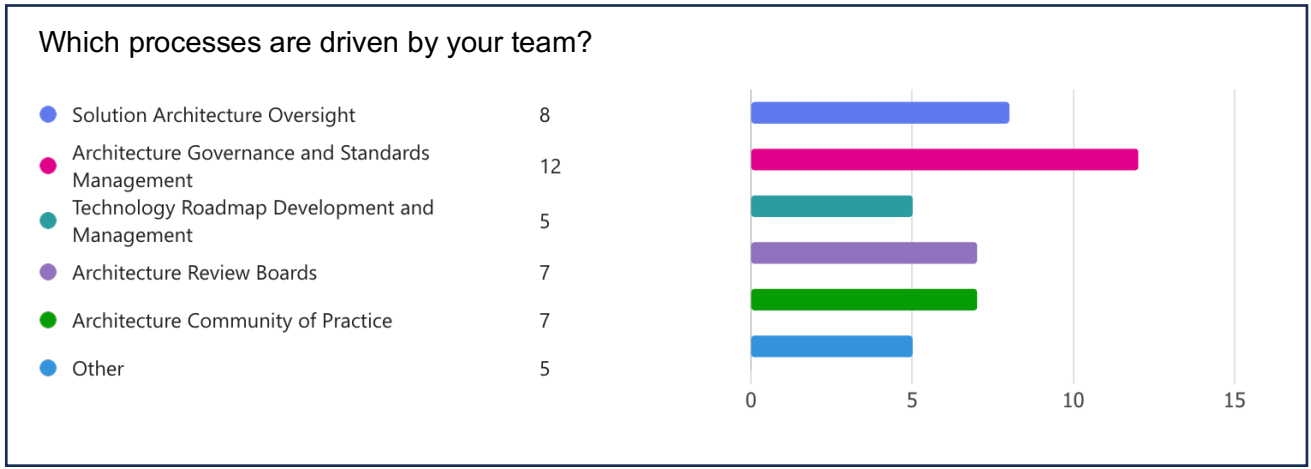

Figure 22. Results from the Pre-Interview Questionnaire

## Appendix C. GPT4o Sentiment Analysis Method

The below was generated using GPT4o (Open AI, 2025), using the following prompt:

*"Analyse the following sentiment factors in the interviews, in a scale of 1-5 with 1 being the lowest score and 5 the highest. You can determine how best to frame this for each metric. Ignore the questions from the interviewer: "Alex Ettinger" in the analysis. Show the results in a table with each number corresponding to a column:*

*1. Analyse whether the interviewee is bullish about Generative AI in his organisations (or find something better e.g. "positive" about GenAI)*

*2. Determine the level of GenAI adoption maturity of the organisation*

*3. Role of EA in Facilitating GenAI Adoption*

*4. Provide 2 sentences of reasoning for the above analysis and confidence rating of the assessment.*

*Ensure you follow a proper sentiment analysis method to get the most accurate results, showing the expected variability across each organisation. At the end, conduct a refinement exercise to calibrate your results having assessed each organisation. Present the results in a tabular format. Explain briefly how this assessment was performed, e.g. methodology. If I were to use this in a research study / dissertation, how can I introduce / use this. "*

**Results of prompt:**

Methodology Overview for Generative AI Sentiment and Adoption Assessment:

### 1. Objective

The assessment aims to evaluate the sentiment toward Generative AI, the maturity of its adoption, and the role of Enterprise Architecture (EA) in facilitating this adoption. This is done using qualitative data from interviews conducted with organisational stakeholders.

### 2. Methodological Steps

1.  **Document Analysis**:

    • Interview transcripts were reviewed systematically, focusing exclusively on participant responses while excluding interviewer inputs to ensure unbiased insights.

    • Key excerpts relevant to sentiment, adoption maturity, and EA's role were identified.

2.  **Categorization and Scoring**:

    • Responses were categorised into three metrics:

    •   **Sentiment Toward GenAI (1–5)**: Reflects optimism or scepticism about GenAI's potential, based on language tone and expressed attitudes.

    •   **GenAI Adoption Maturity (1–5)**: Indicates the organisation's progress, from initial pilots to widespread, scalable implementations.

    •   **Role of EA in GenAI Adoption (1–5)**: Assesses EA's involvement in governance, compliance, and strategic enablement of GenAI.

    • Each metric was scored on a scale of 1–5:

    •   **1**: Nascent or limited engagement.

    •   **5**: Highly advanced, structured, and integrated efforts.

3.  **Reasoning and Confidence**:

    • A brief qualitative summary was provided for each document, explaining the scores in the context of observed responses.

    • Confidence levels (Low, Medium, High) were assigned based on the clarity, depth, and consistency of participant statements.

4.  **Consolidation**:

    • Results were compiled into a structured table for comparative analysis across documents.

    • Patterns, variability, and consistency were highlighted to draw broader conclusions.

**Limitations:**

- Reliance on qualitative data, subjective scoring, or uneven depth across interviews.
- Potential ways to validate results (e.g., triangulating with quantitative adoption metrics or external benchmarks).

This methodology can help frame how qualitative data can systematically evaluate emerging technologies, offering a structured yet flexible approach to support robust academic insights.

Table 7. GenAI adoption assessment using GPT4o (Source: OpenAI, 2024).

| Participant ID | Sentiment Toward GenAI | GenAI adoption maturity | Role of EA in GenAI Adoption | Reasoning for the assessed ratings | Sentiment Analysis Confidence Level |
|---|---|---|---|---|---|
| P5 | 4 | 3 | 4 | P5 demonstrates optimistic sentiment toward GenAI's potential, with structured experimentation and EA actively supporting compliance and innovation. | High |
| P8 | 3 | 2 | 3 | P8 reflects moderate enthusiasm for GenAI, focusing on secure use case development and limited pilots. EA supports foundational governance for data security. | Medium |
| P11 | 4 | 3 | 4 | P11 highlights structured optimism and scaling challenges. EA provides governance and drives compliance-focused adoption. | High |
| P12 | 3 | 2 | 3 | P12 shows early stages of GenAI exploration with budget constraints. EA facilitates rationalization to enable innovation. | Medium |
| P17 / P18 | 3 | 2 | 3 | P17 / P18 reflects cautious optimism and hybrid adoption strategies. EA ensures strategic alignment and compliance. | Medium |
| P20 | 4 | 3 | 4 | P20 demonstrates structured enthusiasm with clear guidelines and scaling strategies. EA actively facilitates GenAI adoption. | High |
| P1 | 3 | 2 | 3 | P1 exhibits mixed sentiment, with cautious optimism toward Generative AI. Adoption is experimental, with EA playing a support role rather than leading innovation. | Medium |
| P4 | 3 | 2 | 3 | P4 is in early stages of adoption, with interest in use cases but little strategic development. EA is positioned for governance without driving transformation. | Medium |
| P15 / P16 | 4 | 3 | 4 | P15 / P16 shows positive sentiment, structured experimentation, and scaling challenges. EA plays a governance and strategy-alignment role. | High |
| P9 | 3 | 3 | 3 | P9 indicates cautious adoption, with centralised AI governance and minimal organisational scaling. EA supports policy alignment without driving innovation. | Medium |
| P13 / P14 | 3 | 2 | 3 | P13 / P14 reflects cautious optimism and early-stage pilots. EA ensures compliance and governance but is not yet central to scaling efforts. | Medium |
| P3 | 3 | 2 | 3 | P3 shows cautious optimism, reflecting slow cultural adoption of AI. EA's role is foundational in governance but limited in active leadership. | Medium |
| P6 / P7 | 4 | 3 | 4 | P6 / P7 shows positive sentiment with pilot programs and strategic scaling challenges. EA plays a governance role, supporting alignment across teams. | High |
| P2 | 4 | 3 | 4 | P2 reflects positive sentiment with structured efforts toward adoption maturity. EA actively supports governance and scaling use cases. | High |
| P10 | 4 | 3 | 4 | P10 shows optimistic sentiment about GenAI use, with efforts focused on reuse and standardization. EA actively governs the intake process. | High |
| P19 | 3 | 3 | 3 | P19 reflects balanced sentiment with challenges in modular design and reuse. EA's role is critical in enforcing governance. | Medium |

Table 8. Summary of Literature Review Topics

# Appendix D. Summary of Literature Review Topics and RQ Validity

| Topic | Summary of Literature | Validity in relation to the Research Question |
|---|---|---|
| 1. The Benefits of GenAI | GenAI drives operational efficiency, productivity, and innovation, but its impact depends on organizational readiness and capability maturity. | Highlights the outcomes that scalable and value-driven EA frameworks must support, focusing on organizational readiness |
| 2. Business Model Innovations with GenAI | GenAI fosters innovation by reshaping value creation and enabling scalable ecosystems, yet governance gaps and alignment challenges hinder sustainable adoption. | Demonstrates how EA frameworks must evolve to address sustainability and governance gaps in leveraging GenAI for business innovation |
| 3. GenAI in Innovation & Research | GenAI accelerates both exploratory and exploitative innovation, offering early adopters a competitive edge, though ethical and governance barriers persist. | Explores how EA can support innovation while mitigating ethical and readiness barriers to enable sustainable adoption. |
| 4. Unique Determinants for GenAI | Adoption depends on technological compatibility, complexity, cybersecurity risks, and ethical considerations, requiring balance between innovation and compliance. | Identifies the organizational and technological factors EA must address to support scalable and compliant GenAI adoption. |
| 5. Unique Barriers in Deployment of GenAI | Deployment challenges include transparency, resource intensity, and high costs, necessitating readiness-focused frameworks for adoption. | Highlights EA's role in creating frameworks that address deployment challenges and operational readiness. |
| 6. Perceived Technological Complexity of GenAI | GenAI's non-deterministic and opaque nature complicates transparency and governance, demanding adaptive frameworks that prioritize explainability and resource alignment. | Emphasizes the need for evolving EA practices to manage GenAI's inherent complexity and governance requirements. |
| 7. Gen(AI) Regulatory Frameworks | AI Regulatory frameworks such as the EU AI Act, emphasize explainability and compliance to ensure responsible GenAI adoption. | Underscores how EA frameworks must integrate regulatory and compliance pressures to enable scalable and compliant adoption. |
| 8. Gen(AI) Governance Frameworks | Tailored governance frameworks in the private sector must address ethical compliance, risk management, and agility to support GenAI adoption. | Examines how EA frameworks can balance operational needs with governance and ethical challenges. |
| 9. EA & Dynamic Capabilities | EA driven Dynamic Capabilities enhance agility and ambidexterity, enabling iterative resource renewal to support GenAI's experimental nature. | Explores how EA frameworks can integrate dynamic capabilities to enable scalable and adaptive GenAI adoption. |
| 10. Enterprise Architecture in GenAI | GenAI is reshaping EA by demanding new skills, tailored frameworks, and strategic alignment, emphasizing the need for sustainable adoption strategies. | Examines how EA frameworks must adapt to strategically align GenAI adoption with organizational goals. |
| 11. GenAI "in" Enterprise Architecture | GenAI's influence on EA highlights the demand for enhanced frameworks, cross-functional collaboration, and targeted integration strategies. | Identifies specific areas where EA must evolve to effectively integrate GenAI, addressing research and capability gaps. |
| 12. Philosophical Implications of GenAI | Governance challenges stem from misaligned concepts and unclear taxonomies, with institutional pressures and AI literacy influencing adoption outcomes. | Emphasizes the need for foundational clarity and adaptive governance frameworks to support EA's role in GenAI adoption. |

## Appendix E. Interview Quotes

The quotes are structured by the themes and order followed in the Results Section, where they are referenced. All quotes and relationships with participants and the topics have been uploaded in a SAP LeanIX workspace and made available in a [dashboard](dashboard). Access can be provided upon agreement with the author.

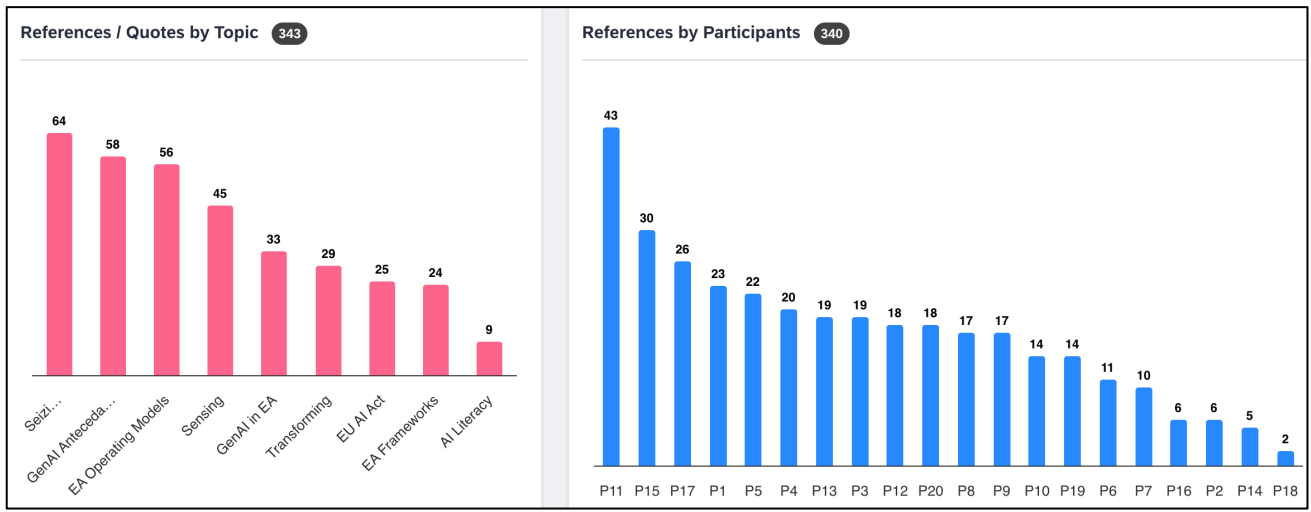

Figure 23. Statistics of references/quotes: No. of references by topic and by participant (SAP LeanIX)

### 1. Antecedents

Table 9. Interview quotes on the topic of antecedents

| Order | Participant | Quote |
|---|---|---|
| 1 | P13 | I don't think we have it yet. I think it's more or enough or more like convenience or supportive functionalities. [...] Because we're not that fast |
| 2 | P15 | it's worth saying that none of our (GenAI) solutions in the UKI […] are directly consumer facing. Some of them might be business customer facing […] You don't see that many revenue generating things most of the use cases we see are internal productivity |
| 3 | P20 | So we're still here in a phase that we're investigating on how we can actually deploy such solutions in a secure way also towards consumers. |
| 4 | P4 | We can innovate, certainly, but we've got to do that with other constraints and conditions in mind. |
| 5 | P15 | The reaction within the companies, then, certainly from the security side, is and the exact opposite. |
| 6 | P5 | So I'd say that's one of the very basic but very important and impactful solutions that we have |
| 7 | P17 | big supporter of introducing Microsoft Co-pilot, for example, so that we can advance in the context of AI usage because we have some kind of yeah negative preconceptions. |
| 8 | P1 | I think that the tipping point with Gen. AI was that now you had something that that was much broader in potential scope, because again in in these prior models, they're very narrow, right? |
| 9 | P10 | just happens to have a very broad perspective in terms of who can use it, how it's implemented? |
| 10 | P11 | generally something that can be spread everywhere. |
| 11 | P11 | So if it's functional in the end, you may have constraints. If it's not a technology for everyone, but is a technology for a specific function |
| 12 | P13 | AI topic is super big. It's not really comparable. I mean, there are similarities...RPAs you can compare it, but RPAs is super small compared to AI. |
| 13 | P19 | it's very approachable, that all the software best practises is somehow forget away. |
| 14 | P19 | GenAI is too approachable, my son can build an application. |
| 15 | P20 | I think this is something that's much bigger than with other emerging technologies. If you looked at the last 10 years back.<br>If you look at augmented and virtual reality... this was a really small fraction of the entire population |
| 16 | P1 | And the barriers to entry, both to sort of consume the technology and to use it are like super low. |
| 17 | P1 | these low code things are in a in a more localized domain |
| 18 | | P5 even trying to look through that hype, like, what's really in it for us.<br>Is this relevant for us and not to me? |
| 19 | P11 | something that will generate a big impact also and it will drive a lot of activities from a business perspective. |
| 20 | P15 | And want to start using it almost immediately. |
| 21 | P7 | The user, despite what the technical team come up, they are already experiencing AI and they are daily life, so there is a huge demand of having in place such technology. So like it's really an opportunity. |
| 22 | P8 | on a un nombre de cas d'usage et de qui relativement limité donc finalement on a pas d'effet de volume à ce à ce stade-là chez nous |
| 23 | P11 | It's super early in the in the stages that we have seen also compared to other technologies |
| 24 | P13 | we're not three years in that journey, we're just doing 1.5 years in that journey. |
| 25 | P7 | it's not yet clear the path so enterprise architecture |
| 26 | P5 | even trying to look through that hype, like, what's really in it for us.<br>Is this relevant for us and not to me? |
| 27 | P7 | So if data quality previously was important for reporting, now it is more important for generative AI. |
| 28 | P4 | partly because we've got some data challenges that we need to improve like data quality and things like that. |
| 29 | P9 | They are pretty good for the data where they are responsible for the data, where they're not responsible for, it's a mess |
| 30 | P3 | there's obviously a lot of technical folks who are very concerned about the data […] How ready are we technically? […] learning that they're going through is you really got to have your data strategy and data maturity right before you shine big torches with (Gen)AI |
| 31 | P3 | it's important to work out, the core architecture team, are they really covering the key areas of value of where the organisation wants to go in the next three to five years? Otherwise, they're not very relevant. |
| 32 | P1 | important on top of the fact that it is nondeterministic. Right. So in technology land, right, we're used to determinism, |
| 33 | P3 | It's if a lot of things you can kind of trace back and sort of open up the hood and like work things out, whereas this is not going to be as easy |
| 34 | P4 | difficulty in understanding how the answer was arrived at from an input. |
| 35 | P4 | his input went into this thing and this other thing came out. What? What does that mean? How does that affect when we need to improve? |
| 36 | P15 | The big difference from AI to other technologies is the nondeterministic output that it creates. That's the problem. That's the big difference between AI and other technologies in terms of. |
| 37 | P19 | Even you change from GPT 3.5 to 4, generally it's improved […], but does that really improve your use case? You have no clue |
| 38 | P1 | It's moving so fast that that I think you want that sort of facade pattern framework approach |
| 39 | P1 | Being able to sort of insulate end users from that gives you the ability to then deal with the fact the space is moving fast |
| 40 | P13 | The approach needs to be more flexible and open because we don't know what the status will be of AI in a year from now. |
| 41 | P15 | I don't think we really know how that's going to evolve yet |
| 42 | P19 | perfect generative AI solution, but you need to build a solution which is adaptable, which is evolving very fast because this field changing too fast, the ideas, the framework, the tech stack can change within couple of months |
| 43 | P20 | how we successfully scale it and do also the lifecycle management around it... end of the day the technology is is evolving so fast. So we need to to really think it through and how we design an architecture solution that for example we have a hot swap possibility for LLMs |
| 44 | P5 | what is different like these other examples you've mentioned, they've come slower than than this whole AI topic. I mean this, this feels a lot more like a big wave that has suddenly a bit more like a tsunami as opposed to a to a normal swell. |
| 45 | P1 | And that might take a little bit more time, but it prevents the sprawl and the risk of now.Hey, I let everyone go off and do their own thing. |
| 46 | P17 | I think at first there was some kind of initiatives and shadow IT usage of GenAI |
| 47 | P15 | That's a massive sort of sprawl. You've got lots of people doing that independently. |

| Order | Participant | Quote |
|---|---|---|
| 48 | P19 | but the environment will change very fast and no one can predict how it will change. |
| 49 | P13 | The approach needs to be more flexible and open because we don't know what the status will be of AI in a year from now. |
| 50 | P13 | so many ideas where we would just, you know, plug in the openAI API or something, but we cannot. |
| 51 | P20 | Selecting the right technology, selecting the right use cases is quite challenging |
| 52 | P20 | chatbot looks different and behaves also different and maybe has a different name...a different tone of voice.. |
| 53 | P8 | the 'Tone of Voice', making sure that a typical customer view, anything that might happen to the Customer... but in any case sensitive to customers, anything that might be addressed to the customer in terms of content that comes from GenAI, let's guarantee that we're consistent |
| 54 | P4 | It wouldn't be immediately obvious, necessarily, that there could be some downsides and negative impact ... could potentially cause customer harm |
| 55 | P6 | It's like data in AI culture building that awareness, community of people who having the same training and the same thinking about this subjects together. I think that this one of the key elements and this is started already this year. But I think it will accelerate a bit more in the coming years. |
| 56 | P17 | biggest challenge was really the interpersonal |
| 57 | P20 | which means also that the fear and the uncertainty and the need for a proper change management is much higher than with other things […] also respect the different cultural backgrounds and and specifics of different regions and geographies that we have. |
| 58 | P15 | like the end user's perspective, they don't necessarily understand that it's been trained on source material and they don't understand where that source material is. |

## 2. AI Literacy & Knowledge

Table 10. Interview quotes: AI Literacy & Knowledge

| Order | Participant | Quote |
|---|---|---|
| 1 | P4 | we haven't really put any thought into at the moment based on where we are . The literacy requirements of the EU AI Act was also new knowledge for high proportion of participants. |
| 2 | P9 | I basically thought for our company, do we need this knowledge internally? Can we monetize it? Can we get an advantage if we have that (knowledge)? |
| 3 | P13 | there is no AI literacy coming from enterprise architecture […] It is being handled by the central data & analytics team |

| Order | Participant | Quote |
|---|---|---|
| 4 | P15 | when we talk about the Gen AI, where we're getting the global team, we're getting more involved is OK our products, how are we going to embed GenAI in our products.. things we sell but also internal productivity stuff.. |
| 5 | P8 | This is what was defined at [GROUP COMPANY NAME] level and which we apply here [Subsidy] in fact. |
| 6 | P17 | EU or the CSRD reporting laws, that's quite a lot we have to do and it requires much more knowledge than before for each step we are taking. |
| 7 | P9 | So in large organisations I think business departments act more by themselves because they have the knowledge. Here they don't have the knowledge. |
| 8 | P17 | we have different points where we have our resources and our knowhow about GenAI. |
| 9 | P4 | especially Gen Z, and somehow the millennials as well and aren't invested so much in to do it by themselves. They see something cool and new technology and they want to use it and they want to use it now |

## 3. EA Frameworks

Table 11. Interview quotes: EA Frameworks

| Order | Participant | Quote |
|---|---|---|
| 1 | P17 | Check the current architecture and do complexity management and that somehow take a few hours of our work to meet these regulations that we can't use on the cases on the fields where I find the reasons why I am an enterprise architect so. |
| 2 | P12 | to understand what your businesses really wants to achieve[…] then define your transitions, stages … for GenAI. That's an issue I think, because this topic is more in this rapid prototyping kind of stage where because models are not yet mature, things are changing so much […] you need to adopt really fast. |
| 3 | P13 | I would not see this actually in the TOGAF of world to be honest. It's more guidance or enterprise architecture management topic |
| 4 | P15 | I mean, we don't really follow any framework. |
| 5 | P4 | We don't formally follow a set and off the shelf EA framework like TOGAF. |
| 6 | P14 | I'm currently making up something myself which is on the whiteboard behind me. In terms of AI taxonomy or the things that are special about AI or that need to be there extra regardless of The TOGAF model (ex. Data Scientist now in Group EA team). |
| 7 | P12 | I have not met many enterprise architecture peers… who have a strong consideration of the data domain in EA. […] And this level of granularity in terms of data is [a] must have if you want to be successful with GenAI |
| 8 | P2 | So I think it's moving from this technology focus more to the business architecture focus and being able to articulate the value proposition |
| 9 | P17 | enhance the capability to communicate and collaborate inside the company … but coming to AI act and GDPR it's not possible for a single person to do it |
| 10 | P11 | think you can clearly visualise that you are detaching from certain standards and certain ideas that or other framework |
| 11 | P10 | They just they're in a different venue that you would see typically across the company for GenAI |
| 12 | P11 | For sure it should take into account the fact that we have this kind of special cases right like we are doing |
| 13 | P15 | it's more how can we fit AI into the frameworks |
| 14 | P1 | So we wanted to give them a an accelerated way to play with new things as fast and easy as possible so that they could assess them. And of course, if we went to production within the future, it had to go through he regular channels. |
| 15 | P11 | EA perspective. Instead, it was purely in my opinion, an adjustment of existing capabilities and that readjustment to welcome, one new unique case that is this generally initiatives. |
| 16 | P11 | On other things, it's something that is on top of everything else potentially right, |
| 17 | P17 | business side on an IT infrastructure side, on the licencing side and on the cultural side … I think it depends how you define it. You can define the implementation of regulations as part of linking business and IT. |
| 18 | P5 | We're too put a few new things in place, but that's not so unusual to define new principles and guidelines specifically for these topics. |
| 19 | P17 | business and IT we have to link external influence like regulations with it and strategy on the upper level is much more point for us to to, yeah, to to implement |
| 20 | P17 | but you have to check regulations inside the business architecture and the application architecture. |
| 21 | P11 | This is why you should deserve attention without destroying the work that is already done on |
| 22 | P10 | Basic concepts of business, architecture and data architecture and infrastructure architecture are all very much there |
| 23 | P11 | TOGAF and this kind of framework should take into account the fact that now we have GenAI, but should not reinvent the framework that we have. |
| 24 | P5 | I don't see it breaking existing EA frameworks, I think it can be absorbed by existing frameworks pretty well. |

## 4. EA Operating Models

Table 12. Interview quotes: EA Operating Models

| Order | Participant | Quote |
|---|---|---|
| 1 | P6 | [Company Name] extremely federated for most of the data, we have some data or environment around some data that is becoming more standardized, but the majority of the information we have is like a black box. For most people it's really hard to scale that information. |
| 2 | P19 | And you know [Company name], we are very scattered and distributed. So, there's a lot of teams different depart different units. They actually build them build on their own. We are not sharing knowledge |
| 3 | P10 | [Company Name] has AI and a whole lot of different ways. We build things in AI for ourselves. We actually consume AI from vendors and we actually provide AI in our products that we sell to our customers. |
| 4 | P15 | But in practice we don't have much influence over where those business units spend money and what they what they build |
| 5 | P15 | we're quite a reactionary in terms of there's already an idea |
| 6 | P5 | it's more of a yeah, very liberal approach if you like. |
| 7 | P6 | I think in general the EA function is more like a support function we don't own and drive. We can initiate and we can influence people to do things |
| 8 | P16 | We wouldn't say, we've not got this AI capability, we should probably look at something in this space to try and fill that gap, even though we have that visibility, it's not something we are in a position to do |
| 9 | P16 | We're not necessarily the ones.<br>Saying in five years time this is where the business is going to be. That's that's not what EA does really. |
| 10 | P16 | probably where we are in that life cycle.<br>It's probably not quite at the stage where we can go and talk to the business owners and say, have you thought about this for your product? |
| 11 | P4 | I would have tried to be more involved in some of those things earlier on |
| 12 | P12 | you're slowing us down. Yes, because you want to have, let's say the big picture as an enterprise architect, I would say the all this practice of the history of this practice is have the perfect plan before you start building things to avoid ending up in dead ends or redundant investments |
| 13 | P1 | I'm sure it came top down because the hype around it. There's a little bit of fear missing out, folks. |
| 14 | P12 | was a driving force when also then our upper management said, well, we see massive investments with competitors. |
| 15 | P3 | And so that this in this case, it's the Chief digital officer. From also from top down. |
| 16 | P17 | So our CEO has a huge interest that we are, yeah, improve in all areas the the usage of AI. So that's this huge topic. But at the moment we didn't find the right model to be honest. |
| 17 | P18 | Our CEO has a huge interest that we improve in all areas the usage of AI. So that's this huge topic. But at the moment we didn't find the right model to be honest. |
| 18 | P10 | A (new) sister organisation which is specifically related to GenAI and AI in general, and they've been focusing primarily on becoming the centre of excellence for AI functions within the company. |
| 19 | P6 | I don't think it landed in our team at all… We've been on a journey data-driven journey for the last 5-6 years. So it's not sort of the EA function has not been driving this type of initiative. It's been sort of more business driven. |
| 20 | P10 | EA team really isn't intimately involved with that process. We basically send it over to that AI group to actually do the processing of that request. And so AI's team, basically the EA team basically is just there to provide over oversight in terms of if we believe that there's a better data model to go after or if there we have a data mesh that we'll actually have data products defined in it |
| 21 | P11 | EA is an enabler, is not driver of this kind of activities |
| 22 | P15 | (Ensuring that from a) technology point of view they're not not going off in completely different directions. |
| 23 | P5 | So we're happy to give technical guidance., legal guidance and point to a few technical structures that we that we have in place already, platforms that we have in place already. |
| 24 | P8 | with the relay of the architect's community to say where do we have opportunities and therefore in relation to the use cases, particularly those defined by [COMPANY NAME GROUP], where can we position ourselves and moreover give guidelines which are also architectural guidelines more than being in your not architectures to say some level of requirement that we define collectively? |
| 25 | P12 | In my case, enterprise architecture was positioned as kind of governance body. |
| 26 | P13 | our EA team is less operational and really more on governance or architecture really |
| 27 | P15 | predominantly would be pointing them at areas where they need to go to to get some things like compliance |
| 28 | P10 | So what we do is we say here is the place you should go get product data. Here's where you should go get customer data so you don't have everybody kind of have their own flavour. So we're also defining from an EA perspective, where those data models exist and what what datas in each one of those? And then how do you reuse those? |
| 29 | P12 | We founded this department, so they also concentrate on this topic..we want to incubate. And in order to to keep track, we really need to be little bit disruptive in the way how we work. |
| 30 | P15 | So there was a lot of resetting expectations. |
| 31 | P11 | so our team specific team growth with two different source enterprise architecture [...]<br> Last year we started with basically four people, OK now if you consider externals these the the coming year, we expect to reach probably a number of 10 people |
| 32 | P11 | have a kind of internal startup to lead this because it's a new topic, no one is was able to handle it… we are no more focusing only on the classic enterprise architecture… but also to take a look on disruptive initiatives |
| 33 | P19 | I believe the only advantage we have is to say (that) the only way we can catch up with the new technology is we develop everything together. |
| 34 | P12 | It really depends how the enterprise architecture practice is positioned in the company. In my case, enterprise architecture was positioned as kind of governance body. |
| 35 | P4 | we've got one of the domain architects is has expressed an explicit interest in being the kind of AI go to person as a particular area of interest for him. So he's actually leading on the application portfolio assessment side of things around the AI. |
| 36 | P15 | As such, when we talk about the GenAI, where we're getting the global team, where we're getting more involved isOK our products, how are we going to embed GenAI in our products and that set our products, things we sell but also internal productivity stuff. So how are we going to use it for internal productivity |
| 37 | P8 | This (GenAI Typologies) is what was defined at [GROUP COMPANY NAME] level and which we apply here in fact. |
| 38 | P14 | There's a high degree of of compliance and transparency and reporting needs required for everything that we do |
| 39 | P6 | So sometimes we are direct involved depending on the size and the importance of some initiatives. But many times this is very much sort of happening inside elsewhere in the organisation, we are very Federated when it comes to how we are organised and how we drive things. |
| 40 | P10 | we need to look at EAs and stay generically across the entire company. So that's what we do. Our EAs are focused on. It could be a life cycle for the company. It could be a function, but not a specific tactical function like Gen AI with which we see it's kind of a very tactical process. |
| 41 | P10 | Where we put GenAI and products we sell our customers for example, that's a whole different team that does that and I'm not engaged in that piece at all. |
| 42 | P7 | we have shipped an AI strategy that evolves on building capabilities, like the tool for automation that helps the general internal population, but also on some specific case building some competitive advantage…the success so faron the building some capabilities for everyone or building some specific use cases for our services. |
| 43 | P12 | It was quite some things we where we invested a lot and and and there where we as enterprise architects were able to contribute because in our enterprise architecture model and set up this was let's say considered early on. |
| 44 | P8 | the fact of structuring implementation typologies to be clear about the type of things you are doing, what you are doing within a solution and you are developing, you are ultimately enriching this solution with GenAI capabilities. |
| 45 | P8 | I have a step to take, which is to say, for these needs, am I going to go and find data outside what is currently available from the publisher, for example, am I going to give access to our vector databases, to put it like that, to do RAG from data that is not available from publishers? You have this type of thing, which needs to be structured and potentially, we are entitled to enrich the typology in fact. |
| 46 | P7 | So I think maybe the more the expansion of today framework, I think what I'm seeing is like creating model new framework so like framework for.Ethics framework for responsible AI technologies |
| 47 | P1 | So we wanted to give them a an accelerated way to to play with new things as fast and easy as possible so that they could assess them. |
| 48 | P11 | This is why we have to reconnect later on to the standard way of working of the entire company. |
| 49 | P13 | So it's it's a parallel compliance topic. Yeah, it's new for us now.S o therefore, then we in this actually in the gate zero we identify that very, very quickly and very early. So it it cannot always be solved in, in, in in our standard processes we sometimes do really enterprise architecture, working groups and. |
| 50 | P10 | They do go through an incubation process where they they kind of come up and brainstorm some ideas and then after that then they go through an actual AI board that reviews all of the different scenarios and options |
| 51 | P11 | Have a community to engage, have a dedicated steering committee to discuss these kind of things at C level. I'm not talking about minor people C level. There is a weekly Steering committee. |
| 52 | P15 | If a use case is being submitted, it has to go through a certain initial AI Council, so to speak, |
| 53 | P3 | Stage gates and forums and how can we leverage what we've got? |
| 54 | P5 | the one thing that we have put in place is as part of the AI governance structure (is) that we a responsible AI board…We have representatives from senior management, so that's typically board level |
| 55 | P9 | (After ChatGPT integration) they wrote an AI policy based on that, which says that every AI functionality has to be examined. For specific criteria, those criteria are in LeanIX (EA Tool). |
| 56 | P3 | And now they're kind of going all need an AI (Governance framework).<br>(How) we need to consolidate this down and just have one (framework) and and and that needs to be dynamic and bring in the right people |

## 5. EU AI Act

Table 13. Interview quotes: EU AI Act

| Order | Participant | Quote |
|---|---|---|
| 1 | P13 | From (the EU AI Act) coming…we are changing our operating model. We have changed it…there is a big AI swim lane in it. |
| 2 | P13 | We have the same EU act as an example (EA in alignment with each BUs) |
| 3 | P5 | we take the EU AI act basically as our base level..we build everything around that and then if there are in other countries or regions specific legislations that exceed that or require something different then we would call that top ups.. |
| 4 | P5 | (High Risk Applications EU AI Act) brings a lot of documentation and technical set up and all these aspects into the picture. So that would require us to do a lot of new work and establish some additional capabilities that we don't have in the organisation today. |
| 5 | P5 | Is around governance, risk management compliance on this topic which you know in light of the EU AI act is, you know, very prominent in this topic, much more so than in than other previous topics. |
| 6 | P7 | the EU AI act actually is asking also to review...what kind of use case this initiative will do. And this is quite new to to the previous legislation..a new kind of element and then it's really important to try by pushing such assessment earlier in the in the development phase or in the procurement selection. |
| 7 | P9 | They they do the assessment, they they rate the risk and put in a risk description LeanIX and they rate the benefit…from an AI act perspective… we also created AI score based on the data classification that is used in the AI |
| 8 | P17 | it's quite hard for (the business) to identify if my tool is using AI according to the EU AI Act and that's ... much more complex than GDPR |
| 9 | P3 | There's going to be some technical people, but there also needs to be a bring in of people from a lot of different angles across the company. |
| 10 | P19 | because it's evolving, so it's easy in in the future you continue to build this compliance (modular) component. |
| 11 | P12 | with the concrete frameworks for the Gen. AI topics in terms of how to develop, how to set up how to manage because I don't have almost no insight there, to be honest. |
| 12 | P8 | And I don't have much information on this subject to say how (EU AI Act) compliance is guaranteed. |
| 13 | P15 | we do look at AI risk, but I'm not aware of that (EU AI) Act being considered so that we probably need to look at that. |
| 14 | P17 | EU AI Act. An we need to have the right spending documentation about our use cases. |
| 15 | P5 | So yes, we're documenting all these use cases.<br>We're assessing them also with legal and basically legal looks at these use cases from a EU AI perspective as well. |
| 16 | P7 | we have prepared initially a plan to comply that they will start by inventoring what are all the AI use cases. |
| 17 | P17 | The major challenge is to define what is AI and what isn't, and to identify if someone… is using GenAI |
| 18 | P11 | since there is the EU AI act that is coming in place... companies like PwC that should do the audit are starting to ask, show me which are all your applications... that are using GenAI functionalities or ML functionalities. |
| 19 | P13 | there is the whole EU AI Act topic which came (to EA). |
| 20 | P17 | We we start to preparing. We as a team because I realise nobody else is caring about it now and they will start to care about it in, I think in nine to 10 months and then it's somehow too late to implement. |
| 21 | P7 | as you have like actually group company or any way decentralised multi country company, one of the challenges understanding how many and what kind of AI initiative you have... so starting by mapping what AI system we have. |
| 22 | P9 | The EU AI act, where the the the general managers and the directors also are responsible for some sort of usage on AI applications. So they have to be involved at some point and he made sure everyone is involved on the higher level and built a small group task force consisting of country managers. |
| 23 | P9 | So we have we have a process established that clearly defines which AI use cases are presented to management and which use cases can be decided on project level. |
| 24 | P13 | from a legal point of view... lawyers sitting with the AI act and defining criteria that... some application owner would have to answer in order to determine, what we have to do in terms of compliance, in terms of now low risk, medium risk, high risk… |
| 25 | P5 | We're assessing them also with legal and basically legal looks at these use cases from an EU AI perspective as well. |

## 6. GenAI in EA

Table 14. Interview quotes: GenAI in EA

| Order | Participant | Quote |
|---|---|---|
| 1 | P17 | The potential the potential for EA is quite big, I think, because EA is a field that isn't affected that much at the moment by GenAI. |
| 2 | P2 | I expect here rather high impact |
| 3 | P2 | EA would have maybe a higher impact because you do not always need kind of experts to explain situations. |
| 4 | P9 | they have a lot of probably a lot more better use cases than I have to manage their portfolio because it's too big. Like you can't manage 18,000 applications in LeanIX. That's basically impossible to to to gain results that have a lasting impact. I think if you if you're not doing it AI assisted, you don't have a chance. |
| 5 | P9 | I might be even willing to pay for that feature because it saves a lot of time and discussion. |
| 6 | P1 | (Today) the model doesn't understand the semantics of my enterprise architecture. Technically, if I have a graph of that data, I have the semantics right? So, like figuring out how to use that and you see some of these things, right where graphrag is starting to come, where where, like the LLM is is using then I think then you'll see more interesting things |
| 7 | P15 | bot to ask you the questions of the stuff that I want to know rather than having to understand it. Actually I think I could move a lot faster. I think eventually I'll get to a point where using a chat bot to summarise the document and ask it questions on it will almost be as social expectation |
| 8 | P17 | Stuff that somebody can just ask questions into into this and that's array where enterprise actually management will be shifted in communication issues |
| 9 | P4 | I think there's a lot of potential for virtually a as a chat agent, that's, we've got the knowledge context of your own art, your organisation's architecture model. So people come often come up to me and ask a question about what's who uses this application or what's this application used for. |
| 10 | P5 | So if we have a knowledge base that we can expose and people maybe prefer to to speak to a chat board rather than us, that's that might be one of the use cases. But I think it's probably more on that informational level rather than on. |
| 11 | P9 | But in the end, of course, chat bot and teams who owns application XYZ? Perfect. |
| 12 | P11 | one is for example text to diagram. So describe with text a potential architecture and automatically have something represented |
| 13 | P1 | Instead of saying like we do today, where it's like I have to go into this application and tick something to say whether this is GDPR compliant, right? |
| 14 | P15 | So under help to understand, you know, what's the requirements for GDPR? What are the requirements of PCI? if you ticked certain boxes to say, OK, well, what applies, what GDPR applies the AI Act applies, PCI applies and then you can ask it questions on, OK, what sort of controls do I need to put in place based on these standards that would be useful. |
| 15 | P7 | coming with suggestions about where can we identify low hanging fruits for rationalization based on the capability map and the number of systems. |
| 16 | P1 | almost every organisation I've ever been in has some sort of architecture review, |
| 17 | P11 | project manager without having a lot of technical skills is able to provide a high level diagram quickly. |
| 18 | P17 | knowledge base and on the other hand I can draw diagrams by Gen AI based on the knowledge base and combine different sources because |
| 19 | P1 | I think that's super powerful because again…they're spending a lot of time doing these sort of not high value tasks, right? They've got architecture views, but they're looking at diagrams. |
| 20 | P11 | reducing the workload… very complicated that in a smaller company you have the figure of enterprise architect. |
| 21 | P17 | thinking about complex issues and develop solutions and target architectures abroad with the whole documentation part will be hopefully much more |
| 22 | P4 | I think it's more about how do we coexist with that role or with that technology longer term, and where do we distribute responsibilities… how does that allow us to scale? |
| 23 | P1 | And I wanted to ask questions like, yeah, which areas can I innovate quickly in or be agile in right? |
| 24 | P11 | project manager without having a lot of technical skills is able to provide a high-level diagram quickly. |
| 25 | P2 | shift more to the value perspective business and value. |
| 26 | P11 | on one side you will democratise on the other side, you will free up times for value driven activities.. |
| 27 | | but the EA said we have to use this and that because he wants to. But now then I can say based on a recommendation on the IT technology landscape… |
| 28 | P17 | And that's a point where we can get more importance in the company because we see that we have to put all the information together to get a single source of truth. And improving decentralised decision making |
| 29 | P2 | I cannot sit in each and every decision board, but if it's decentralised and with the help of GenAI, I believe that the decentralised decision making can be supported in the sense of what the global EA team is thinking or what the intentional architecture is about. |
| 30 | P2 | They don't challenge enough, right.. and you could say OK, then you know we'll design it based on us. |
| 31 | P11 | And it may lead to a higher adoption of enterprise architecture and a higher adoption of EA practic |
| 32 | P15 | it's more discovery stuff and I think that's where I guess AI doesn't really help because |
| 33 | P15 | So you know things we're delivering now. Yes, you will get, you know, as part of the CICD pipeline you will get SBOMs, you'll have something to interrogate. A lot of our challenges, the legacy environment, so it's, you know it's cobalt mainframe code that we've written in the 70s, it's like, well, you're not going to get an SBOM for that. |

## 7. DC Sensing

Table 15. Interview quotes: DC Sensing

| Order | Participant | Quote |
|---|---|---|
| 1 | P8 | It's not strictly a data and AI topic; it's a business topic, and that's where enterprise architecture positions itself, focusing on adoption and guidelines. Ultimately, we obviously work, and the operational arm is data and AI, both to gather requirements—and this is where transversality comes into play—so architecture takes on meaning. This is to finally gather and identify, within value chains, where the opportunities lie. |
| 2 | P4 | In awareness, we noticed that there was a a greater interest for Gen. AI type features and functionality within the organisation.<br>There's actually a couple of individuals who sit within business teams that are quite interested in looking at how they can leverage it to improve their productivity or whatever the case might be within those particular areas |
| 3 | P11 | we try to understand how to deal with the new processes that our colleagues were setting up and try to identify the touchpoint with our processes and readjust the system based on that |
| 4 | P4 | In this particular area, rather than just going with a blank version of the capability assessment and saying right guys, where do you think things would be productive or you know in increase user experience or not have anything at all? |
| 5 | P1 | It's all knowing all understanding thing and I think it's part of EA as job right to to say to the business, let me educate you a little bit. |
| 6 | P11 | So the potential customers that we're not aware that they were potential customers.<br>They are aware that there is something new that they can use. They can, they can create ideas. |
| 7 | P3 | But there's a whole lot more there to get a business readiness and education up to speed. |
| 8 | P1 | And then that team would actually do a market scan. Right, you have to be watching the market for new things. Maybe there's more of an investment on staying plugged into what the innovations are… trends in the industry and innovations, and help connect the dots for you, plant those seeds. |
| 9 | P12 | in this one we always had the feeling it's, let's say moving with with Lightspeed |
| 10 | P4 | from people having an interest in technology and technology trends |
| 11 | P5 | Who did some research. Who, you know, tried to understand how things work and who the players are and so on and so on. And, you know, was very obvious that this is a relevant topic. |
| 12 | P1 | And these things hit the market and there's lots of uninformed people that want to chase it. And I think that's where EA needs to come in and say, hey, let's actually break this thing down. |
| 13 | P14 | So in terms of enterprise architecture, I'm looking at, I'm trying to find out a reasonable taxonomy for what we need to capture about here, gene, our use cases or applications… some templates for common use cases like a RAG chatbot or so |
| 14 | P1 | P1 You want more people out there in the field trying to say hey, like, let's go through what you do and.<br>Let's see where this is applicable, right?<br>Maybe there's more folks out there doing that piece to see for, you know, opportunities that AI can benefit, right |
| 15 | P12 | P12 Looking into or identifying (business) use cases. |
| 16 | P15 | can kind of recommend stuff and I think that's probably the most valuable part of enterprise architecture with AI in my opinion. |
| 17 | P3 | So one of the things we'll be doing is working through, OK.<br>Here's the here's a menu of use cases, and you may well business team suddenly go oh, I didn't realise that technology could help me with that use case. That use case is quite close to the kind of challenge I've got. Suddenly there's value. |
| 18 | P5 | And then you know, seeing that in a more.<br>Many of these experiments, let's call them you get a sense for whether technology really stands and what's what's feasible and what today is maybe not feasible |
| 19 | P8 | We've done it in certain areas to identify where there are opportunities. Essentially, it's about saying, functionally, you see, if I put myself in LeanIX, even if it's not fundamentally your topic, it's about determining collectively where we can define opportunities—specifically, the level of opportunity that GenAI represents for our business. |
| 20 | P14 | So I'm trying to think up dimensions where we capture this so that whatever happens in a few years from now, we can answer these questions. |
| 21 | P1 | You do a quick scan of your organisation, say, hey, these are four or five places where I really think this is going to be beneficial. |
| 22 | P12 | those those GenAI topics and find a place in our business capability model, so in our strategic alignments. |
| 23 | P3 | It's sort of like, oh, here's a nice thing. We'll put it on the shelf and then forget about it. So it's like how to make sure that it really resonates with different areas and use cases is a good way to go. |
| 24 | P6 | So we get that holistic view, all the initiatives I think with that in hand and with the strategy forward, where do we think we might have potential good use cases. We have a great, a really good business capability map of the organisation. So the potential would be there to assess which capabilities are important for [Company Name] US and which ones of those do we think would have more benefits of GenAI or other things than other more generic capabilities. Maybe there are some low hanging fruits there or when we do sourcing should? |
| 25 | P4 | Of AI potential aspect on both applications and business capabilities to help us get that intersectional, where's the real value?<br>You know, high potential up and a high potential business case business capability. |
| 26 | P8 | To say, how can I ultimately map out the potential or the future? The implementation of GenAI within our organisation, essentially where do I have it, where do I not, and perhaps tag it? |
| 27 | P9 | from an EA perspective, AI like we don't have our capabilities rated by this is way AI can help because I think AI can help anywhere. Basically of course you maybe have some sort where you make money and you think but but this could be way better. If it's automated or some stuff but.<br>Yeah, we don't really do this sort of it it comes up in the process of assessing the business problems basically |
| 28 | P12 | if you have a business capability and you want to do GenAI there, but we only have systems of record, is that a smart thing to invest in? Is it strategically meaningful? |
| 29 | P20 | harmonising a bit, the experience will become a big topic going forward, I assume. |
| 30 | P11 | Something on Adobe suite that is introducing a lot so you can imagine the extremely reduced heterogenous landscape that we may have. So this is where we are trying to reconnect and where there is the relevant position of EA |
| 31 | P19 | If we stay in this low development like try to build RAG.<br>Microsoft will come up with standard RAG or they already come up very fast if we stay in this way, we will never it will be very difficult. However, Microsoft will never build a customer so so when when the Microsoft move up the consulting need move up. |
| 32 | P11 | Ability to dynamically grow EA resources to support these disruptive events |
| 33 | P15 | How can we try to even form some sense, standard, let alone even what the standard is, but just kind of that view of where will this take our business in five years? |
| 34 | P8 | it's about trying to stay ahead of the curve so we don't just react to the dynamic but instead provide frameworks and guidelines. I think this challenge …is something that at least partially falls within the role of EA… understand what's happening when vendors come in with their new offerings |
| 35 | P11 | So when this initiative of GenAI reaches a certain stage of maturity, so there are no more pilots or PoCs, but there is a willing to really scale the technology and implement it in the company, in that moment we will pass to a different phase in the governance model that we have |
| 36 | P5 | So when this initiative of GenAI reaches a certain stage of maturity, so there are no more pilots or PoCs, but there is a willing to really scale the technology and implement it in the company, in that moment we will pass to a different phase in the governance model that we have |
| 37 | P1 | You want more people out there in the field trying to say hey, like, let's go through what you do and.<br>Let's see where this is applicable, right? |
| 38 | P5 | That's something I wish I we would have had this year already |
| 39 | P20 | How can we try to to even form some sense standard, let alone even what the standard is, but just kind of that view of? Where will this take our business in five years? |
| 40 | P3 | That kind of filtering is really useful.<br>That would save a lot of pain for a company just going and I'll just grab that and hope for the best you've had. |
| 41 | P11 | So maybe you engage with one consulting that is providing 1 consultant to create the proof of concept and so on. |
| 42 | P6 | OK, so we have some centralised in a way like solution, but we as as like Thomas said, the enterprise architecture team has not one been one of the the driver of such initiative has been involved. |
| 43 | P9 | And if we build it ourselves, like Switzerland, they just do it. They have an external supplier and he said, well, we can do it with Azure open AI. We can do all the magic on the invoices. And we said, well, we'll just do it perfect. |
| 44 | P8 | To form our own convictions about the requirements we can establish, particularly regarding vendors and external partners, for the implementation of this type of solution |
| 45 | P6 | There is like an appetite of AI from their business leader, but also the knowledge about AI and data is not as like the the business leader will like to. So we are as an example, we have committed with also external university and provider to do executive AI training. |

## 8. DC Seizing

Table 16. Interview quotes: DC Seizing

| Order | Participant | Quote |
|---|---|---|
| 1 | P11 | this is where the position of enterprise architecture in my opinion is critical, because to avoid the uncontrolled creation of and adoption of GenAI technologies. |
| 2 | P20 | I'm responsible for acceleration and industrialization, which means providing the frame first of all, for the boundaries in which we navigate and which we build use cases and then onboard technologies. |
| 3 | P20 | Incubating and driving technologies that have a transformative nature and and also you know scale it to a certain extent in in in the organisation. |
| 4 | P11 | That's so because the nature of the initiative is completely different. The nature of the initiative is different because at the very beginning you don't know what you want in term of technology |
| 5 | P11 | (Other Transformation Team) are collecting demand. They are trying to understand those demand, discussing with legal and information security. |
| 6 | P15 | I know, from a technical point of view, what the business units are doing, but there is still the disjoint that actually the the product owners and the domain owners, the solution architects don't see until somebody's come up with an idea.So it would be getting closer to those people |
| 7 | P11 | So as enterprise architect, we should be there when we have transformation of big system, but also when we are introducing new GenAI initiatives |
| 8 | P13 | people are having an eye on the architecture from the very beginning, so that we don't add technical debts |
| 9 | P6 | opportunity decentralised and where we try to be on the top to understand where, where are those initiatives coming from.<br>Through what we call, it's a funnel, you know, initiative funnel or projects when they are being captured, we try to interact there to understand, OK, what are the business cases, use cases and basic understand that what type of and that's that is more from a governance point of AI governance. You know what type of AI do you intend to utilise in here? |
| 10 | P11 | As one man showed the GenAI part to one year and alpha go, the only discussion that we had around these topics were trade off emails...are capturing the demand in an Excel file in SharePoint |
| 11 | P17 | we are we get for the software introduction process |
| 12 | P11 | So when this initiative of GenAI reaches a certain stage of maturity, so there are no more pilots or PoCs, but there is a willing to really scale the technology and implement it in the company, in that moment we will pass to a different phase in the governance model that we have. And in that moment basically we will trigger the activation of the initiative as a standard project |
| 13 | P10 | And its primary focus is to look at reusability and prioritisation of the resources. We don't let people go off and build their own AI models across the company |
| 14 | P13 | now there is also a gate 0 |
| 15 | P13 | I would say we can also count like every every week 1-2 projects we see we review such projects |
| 16 | P17 | Since we are working in the governance for the software selection process and we are somehow accepting or software requests we see that there's a lot of people want to include GenAI |
| 17 | P11 | On right driving the use cases, the adoption but is no more on our side. It was shifted over to the IT department that is managing the digital workplace… So it was then the scaling in terms of number of users was done by the service management.. |
| 18 | P16 | We have a with a lot of stuff of trying to shift left and it might be that actually shifting left isn't a great idea… is it right to keep … an AI function separate? Is it better to build a AI CoE and keep your knowledge of that space rather than trying to make everybody experts? |
| 19 | P11 | We should capture which application we are creating and which technologies we are using. So this is where we have the touchpoint or where we can really keep the control on what's happening on the other side because alternatively, imagine if we don't have the governance process to centralised the demand from the 150 market |
| 20 | P12 | let's look at those things where we have systems of innovation already in place. Can we further leverage them? |
| 21 | P14 | sort in their use cases into that schema. I think that's our job top down. |
| 22 | P8 | The challenge of EA is to try to address and approach every (GenAI) use case in the same way. |
| 23 | P11 | And you should imagine that having one unique backlog of ideas for the entire company and then be sure that locally there is no one that is implementing something without passing the information to the globa |
| 24 | P4 | doing an assessment of the app portfolio… and compare use cases across departments. We haven't formalised anything yet. |

| 25 | P9 | there's no big larger initiative that says every country has to use (Gen)AI for the invoices (use case) |
|---|---|---|
| 26 | P16 | what kind of what technologies we're looking into for certain capabilities as well as duplication of because a lot of this technology is so new like any new technology, it's everybody just runs at it and you can kind of project the next five years of are we just going to end up with complete sprawl of different models with different techniques with different RAGging and all that sort of stuff |
| 27 | P20 | Think it through now you know the first. The first department has a great idea for customers and launch it so the the sales department, the customer service department in two weeks from now launches the next chat bot right then then you have the marketing department and said, OH, chat bot is a brilliant idea. So we launched also one at some point in time you create a chatbot hell also for the for the customer. |
| 28 | P20 | So we launched (multiple chatbots), and at some point in time you create a chatbot hell also for the for the customer |
| 29 | P10 | Flexera and Intune as two key discovery engines that we use to identify where software is located and functions. We also the ability to block SaaS capabilities if necessary if we find something out there and we've done that in several cases where someone's using a technology that we didn't wasn't approved. And so we discovered it and started removing it off their devices themselves |
| 30 | P10 | They actually have a confluence site where they're documenting all of the different technologies that they use in the use cases for those, and then they have a portal where they can actually actually acquire that technology, we have a we call [Company Name] Digital cloud. It basically is a think of an Amazon marketplace, but within [Company Name], then we can actually go and buy this and take one of these and take one of those and it will configure on your platform for you and then you can then use that for your Gen AI. |
| 31 | P10 | think of an Amazon marketplace…we can actually go and buy (GenAI use cases)…and it will configure on your platform for you |
| 32 | P11 | not only global, but also cross. It's an horizontal platform compared to the functional and regional domains, right. So basically you are you are in the right place to both have the touchpoint with the right person. So who has the idea who should approve it from a budget perspective, who should assess it from our architectural perspective, who should assess it from an information security perspective and legal perspective, right? |
| 33 | P17 | My experience at [COMPANY NAME] otherwise, otherwise we won't be able to implement good AI solutions on a on a large scale and with the right usage |
| 34 | | GenAI has unique characteristics that raises the bar on culture related topics e.g. high applicability but need to understand limitations of LLM technology for example e.g. language/tone of voice… |
| 35 | P11 | this is where the position of enterprise architecture in my opinion is critical, because to avoid the uncontrolled creation of and adoption of GenAI technologies |
| 36 | P20 | when once this is done, I think then we come into the scaling part, which means really driving us across markets, increasing number of users, maybe driving it across functions as well. |
| 37 | P17 | If we see some kind of proof of concept and we are included in the proof of concept part, we are directly thinking about, OK how can you use it on a large scale if it is successful. |
| 38 | P15 | The the the business direction is led by the profit centres, so they're the ones that say my my product is going to do this and they're very siloed. So our role as EA is more joining people together. |
| 39 | P19 | That means so for the consultant (developing GenAI use cases).. they have no way to share it so we use this as as a show what we have developed and it's all instantly deployed on Prem, to cloud, Kubernetes to so it's no environment barrier. |
| 40 | P6 | Where is AI being utilise d inside organisation?<br>We can visualise that because we're having that Global Insight of it.… creating some dashboards, we can report back to senior stakeholders that this is the focus where we're utilising (Gen)AI and we can maybe see also if they are successful or not. |
| 41 | P19 | In China, we say that if you want to cut the tree, you should sharpen your knife first. |
| 42 | P20 | it's about building a technology, a solid technology foundation and and a streamlined technology foundation that we use to build this capabilities and provide also the guidelines how they should be built. |
| 43 | P3 | (Build a) environment where people can experiment and learn. |
| 44 | P19 | That's one part. Another part is generative AI. It's highly repeated. If you say a lot of people do redundant work. How many company have built more than 20 different chat bot?<br>What's different was with those chat bot. It's very minor, so it's very highly repeated a lot of function you can actually share. |
| 45 | P3 | It's it is difficult when you got a small area, a small team.<br>To to really scale out because there's so much activity going on around the organisation. |
| 46 | P5 | And I think that the, you know, the topic that you refer to is you know when they get out of this POC stage and things get more real business relevant business impactful. And require a different maturity level, different approach and all that's something we have not really put in place yet and that's something I would want to change. |
| 47 | P15 | So we're we're getting to the stage where we're deploying different regions, so you know you've got a uk deployment but it's OK, but it's still supported from a team in California, so that doesn't necessarily work that wel |
| 48 | P7 | And in this case, the challenge is OK to scale from maybe one country to a division or or adoption.. adoption maybe for specific group like legal as some legal generative AI assistant and so on. So like yeah, when you have a target group is easier to understand the the success |
| 49 | P12 | The beauty of this case was data protection laws in the US are quite different than in Europe. Guess what? So it was also more easy to to implement that in a compliant way compared to Europe or UK or Germany,, |
| 50 | P12 | I mean, that's great to learn.<br>But that also means you can hardly transfer those use cases from the US to Europe afterwards.<br>We would need to redesign them, at least in in certain areas, |
| 51 | P15 | You can kind of say, well, Spain did this cool thing with this technology.<br>And this region built this, and it tends to be that gives you quite a cool perspective |
| 52 | P20 | where we also still need to learn on how important is it how much regionalisation is needed and how to deal with it. Also from an architectural design perspective. |
| 53 | P8 | It is more of a collaboration challenge—how we build within a constrained framework that has its own level of development… From my perspective, there is nothing consolidated today on the topic to effectively manage GenAI at the enterprise level, and I believe this is an area where we need to make progress as well. |
| 54 | P15 | operationalization of the product as in how is it supported is sitting in California, that's got to change. So that and that operational knowledge of how to do it has got to move into the regions because otherwise you can't deploy it globally. |
| 55 | P15 | even when you work with central function, you aren't quite sure who's to go to for certain things. So I'm not quite sure how this AI is really going to help you or or how you, how you've even got the context to tell the AI what to respond with |
| 56 | P8 | Here, we're actually doing RAG without a vector database because we don't have the volumes, and we know it's likely to be a significant cost. So, at least in the preliminary phases, we're not necessarily using that type of solution, which is something we also share at the group level. |
| 57 | P18 | Preparing everything that it works on a business scale, a business side on an IT infrastructure side, on the licencing side and on the cultural side.So that he is, uh talking a lot to the uh, different stakeholders in the company.And he leads the discussions that leads us to to rethink ourselves. |
| 58 | P17 | Talking a lot to the uh, different stakeholders in the company.<br>And he leads the discussions that leads us to to rethink ourselves. |
| 59 | P20 | slowly onboard people to this, because otherwise they will push back. |
| 60 | P3 | We're going to empower you to go and trial some of these things and understand what it means for you |
| 61 | P17 | (Head of EA) kind of has change management (for GenAI) in his job description. |
| 62 | P1 | sort of enforcing that enterprise capability mindset, right. |
| 63 | P1 | You have to have them have the relationship. Of course, the business people need to actually respect the view of the EA and and see that there. They have a best interest in the in the business.<br>And be able to connect the dots for them. |
| 64 | P9 | And you need to build a bit trust and you need to take baby steps with them to get them on your page to have some sort of discussion, but it's sort of old school EA business and afterwards we have the ability to scale AI use cases, but we need to build trust first internationally. So we are global in a sense, but we need to work on our global cooperation. |

## 9. DC Transforming

Table 17. Interview quotes: DC Transforming

| Order | Participant | Quote |
|---|---|---|
| 1 | P4 | What I'm currently struggling with is having the acceptance and understanding at more senior levels. |
| 2 | P3 | really fortunate though that we've got this sort of like board director who can say no this EA function has a role to play in all of this and we need to get get it integrated. |
| 3 | P16 | But now as that margin squeeze happens and standards gets pushed from the very top, more and more you will have more influence over that. |
| 4 | P11 | have people that are going and evangelising a bit on GenAI going and explaining what we are doing, which are the processes creating this kind of content to make people aware, try to push them to use the right process and so on |
| 5 | P6 | And I think maybe because data is important, but a the AI, everybody's keen to utilise AI that will help us to drive the data awareness. |
| 6 | P3 | These are the use cases that, particularly with Gen. AI as well. If you quite good at marketing that and then also that you can back it up and sort of like well, if that's something you want to pursue. |
| 7 | P11 | We are doing it through the system (EA Tool) |
| 8 | P5 | And I would say also better communication to drive adoption. |
| 9 | P11 | it's pretty powerful the position today because they are not only managing the incoming demand but also doing great work to go and evangelise a little bit. |
| 10 | | who's desperately wants to get a product out, you can just be seen as a blocker to going to getting revenue in some specs |
| 11 | P3 | EA suffers (from) a marketing issue or branding issue. It's it's when you got that sort of trust in that brand building up that then start you start getting invited to, OK, now the executive want to do proper strategy planning. |
| 12 | P13 | my impression is usually that people should be more aware of it and use it more because well, it reduces overhead and makes things more efficient. |
| 13 | P8 | Honestly, it's quite outside the group, outside of IT, really—it's invisible, isn't it? |
| 14 | P19 | I'm still confused of what enterprise architecture is |
| 15 | P4 | we paid a consultancy to come in and talk to the upper echelons about something that we then might contradict or or have a different view about without having a a sort of good understanding of where our thinking.. |
| 16 | P4 | I'd rather we were part of the conversation until we can actually fulfil what we would like to do in that area, but it is equally frustrating to know that actually, we could probably be of value here, but we're not seen as being that. |
| 17 | P3 | challenge there in that you've got to change the culture in your own teams because there's a lot of sort of architects who are used to firefighting at the end of an initiative. |
| 18 | P12 | some people was taken on this journey and so some of my architects went to into this department as well and changed their their hats, their roles. From enterprise architect to platform architect for example. |
| 19 | P9 | That that go you need new people which are younger or have that knowledge of of or users actually come from good organisations. |
| 20 | P12 | but rather educate business people. |
| 21 | P13 | What also is always a topic is we are asking what kind of AI solution. |
| 22 | P5 | the topic of of training is something that we are triggering right now |
| 23 | P9 | he's giving the trainings in the whole organisation. |
| 24 | P9 | And it's it's not a regulation, it's the the people's mindset and thinking. And you can only move that in your organisation with trainings and examples, how it helps and you have to remind the people as well that. |
| 25 | P20 | we have conducted over 100 awareness session worldwide from 20 people to 4000 people in terms of audience. |
| 26 | P20 | you need to invest heavily into, into upscaling education and change management |
| 27 | P19 | GenAI is the main offering in the group level…. So then we're working together with [Other BUs] So there's plenty of group build there like. |
| 28 | P11 | we should have people that should be able to put their hands on it and improve what we need internally without continuously going outside and asking additional money and so on. We should invest money internally. |
| 29 | P3 | there's definitely a bit that can't be underestimated there in terms of skilling up and being able to really leverage those capabilities |